\documentclass[3p]{elsarticle}

\usepackage{lineno,hyperref}
\usepackage{placeins}\usepackage{tikz}\usetikzlibrary{shapes,shadings}
\usepackage{algorithm}
\usepackage[noend]{algpseudocode}
\usepackage{pgfplots}
\pgfplotsset{compat=newest}
\pgfplotsset{plot coordinates/math parser=false}
\newlength\figureheight
\newlength\figurewidth

\usepackage{subcaption}     
\usepackage{tabularx}
\usepackage{arydshln}

\usepackage{amsmath}
\usepackage{amsthm}
\usepackage{amsfonts}
\usepackage{setspace}

\definecolor{myred}{RGB}{128,0,0}
\definecolor{myblue}{RGB}{0,0,128}
\definecolor{mygreen}{RGB}{0,128,128}
\definecolor{myorange}{RGB}{155,100,0}
\definecolor{mygray}{RGB}{155,155,155}

\modulolinenumbers[1]
\usepackage[normalem]{ulem}

\newtheorem{remark}{Remark}

\newcolumntype{H}{>{\setbox0=\hbox\bgroup}c<{\egroup}@{}}

\journal{Elsevier}

\begin{document}
\begin{frontmatter}

\title{Time-periodic steady-state solution of fluid-structure interaction and cardiac
flow problems through multigrid-reduction-in-time\footnote{This
work was performed under the auspices of the U.S. Department of Energy
by Lawrence Livermore National Laboratory under Contract DE-AC52-07NA27344,
LLNL-JRNL-820515.}}

\author[Stuttgart,SimTech]{Andreas Hessenthaler\corref{corauth}}
\ead{andreas.hessenthaler@gmail.com}

\author[LLNL]{Robert D.\ Falgout}

\author[UNM]{Jacob B.\ Schroder}

\author[KCL]{Adelaide de Vecchi}

\author[KCL,UM]{David Nordsletten}

\author[Stuttgart,SimTech]{Oliver R\"ohrle}

\address[Stuttgart]{Institute for Modelling and Simulation of Biomechanical Systems,
University of Stuttgart, Pfaffenwaldring 5a, 70569~Stuttgart, Germany}

\address[SimTech]{Stuttgart Center for Simulation Technology, University of Stuttgart,
Pfaffenwaldring 5a, 70569 Stuttgart, Germany}

\address[LLNL]{Center for Applied Scientific Computing, Lawrence Livermore National Laboratory, Livermore, CA 94551}

\address[UNM]{Department of Mathematics and Statistics, University of New Mexico, Albuquerque, NM 87131}

\address[KCL]{School of Biomedical Engineering and Imaging Sciences, King's College London, 4th FL Rayne Institute, St.~Thomas~Hospital, London, SE1 7EH}

\address[UM]{Department of Biomedical Engineering and Cardiac Surgery, University of Michigan, NCRC B20, 2800 Plymouth Rd, Ann~Arbor, 48109}

\cortext[corauth]{Corresponding author.}

\begin{abstract}
    In this paper, a time-periodic MGRIT algorithm is proposed
    as a means to reduce the time-to-solution of numerical algorithms
    by exploiting the time periodicity inherent to many applications
    in science and engineering.
    The time-periodic MGRIT algorithm is applied
    to a variety of linear and nonlinear single- and multiphysics problems that are periodic-in-time.
    It is demonstrated that the proposed parallel-in-time algorithm
    can obtain the same time-periodic steady-state solution as sequential time-stepping.
    It is shown that the required number of MGRIT iterations
    can be estimated a priori and that the new MGRIT variant can significantly and consistently reduce
    the \emph{time-to-solution} compared to sequential time-stepping,
    irrespective of the number of dimensions, linear or nonlinear PDE models,
    single-physics or coupled problems and the employed computing resources.
    The numerical experiments demonstrate that the time-periodic MGRIT algorithm
    enables a greater level of parallelism yielding faster turnaround,
    and thus, facilitating more complex and more realistic problems to be solved.
\end{abstract}


\end{frontmatter}
\newif\iffigure
\figuretrue

\section{Introduction}\label{introduction-sec}
Modern computer architectures enable complex and detailed numerical investigations
of various problems of interest in science and engineering.
Many of these problems include spatial and temporal components
that need to be sufficiently resolved to answer important research questions.
A significant subset of such models share the presence of periodicity in the time dimension.
For example, a simulation model of the human heart may model the time-periodic contraction
of the heart tissue due to electrophysiological signals, while a flow model governs
the time-periodic evolution of fluid flow in the heart's chambers
due to influx, e.g., from the pulmonary veins
and the ejection of blood through the aortic valve.

While parallelization techniques (such as the domain decomposition method,
e.g., \cite{Roache1978,Roache1995,QuarteroniValli1999})
for reducing the time-to-solution for such numerical models
are well-known,
well-established and (for the most part) straightforward to apply for the space dimension,
parallelization techniques for the time dimension are less wide-spread
and less well-known as their spatial counterpart, especially in the application-driven research communities.
With the first explorations into time-parallelism in the work of Nievergelt in 1964~\cite{Nievergelt1964},
so-called parallel-in-time methods have seen a rise and fall in interest for the decades to follow,
e.g., see the excellent review articles by Gander~\cite{Gander2015}
and Ong and Schroder~\cite{OngSchroder2020}.
More recently, the introduction of massively parallel hardware
has led to situations where spatial speedups saturate before all available hardware can be employed,
such that interest in parallel-in-time algorithms has started increasing steadily.

Nowadays, time-parallel algorithms have seen uptake in a wider range of research fields.
Many such parallel-in-time methods exist, e.g.,
\cite{LubichOstermann1987,VandewalleVandevelde1994,HortonVandewalle1995,
FarhatChandesris2003,FarhatCortialDastillungBavestrello2006,
ChristliebMacdonaldOng2010,SpeckRuprechtKrauseEmmettMinionWinkelGibbon2012,
EmmettMinion2012,HamonSchreiberMinion2019,LionsMadayTurinici2001,
FriedhoffFalgoutKolevMaclachlanSchroder2012,FalgoutFriedhoffKolevMaclachlanSchroder2014}
and researchers continue to develop and propose new techniques.
On the other hand (and similar to numerical algorithms that fall into other categories than time-parallelism),
a growing number of researchers develop convergence analysis for their algorithms
to enable well-informed parameter choices and thus improve parallel efficiency,
e.g., see \cite{GanderJiangSongZhang2013,BenedusiHuppArbenzKrause2016,
FriedhoffMaclachlan2015,FriedhoffSouthworth2019_preprint,KrzysikDesterkMaclachlanFriedhoff2019,DesterckFriedhoffHowseMaclachlan2019,
DobrevKolevPeterssonSchroder2017,Southworth2019,HessenthalerSouthworthNordslettenRoehrleFalgoutSchroder2020,
GanderLunet2020,SouthworthMitchellHessenthalerDanieli} and many others.

The number of parallel-in-time techniques that were successfully applied to time-periodic
problems is smaller; however,
a time-periodic waveform relaxation-type method was introduced
in the work of Vandewalle and Piessens~\cite{VandewallePiessens1992}.
Other algorithms fall into two principal categories:
algorithms that employ a periodic-in-time
coarse-problem, e.g.,~\cite{BenedusiHuppArbenzKrause2016,GanderKulchytskaruchkaSchoeps2018,
ChristopherFalgoutSchroderGuzikGao2020,ChristopherGaoGuzikFalgoutSchroder2020},
or a non-periodic-in-time
coarse-problem, e.g.,~\cite{SongJiang2014,GanderKulchytskaruchkaSchoeps2018,GanderKulchytskaruchkaSchoeps2020}.

In this work, we extend the multigrid-reduction-in-time (MGRIT) algorithm
for the class of applications that exhibit periodicity in the time dimension
using a non-periodic coarse-problem approach.
While the proposed method is agnostic to the underlying PDE,
the applications in this work stem from the biomedical engineering field.
In Section~\ref{methodology-sec},
we commence by introducing the (non-periodic) MGRIT algorithm
and develop its time-periodic variant.
In Section~\ref{numerical-experiments-sec},
we introduce three application classes covering linear and nonlinear flow and fluid-structure interaction problems
in two and three space dimensions.
We assess the time-periodic MGRIT algorithm for each of these applications in Section~\ref{results-sec},
where we compare the time-periodic steady-state solution
obtained by using sequential time-stepping and time-periodic MGRIT.
We further explore the cost-to-accuracy ratio using analytic solutions
and demonstrate the achieved runtime reduction using time-periodic MGRIT
(up to $37$x speedup)
over a range of small- to medium-sized parallel computing hardware
(up to $256$~processors).
We conclude by discussing the benefits of the proposed algorithm and its limitations in Section~\ref{discussion-sec}.
\section{Methodology} \label{methodology-sec}
In this section, we discuss the sequential time-stepping solution of a general time-dependent PDE
and introduce MGRIT as a parallel-in-time solver that solves the same space-time problem iteratively.
We then introduce an extension to standard MGRIT
for the solution of general nonlinear systems of PDEs that exhibit time-periodic behavior.
By adapting the MGRIT solution process to the time-periodic property of the considered system,
we propose to exploit this property at the solver level
as a means to exploit the full potential for parallel speedup.
In general, MGRIT is a true multilevel method (compare with, e.g.,
spatial and algebraic multigrid methods~\cite{BrandtMccormickRuge1985,BriggsHensonMccormick2000,HackbuschTrottenberg2006}).
For ease of presentation, however, we restrict ourselves to the two-level setting.
Nevertheless, the time-periodic MGRIT algorithm's design encompasses
the true multilevel setting~\cite{Hessenthaler2020_PhD}.
\subsection{Sequential time-stepping}
\label{mgrit-sequential-time-stepping-sec}
First, consider the general form of a time-dependent PDE in one spatial dimension:
\begin{equation}
    f (x, t; u, \partial u / \partial x, \partial u / \partial t, \partial^2 u / \partial x^2, \ldots) = 0
    \qquad \text{for } (x, t) \in \Omega \times [0, T].
    \label{general-form-pde-eqn}
\end{equation}

Discretizing Equation~\eqref{general-form-pde-eqn} in space and time, sequential time-stepping
can be written as,
\begin{equation}
    \boldsymbol{u}_n = \boldsymbol{\Phi} \left( \boldsymbol{u}_n, \boldsymbol{u}_{n-1} \right) + \boldsymbol{g}_n
    \qquad \text{for } n = 1, \ldots, N_0 - 1,
    \label{general-pde-phi-form-nonlinear-eqn}
\end{equation}
with approximation $\boldsymbol{u}_n$ at time $t_n$,
and forcing function $\boldsymbol{g}_n$
for $N_0$ time points, $0 = t_0 < t_1 < \ldots < t_{N_0 - 1} = T$, and final time $T$.
Here, $\boldsymbol{u}_n, \boldsymbol{g}_n \in \mathbb{R}^{N_x}$
with $N_x$ spatial degrees-of-freedom.
Further, $\boldsymbol{\Phi}$ is referred to as the \emph{time-stepping operator}.

For linear problems, Equation~\eqref{general-pde-phi-form-nonlinear-eqn} can be written as,
\begin{equation}
    \boldsymbol{u}_n = \boldsymbol{\Phi}_n \boldsymbol{u}_{n-1} + \boldsymbol{g}_n
    \qquad \text{for } n = 1, \ldots, N_0 - 1.
    \label{general-pde-phi-form-linear-eqn}
\end{equation}

Further, $\boldsymbol{\Phi}_n = \boldsymbol{\Phi}_0$, if a fixed spatial discretization
and equidistant time points $t_n = n \cdot \delta_0$ are assumed for $n = 0, \ldots, N_0 - 1$
and time step size $\delta_0 = T / (N_0 - 1)$.
Then, Equation~\eqref{general-pde-phi-form-linear-eqn} can be written in matrix form,
\begin{equation}
    A_0 \boldsymbol{u} =
    \begin{bmatrix}
        I \\
        -\boldsymbol{\Phi}_0 & I \\
              & -\boldsymbol{\Phi}_0 & I \\
              &       & \ddots & \ddots \\
              &         &       & -\boldsymbol{\Phi}_0 & I \\
    \end{bmatrix} \begin{bmatrix}
        \boldsymbol{u}_0 \\
        \boldsymbol{u}_1 \\
        \boldsymbol{u}_2 \\
        \vdots \\
        \boldsymbol{u}_{N_0 - 1}
    \end{bmatrix}
    = \boldsymbol{g},
    \label{mgrit-intro-space-time-system-eqn}
\end{equation}
with space-time matrix $A_0 \in \mathbb{R}^{N_x N_0 \times N_x N_0}$
and space-time solution vector $\boldsymbol{u} \in \mathbb{R}^{N_x N_0}$.
Now, sequential time-stepping can be identified as a \emph{block-forward solve}
of~\eqref{mgrit-intro-space-time-system-eqn}.
\subsection{Multigrid-reduction-in-time (MGRIT)}
\label{mgrit-time-grid-hierarchy-sec}
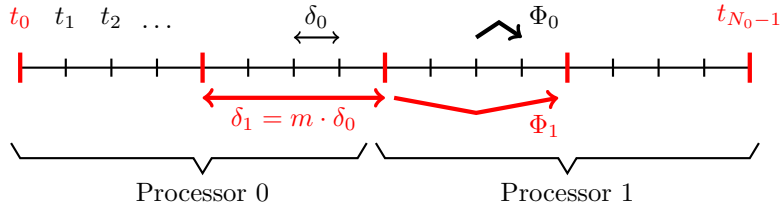
\begin{figure}[ht!]
    \centering
        \begin{tikzpicture}[scale=0.8]
        \def\gf{1.75};
        \def\gc{0};
        \def\nx{16}
        \def\df{0.75}
        \def\m{4.0}
        \def\dc{\df * \m}
        \def\add{0.1*\df}
                \draw[thick,black] (0,\gf) -- (\nx*\df,\gf);
                \draw[ultra thick,red] (0*\dc,\gf-0.25) -- (0*\dc,\gf+0.25);
        \draw[ultra thick,red] (1*\dc,\gf-0.25) -- (1*\dc,\gf+0.25);
        \draw[ultra thick,red] (2*\dc,\gf-0.25) -- (2*\dc,\gf+0.25);
        \draw[ultra thick,red] (3*\dc,\gf-0.25) -- (3*\dc,\gf+0.25);
        \draw[ultra thick,red] (4*\dc,\gf-0.25) -- (4*\dc,\gf+0.25);
                \draw[thick,black] (1*\df,\gf+-0.15) -- (1*\df,\gf+0.15);
        \draw[thick,black] (2*\df,\gf+-0.15) -- (2*\df,\gf+0.15);
        \draw[thick,black] (3*\df,\gf+-0.15) -- (3*\df,\gf+0.15);
        \draw[thick,black] (5*\df,\gf+-0.15) -- (5*\df,\gf+0.15);
        \draw[thick,black] (6*\df,\gf+-0.15) -- (6*\df,\gf+0.15);
        \draw[thick,black] (7*\df,\gf+-0.15) -- (7*\df,\gf+0.15);
        \draw[thick,black] (9*\df,\gf+-0.15) -- (9*\df,\gf+0.15);
        \draw[thick,black] (10*\df,\gf+-0.15) -- (10*\df,\gf+0.15);
        \draw[thick,black] (11*\df,\gf+-0.15) -- (11*\df,\gf+0.15);
        \draw[thick,black] (13*\df,\gf+-0.15) -- (13*\df,\gf+0.15);
        \draw[thick,black] (14*\df,\gf+-0.15) -- (14*\df,\gf+0.15);
        \draw[thick,black] (15*\df,\gf+-0.15) -- (15*\df,\gf+0.15);
                \draw[above,red] (0*\df,\gf+0.5) node {$t_0$};
        \draw[above,black] (1*\df,\gf+0.5) node {$t_1$};
        \draw[above,black] (2*\df,\gf+0.5) node {$t_2$};
        \draw[above,black] (3*\df,\gf+0.5) node {$\ldots$};
        \draw[above,red] (16*\df,\gf+0.5) node {$t_{N_0 - 1}$};
                \draw[<->,thick,black,below] (2*\dc-1*\df,\gf+0.5) -- (2*\dc-2*\df,\gf+0.5);
        \draw[above,black] (2*\dc-1.5*\df,\gf+0.5) node {$\delta_0$};
        \draw[<->,ultra thick,red,below] (1*\dc,\gf-0.5) -- (2*\dc,\gf-0.5);
        \draw[below,red] (1.5*\dc,\gf-0.5) node {$\delta_1 = m \cdot \delta_0$};
                \draw[<-,ultra thick,black,below,bend right] (3*\dc-1*\df,\gf+0.5) -- (3*\dc-1.5*\df,\gf+0.75) -- (3*\dc-2*\df,\gf+0.5);
        \draw[above,black] (3*\dc-0.5*\df,\gf+0.5) node {$\Phi_0$};
        \draw[<-,ultra thick,red,below,bend right] (2.95*\dc,\gf-0.5) -- (2.5*\dc,\gf-0.75) -- (2.05*\dc,\gf-0.5);
        \draw[below,red] (3*\dc-0.5*\df,\gf-0.625) node {$\Phi_1$};
                \draw[-,thick,black,below] (-0.05*\dc,\gf-1.25) -- (0*\dc,\gf-1.5) -- (0.95*\dc,\gf-1.5) -- (1.0*\dc,\gf-1.75) -- (1.05*\dc,\gf-1.5) -- (1.85*\dc,\gf-1.5) -- (1.9*\dc,\gf-1.25);
        \draw[below,black] (1*\dc,\gf-1.75) node {Processor~$0$};
        \draw[-,thick,black,below] (1.95*\dc,\gf-1.25) -- (2.0*\dc,\gf-1.5) -- (2.95*\dc,\gf-1.5) -- (3*\dc,\gf-1.75) -- (3.05*\dc,\gf-1.5) -- (4*\dc,\gf-1.5) -- (4.05*\dc,\gf-1.25);
        \draw[below,black] (3*\dc,\gf-1.75) node {Processor~$1$};
        \end{tikzpicture}
        \caption{Two-level decomposition of the time grid with time points $t_n$
            into F-points (black, only on fine grid) and C-points (red, only on coarse-grid)
            with fine grid time step size $\delta_0$ (level~$0$)
            and coarse grid time step size $\delta_1$ (level~$1$).
            Here, the temporal coarsening factor is given as $m = 4$
            and the time stepping operators are denoted as $\Phi_0$ and $\Phi_1$.}
    \label{time-grid-hierarchy-fig}
\end{figure}

MGRIT solves~\eqref{mgrit-intro-space-time-system-eqn} iteratively using a linear or near-linear number of block
operations~\cite{HessenthalerSouthworthNordslettenRoehrleFalgoutSchroder2020}. That is, MGRIT is an $O (N_0)$
method~\cite{FriedhoffFalgoutKolevMaclachlanSchroder2012,FalgoutFriedhoffKolevMaclachlanSchroder2014},
similar to sequential time-stepping.
To achieve parallelism in the temporal domain, MGRIT introduces two\footnote{For a detailed introduction
of MGRIT for an arbitrary number of time grids, the interested reader is referred
to~\cite{FriedhoffFalgoutKolevMaclachlanSchroder2012,FalgoutFriedhoffKolevMaclachlanSchroder2014}
and the developer's manual of the {XBraid} library~\cite{XBraid},
an open-source implementation of the MGRIT algorithm.}
(or more) time grids,
accompanied by a coarse-grid correction scheme based on multigrid reduction.
The fine grid (or time grid on level $\ell = 0$) is composed of all time points
$t_n$ (for $n = 0, \ldots, N_0 - 1$) and the coarse grid (or time grid on level $\ell = 1$)
is derived from a uniform coarsening of the fine grid, see Figure~\ref{time-grid-hierarchy-fig}.
The temporal coarsening factor is denoted as $m \in \mathbb{Z}^+$,
such that the number of time points on the coarse grid is given by $N_1 = ( N_0 - 1 ) / m + 1$,
with corresponding time step size~$\delta_1 = m \delta_0$ and time-stepping operator $\Phi_1$.
On the fine grid, time points are partitioned into F-points and C-points, where C-points refer to the time points
also present on the coarse grid and F-points only exist on the fine grid, see Figure~\ref{time-grid-hierarchy-fig}.

Instead of applying the fine-grid time-stepping operator $m$ times per coarse-grid time step,
MGRIT approximates the exact coarse-grid time-stepping operator,
\begin{equation}
    \Phi_1 \approx \Phi_0^m,
\end{equation}
in order to reduce complexity.
A Schur complement decomposition of $A_0$ yields \emph{ideal} multigrid restriction and interpolation operators.
That is, ideal restriction from level $0$ to $1$  is given as,
\begin{alignat}{4}
    R &= \begin{bmatrix}
        I \\
        & \Phi_0^{m - 1} & \Phi_0^{m - 2} & \cdots & I \\
        &                        &                                &           &   & \ddots \\
        &                        &                                &           &   &        & \Phi_0^{m - 1} & \Phi_0^{m - 2} & \cdots & I \\
    \end{bmatrix} \quad \in \mathbb{R}^{N_x N_1 \times N_x N_0},
    \label{Rl-eqn}
\end{alignat}
and restricts the approximate solution at C-points
from the fine grid to the coarse grid.
Ideal interpolation from level $1$ to level $0$ is given as,
\begin{alignat}{4}
    P &= \begin{bmatrix}
        [I, \Phi_0, \cdots, \Phi_0^{m - 1}]^T \\
        & [I, \Phi_0, \cdots, \Phi_0^{m - 1}]^T \\
        && \ddots \\
        &&& I \\
    \end{bmatrix} \quad \in \mathbb{R}^{N_x N_0 \times N_x N_1},
    \label{Pl-eqn}
\end{alignat}
and interpolates the approximate solution
from the coarse grid to the C-points and its neighboring F-points
on the fine grid.

Note, that in general $\Phi_\ell \neq \Phi_\ell^T$
and thus $P \neq R^T$~\cite[Equation (28)]{HessenthalerNordslettenRoehrleSchroderFalgout2018}.
Then, the ideal\footnote{The coarse-grid operator $R A_0 P$ is ideal
since one step on the coarse grid is performed by $m$~fine grid time steps,
which yields the exact solution with respect to the fine grid.
Without approximating the operator $R A_0 P$, the method would be as expensive,
which motivates the approximation in Equation~\eqref{RlAlPl-eqn}.}
multigrid coarse-grid operator, $R A_0 P$ approximates the space-time matrix $A_1$
on level~$1$~\cite{DobrevKolevPeterssonSchroder2017},
\begin{align}
    R A_0 P
    &= R_I A_0 P
    &= \begin{bmatrix}
        I \\
        - \Phi_0^m & I \\
        & - \Phi_0^m & I \\
        && \ddots & \ddots \\
        &&& - \Phi_0^m & I \\
    \end{bmatrix}
    \approx \begin{bmatrix}
        I \\
        -\Phi_1 & I \\
                   & -\Phi_1 & I \\
                &               & \ddots & \ddots
    \end{bmatrix}
    = A_1 \in \mathbb{R}^{N_x N_1 \times N_x N_1}
    \label{RlAlPl-eqn}
\end{align}
where the first equality is given by Equation~\eqref{RlAlPl-exact-eqn}
and restriction by injection is given by the operator,
\begin{alignat}{4}
    R_I &= \begin{bmatrix}
        I \\
        & 0 & 0 & \cdots & I \\
        &   &   &        &   & \ddots \\
        &   &   &        &   &        & 0 & 0 & \cdots & 0 & I \\
    \end{bmatrix} \quad \in \mathbb{R}^{N_x N_1 \times N_x N_0}.
    \label{RIl-eqn}
\end{alignat}

$R_I$ in Equation~\eqref{RIl-eqn} has a similar block structure as $R$ in Equation~\ref{Rl-eqn},
but with all blocks $\Phi_1^k$ (for $k = 1, \ldots, m - 1$) set to zero.
Thus, $R_I$ restricts the C-points from level~$0$ to level~$1$, omitting the respective F-points.

Coarse-grid correction reduces error at C-points
and is thus coupled with a process referred to as \emph{F-relaxation} to reduce error at F-points.
Algebraically, this is an application of the idempotent operator,
\begin{equation}
    F = P R_I.
    \label{F-relaxation-eqn}
\end{equation}

When F-relaxation does not sufficiently reduce error at F-points,
a stronger relaxation scheme can be used, called FCF-relaxation.
FCF-relaxation refers to subsequently applied F-, C- and F-relaxation steps.
C-relaxation updates every C-point based on taking one time step from the corresponding previous F-point.
Algebraically, this corresponds to an application of $C = I - R_I^T (R_I A_0 R_I^T)^{-1} R_I A_0$.
FCF-relaxation can then be written as,
\begin{alignat}{4}
    F C F = P (I - R A_0 P) R_I.
    \label{FCF-relaxation-eqn}
\end{alignat}

F-relaxation reduces error at the F-points following a given C-point,
whereas C-relaxation reduces error at the C-point following the previous F-point;
thus, both can be seen as block Jacobi applied to the respective rows in $A_0$.

Now that we have the basic ingredients for MGRIT, we can describe a standard two-level cycle in the following pseudocode:
\FloatBarrier
\begin{algorithm}
  \caption{Pseudocode for two-level MGRIT algorithm, adapted from \cite{DobrevKolevPeterssonSchroder2017}.}
  \label{linelas-MGRIT-V-alg}
  \begin{algorithmic}[1]
    \item[]
    \Repeat
      \State Relax on $A_0 \boldsymbol{u} = \boldsymbol{g}$ using F-/FCF-relaxation with $\boldsymbol{\Phi}_0$. \Comment{{\small In parallel}}
      \State Compute coarse-grid residual $\boldsymbol{r}^\Delta  = R (\boldsymbol{g} - A_0 \boldsymbol{u})$.\label{linelas-MGRIT-V-restriction-alg}
      \State Solve coarse grid correction problem $A_1 \boldsymbol{e}^{\Delta} = \boldsymbol{r}^\Delta$
      using $\boldsymbol{\Phi}_1$. \label{linelas-MGRIT-V-recursive-step-alg}
      \State Correct solution at fine-grid C-points with $\boldsymbol{e}^{\Delta}$. \Comment{{\small In parallel}}
      \State Update the solution at the F-points with F-relaxation and $\boldsymbol{\Phi}_0$. \Comment{{\small In parallel}}
    \Until{norm of residual is small enough}
  \end{algorithmic}
\end{algorithm}
\FloatBarrier

It is clear from the pseudocode in Algorithm~\ref{linelas-MGRIT-V-alg},
that more work needs to be done compared to sequential time-stepping;
however, this work can be done in parallel, which highlights MGRIT's potential for parallel speedup.

\begin{remark}
    MGRIT does not require the same time-stepping operator
    on the fine and coarse time grids.
    In fact, employing a \emph{cheaper} (w.r.t.\ computational work)
    time-stepping operator and / or spatial solver can improve parallel performance,
    e.g., see \cite{BrummelenZeeBorst2008,Richter2015}.
\end{remark}

\begin{remark}
    MGRIT is neither restricted to the linear PDE case nor to simple two-level cycles.
    Similar to other multigrid methods, MGRIT can accomodate various cycling strategies, such as V- and F-cycles.
    Multilevel V-cycles are achieved by calling the algorithm recursively in Step~\ref{linelas-MGRIT-V-recursive-step-alg}
    in Algorithm~\ref{linelas-MGRIT-V-alg}.
    Multilevel F-cycles can be achieved by performing (at least) one V-cycle as postrelaxation
    at each level~\cite{TrottenbergOsterleeSchueller2001}.
    Similar to other multigrid reduction problems, solving the nonlinear setting is achieved
    by solving Equation~\eqref{general-pde-phi-form-nonlinear-eqn}
    and employing a full approximation storage approach. For more details, refer to
    \cite{FriedhoffFalgoutKolevMaclachlanSchroder2012,FalgoutFriedhoffKolevMaclachlanSchroder2014}.
\end{remark}

\begin{remark}
    To simplify terminology, we refer to the exact solution as the solution
    obtained from sequential time-stepping on the fine grid.
    Note, that sequential time-stepping can be identified as a direct solution approach
    with respect to Equation~\eqref{mgrit-intro-space-time-system-eqn}.
    In that regard, MGRIT can be interpreted as an interative solver,
    which obtains the same solution as sequential time-stepping on the fine grid.
\end{remark}
\subsection{The XBraid library}
\label{mgrit-xbraid-sec}
For the work presented in this article, the open-source implementation {XBraid}~\cite{XBraid}
(release v2.3.0) was employed and extended, e.g., to implement the time-periodic MGRIT
variant proposed in the following.
To successfully link an existing simulation package with XBraid, the user has to implement a number
of wrapper routines.
The wrapper steers and calls the simulation code to perform a number of basic tasks
(e.g., norm computation, data I/O, etc.) and to perform time integration.
Time integration routines from the simulation code are exposed through the so-called XBraid \emph{step function}
that represents the application of the time stepping operator~$\boldsymbol{\Phi}_\ell$,
see Equation~\eqref{general-pde-phi-form-nonlinear-eqn} and Equation~\eqref{general-pde-phi-form-linear-eqn}.

The \emph{step function} can be summarized as the following pseudocode:
\FloatBarrier
\begin{algorithm}
    \caption{Pseudocode for XBraid's \emph{step function} on level $\ell$.}
    \label{mgrit-step-function-alg}
    \begin{algorithmic}[1]
        \Function{step\_function}{$\boldsymbol{u} (\cdot, t)$, $t$, $\delta_\ell$}
            \State Set $\boldsymbol{u} (\cdot, t)$,
                time $t$ and time step size $\delta_\ell$
            \State Compute solution
                $\boldsymbol{u} (\cdot, t + \delta_\ell)
                = \boldsymbol{\Phi}_\ell \left( \boldsymbol{u} (\cdot, t + \delta_\ell), \boldsymbol{u} (\cdot, t); t, t + \delta_\ell \right)
                + \boldsymbol{g} (\cdot, t + \delta_\ell)$
            \State\Return $\boldsymbol{u} (\cdot, t + \delta_\ell)$
        \EndFunction
    \end{algorithmic}
\end{algorithm}
\FloatBarrier

A realistic \emph{step function} may have many more steps or customizations
depending on the considered setting; however, Algorithm~\ref{mgrit-step-function-alg} highlights
the fundamental algorithmic steps.
Overall, the required implementation overhead to enable parallel-in-time integration
through MGRIT is relatively low.
For example, the finite element package {CHeart}~\cite{LeeEtAl2016,CHeart}
has about $98500$ lines of Fortran code (includes comments),
whereas the wrapper routines only amount to about $2050$ lines of Fortran code (includes comments).\footnote{Note, that additional changes or additions to the {CHeart} source code (e.g., I/O in the time-parallel setting) have been made
since this computation.}
\subsection{Extension for time-periodic problems}
\label{mgrit-extension-for-time-periodic-problems-sec}
In this section, we describe how the MGRIT algorithm can be extended
to solve time-periodic problems more efficiently by making the fine grid periodic.\footnote{By modifying
the algorithm only on the fine grid, it naturally extends to the true multilevel setting.}
Instead of solving the problem in a traditional manner using sequential time-stepping
and simulating $q$ cycles (i.e.\ $t \in [0, q T]$) to achieve a time-periodic steady-state,
the proposed time-periodic MGRIT algorithm approximates the space-time solution
over \emph{one} periodic cycle only (i.e.\ $t \in [0, T]$).

First, an initial space-time guess is obtained ({XBraid} iteration~$0$)
by performing sequential time-stepping on the coarse-grid,\footnote{The
cost for obtaining the initial space-time guess for a linear problem
is roughly one $m^{th}$ of the cost
of running sequential time stepping on the fine grid, where $m$
is the temporal coarsening factor.}
which can be configured in the {XBraid} library
(refer to the \emph{skip-first-down} option in the {XBraid} manual~\cite{XBraid}).
{XBraid} cycling then continues as normal.
Convergence to the time-periodic steady-state is achieved
by \emph{updating} the initial condition at $t = 0$ on the \emph{fine grid},
whenever a new value is computed / available at $t = T$ on the \emph{fine grid}.
That is, the step function from Algorithm~\ref{mgrit-step-function-alg} is used on the coarse grid,
whereas a modified step function is employed on the fine grid:
\FloatBarrier
\begin{algorithm}
    \caption{Pseudo-code for {XBraid}'s \emph{step function} for the time-periodic case on level $0$.}
    \label{mgrit-step-function-time-periodic-alg}
    \begin{algorithmic}[1]
        \Function{step\_function}{$\boldsymbol{u} (\cdot, t)$, $t$, $\delta_0$}
            \State Set $\boldsymbol{u} (\cdot, t)$, time $t$ and time step size $\delta_0$
            \If{time $t = 0$ and iteration $i > 0$}\label{mgrit-step-function-iter0-step}
                \If{initial condition at $t = 0$ not converged}
                    \State Receive update as $\boldsymbol{u} (\cdot, t)$
                        using blocking \href{https://www.mpich.org/static/docs/v3.2/www3/MPI_Recv.html}{MPI\_Recv}
                        \label{mgrit-step-receive-first-update-step}
                    \If{iteration $i = 1$ and using F-relaxation}\label{mgrit-step-second-update-step}
                        \State Discard first update and receive second update as $\boldsymbol{u} (\cdot, t)$
                            using blocking \href{https://www.mpich.org/static/docs/v3.2/www3/MPI_Recv.html}{MPI\_Recv}
                    \EndIf
                \EndIf
            \EndIf
            \State Compute solution
                $\boldsymbol{u} (\cdot, t + \delta_0)
                = \boldsymbol{\Phi}_0 \left( \boldsymbol{u} (\cdot, t + \delta_0), \boldsymbol{u} (\cdot, t); t, t + \delta_0 \right)
                + \boldsymbol{g} (\cdot, t + \delta_0)$ \label{mgrit-step-integration-step}
            \If{time $t + \delta_0 = T$}
                \If{initial condition at $t = 0$ not converged}
                    \If{$\| \boldsymbol{u} (\cdot, T) - \boldsymbol{u} (\cdot, 0) \|_2 < tol$}\label{mgrit-step-function-tol-step}
                        \State Set initial condition as converged
                    \EndIf
                    \State Send update of $\boldsymbol{u} (\cdot, T)$ using nonblocking
                    \href{https://www.mpich.org/static/docs/v3.2/www3/MPI_Isend.html}{MPI\_iSend}
                    \label{mgrit-step-send-update-step}
                \EndIf
            \EndIf
            \State\Return $\boldsymbol{u} (\cdot, t + \delta_0)$ \label{mgrit-step-return-step}
        \EndFunction
    \end{algorithmic}
\end{algorithm}
\FloatBarrier

In the following, we provide a more detailed description
of the modified fine grid step function for the time-periodic case
to accompany the pseudocode in Algorithm~\ref{mgrit-step-function-time-periodic-alg}:
First, if $t = 0$ and the initial condition at $t = 0$ is not converged,
an update of the initial condition is received
in Step~\ref{mgrit-step-receive-first-update-step}.
During the first iteration and if F-relaxation is employed,
a second update of the initial condition is received
in Step~\ref{mgrit-step-second-update-step}.
Then, perform time-integration in Step~\ref{mgrit-step-integration-step} as usual.
If the current time equals the cycle time, i.e.\ $t + \delta_0 = T$,
check if the initial condition has converged (Step~\ref{mgrit-step-function-tol-step})
and send an update in Step~\ref{mgrit-step-send-update-step}, if required.
Finally, return the new value in Step~\ref{mgrit-step-return-step}.

\begin{remark}
    Since the time-periodic MGRIT algorithm starts a forward-solve on the coarsest-grid,
    no updates are available for iteration~$0$ in Step~\ref{mgrit-step-function-iter0-step}
    of Algorithm~\ref{mgrit-step-function-time-periodic-alg}.
    In the case that no time-parallelism is employed (i.e.\ one processor in time),
    the initial condition is directly updated from the value computed for $t = T$ without MPI communication.
            \end{remark}

\begin{remark}
    In iteration~$1$, processor~$0$ waits for two messages in the case of F-relaxation,
    see Algorithm~\ref{mgrit-step-function-time-periodic-alg}, Step~\ref{mgrit-step-second-update-step}.
    Preliminary experiments showed that this was required to achieve a robust residual reduction
    with a convergence factor smaller than one.
    The following iterations, however, proceed similarly for F- and FCF-relaxation
    and ensure that updates are incorporated as they become available.
\end{remark}

\begin{remark}
    Step~\ref{mgrit-step-function-tol-step} of Algorithm~\ref{mgrit-step-function-time-periodic-alg}
    allows for an approximate convergence to the time-periodic steady-state with a given tolerance \emph{$tol$}.
    To achieve this, a \emph{callback} function was implemented in {XBraid} that lets the simulation code
    compute the size of the jump and therefore communicate to {XBraid}
    whether it may terminate (initial condition converged)
    or not (initial condition \emph{not} converged).
    In this work, convergence of the initial condition is evaluated using
    an absolute convergence criterion, however,
    a relative tolerance could be similarly appropriate.
\end{remark}

\begin{remark}
    A similar idea to exploit time-periodicity was applied
    for the Parareal algorithm~\cite{GanderJiangSongZhang2013,GanderKulchytskaruchkaSchoeps2018}.
    In our approach, however, we only require an \emph{approximate} update of the initial condition.
    In this work, the solution at $t = 0$ is considered converged (indicating a time-periodic steady state)
    if the absolute difference of subsequent approximations is less than $10^{-10}$ in the Euclidean norm.
\end{remark}

In the following, we only consider two-level MGRIT using F- and FCF-relaxation,
a temporal coarsening factor of $m = 8$ and an initial space-time guess
based on sequential time stepping on the coarse grid, unless noted otherwise.
\section{Numerical experiments}\label{numerical-experiments-sec}
In this section, we introduce three application classes
for which we will assess the proposed time-periodic MGRIT algorithm in Section~\ref{results-sec}.
To assess the new algorithm for a simple test problem,
we commence with a simple linear PDE model defined
over a rigid two-dimensional flow domain (resembling flow in the ascending aorta)
in Section~\ref{stenosed-valve-sec}.
We proceed by adding complexity in a multiphysics problem
in Section~\ref{analytic-fsi-sec}:
linear and nonlinear fluid-structure interaction models
in a 2D channel and 3D vessel, where analytic solutions are available
to highlight the space-time error reduction properties of the algorithm.
We complete the assessment by applying a nonlinear flow model
over a realistic left atrium / left ventricle (LA / LV) geometry
taken from 3D MRI data
in Section~\ref{lv-flow-application-model-problem-sec},
to demonstrate that the new algorithm is applicable
to more complex applications without further modifications.
\subsection{Simplified flow in the ascending aorta}
\label{stenosed-valve-sec}

\begin{figure}[ht!]
    \centering
\resizebox{\textwidth}{!}{
        \begin{tikzpicture}
                \node [anchor=south west] at (0,0) {
            \includegraphics[angle=180,width=0.8\linewidth]{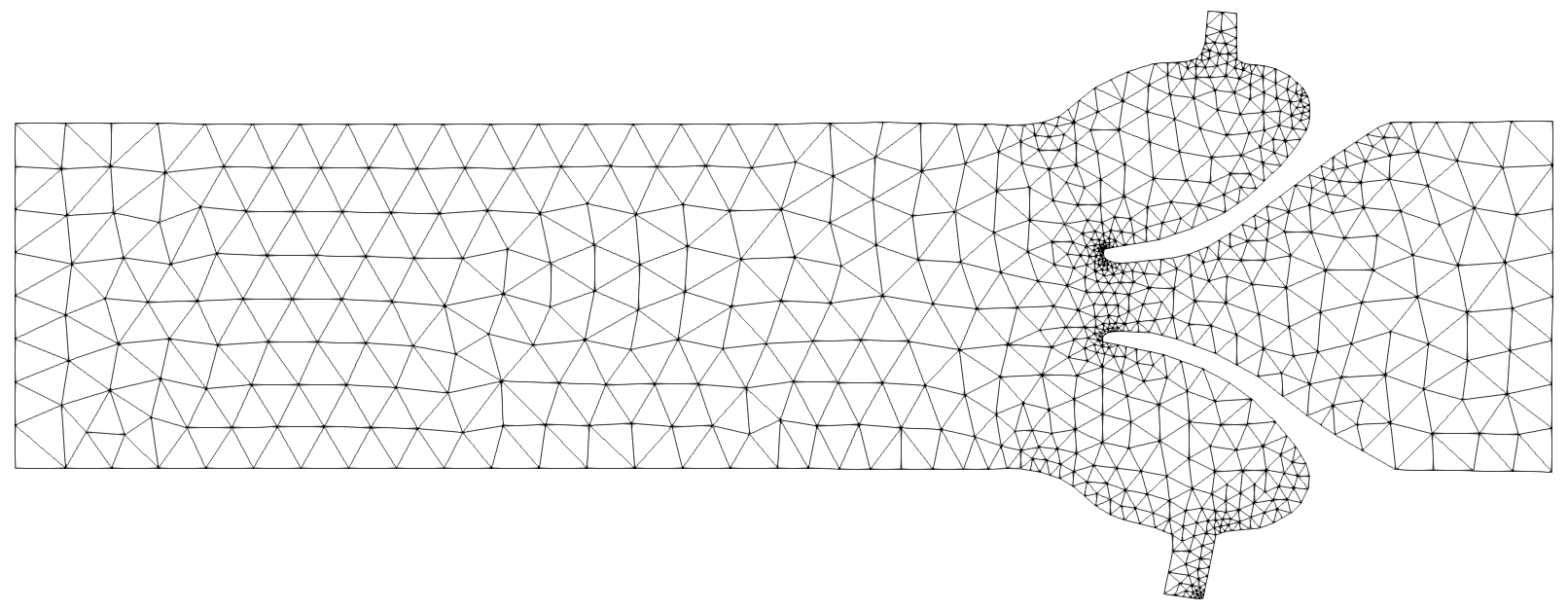}
            };
                \draw[black, right] (8,2.5) node[fill=white] {$\Omega_f$};
                \draw[->, line width=2pt,black] (-2,2.5) -- (0,2.5);
        \draw[black, below] (-2,3.5) node {inflow on $\Gamma_f^I$};
                \draw[->, line width=2pt,black] (13.5,2.5) -- (15.5,2.5);
        \draw[black, below] (15.5,3.5) node {outflow on $\Gamma_f^O$};
                \draw[->, line width=2pt,black] (3.4,5.2) -- (3.6,6.5);
        \draw[black, right] (3.7,6.5) node {~~outflow on $\Gamma_f^T$};
                \draw[->, line width=2pt,black] (3.0,0.1) -- (2.9,-1.3);
        \draw[black, right] (2.95,-1.3) node {~~outflow on $\Gamma_f^B$};
                \draw[black, right] (7,4.3) node {~~no-slip on $\Gamma_f^W$};
                \draw[->, line width=1pt,black] (-2,-1.75) -- (-0.5,-1.75);
        \draw[black, above] (-0.5,-1.75) node {$x$};
        \draw[black, below] (-0.5,-1.75) node {$~$};
        \draw[->, line width=1pt,black] (-2,-1.75) -- (-2,-0.25);
        \draw[black, right] (-2,-0.25) node {$y$};
    \end{tikzpicture}
}
    \caption{Simplified ascending aorta geometry
        in two dimensions with rigid maximally dilated valves.
        Inflow occurs from the left ventricle on the left.
        Minor outflows are observed toward the left and right coronary arteries at the top and bottom,
        and main outflow toward the aortic arch on the right.}
    \label{stenosed-valve-geometry-fig}
\end{figure}
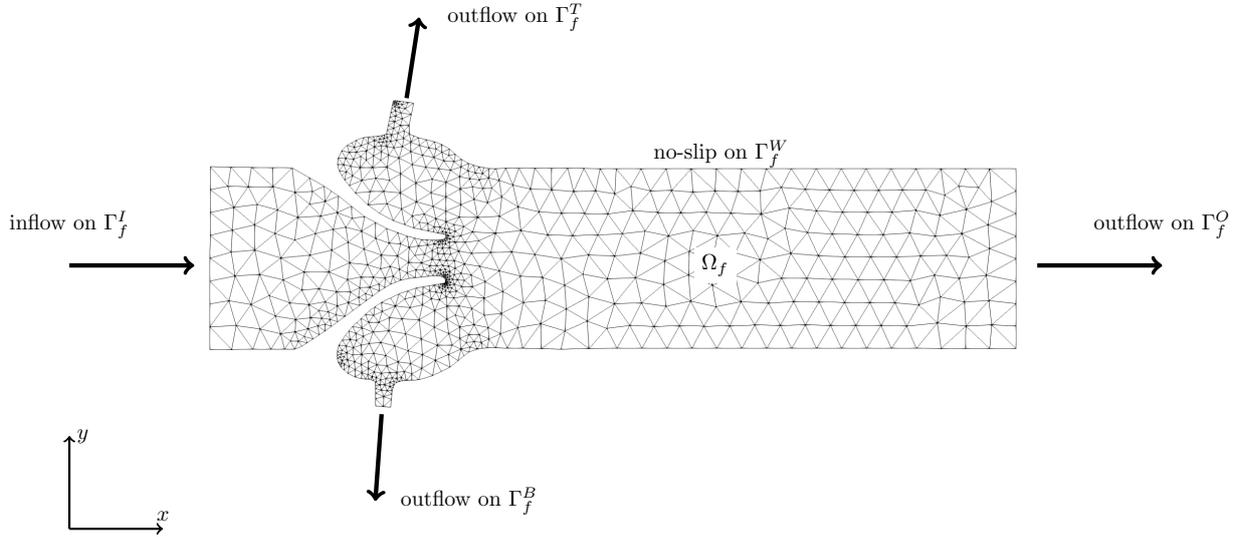

Consider flow in the ascending aorta in a simplified two-dimensional geometry,
see Figure~\ref{stenosed-valve-geometry-fig}:
Pulsatile blood flow is entering the domain from the left ventricle.
The blood flows past two rigid valves,\footnote{Here,
we simplify the problem by making the geometry rigid
to be able to test the algorithm for a simple flow model
over a complex geometry.}
the aortic sinus (anatomic dilations of the ascending aorta)
and enters the ascending aorta.
The outflow domain is slightly extended where the aortic arch would be located.

Further, consider the Stokes\footnote{For this type of application,
a simple Stokes flow model may not be the most accurate choice.
This choice, however, was motivated by the aim of investigating the time-periodic MGRIT
algorithm in a number of different settings,
e.g., by starting off from a simple linear model and adding complexity
in a step-by-step manner, to ultimately highlight that the algorithm performs
well in various settings.}
equations for incompressible Newtonian flow:
\begin{align}
    \rho_f \partial_t \boldsymbol{v}_f
    - \nabla \cdot \left[ \mu_f \left( \nabla \boldsymbol{v}_f + \nabla^T \boldsymbol{v}_f \right) - p_f \boldsymbol{I} \right]
    &= 0 \qquad \text{in } \Omega_f, \label{se-model-stenosed-valve-momentum-eqn} \\
    \nabla \cdot \boldsymbol{v}_f
    &= 0 \qquad \text{in } \Omega_f,
    \label{se-model-stenosed-valve-divu-eqn}
\end{align}
with velocity $\boldsymbol{v}_f$,
pressure $p_f$,
density $\rho_f = 10^{-3}~g / mm^3$
and viscosity $\mu_f = 0.04~g / (mm \cdot s)$.
Pulsatile parabolic inflow with cycle length $T = 1.024~s$ is prescribed at the inflow,
\begin{equation}
    \boldsymbol{v}_f = \begin{bmatrix}
        200 \\
        0
    \end{bmatrix} \frac{mm}{s}
    \left( 0.5 + 0.5 \cdot \cos{\left( \frac{2 \pi t - \pi}{T} \right)} \right)
    \left[ 1 - \left( \frac{y - 0.685~mm}{14.415~mm} \right)^2 \right],
    \label{stenosed-valve-inflow-eqn}
\end{equation}
and pulsatile outflow\footnote{A constant flow profile was applied across the outflow boundary
for ease of setting up the numerical problem, as the overall flow problem was already simplified.}
is prescribed at the left and right coronary arteries,
\begin{equation}
    \boldsymbol{v}_f = \begin{bmatrix}
        0 \\
        10
    \end{bmatrix} \frac{mm}{s}
    \left( 0.5 + 0.5 \cdot \cos{\left( \frac{2 \pi t - \pi}{T} \right)} \right)
    \text{sign} (y).
    \label{stenosed-valve-outflow-eqn}
\end{equation}

Zero Neumann conditions are prescribed at the outflow and zero Dirichlet conditions
are prescribed at all other boundaries.
The initial condition is set as zero.
\subsubsection{Space-time discretization and weak formulation}
\label{stenosed-valve-space-time-discretization-sec}
The fluid domain $\Omega_f$ is discretized
using finite elements with $1388$ inf-sup stable $\mathbb{P}^2-\mathbb{P}^1$
Taylor-Hood elements and $N_x = 6879$ degrees-of-freedom, see Figure~\ref{stenosed-valve-geometry-fig}.
A backward Euler time discretization scheme is employed to discretize the temporal domain
with equidistant time points $t_i = i \delta_0$ for $i = 0, 1, \ldots, 1024$,
such that the time step size is given as $\delta_0 = T / 1024 = 1~ms$.
Then the general discrete weak form of Equation~\eqref{se-model-stenosed-valve-momentum-eqn}
and Equation~\eqref{se-model-stenosed-valve-divu-eqn} can be written as follows:

Find $\boldsymbol{s}^{n+1}
:= ( \boldsymbol{v}_f^{n+1}, p_f^{n+1} ) \in \boldsymbol{\mathcal{V}}_D^h \times \mathcal{W}_f^h$,
such that for every
$\boldsymbol{d} := ( \boldsymbol{w}_f, q_f ) \in \boldsymbol{\mathcal{V}}_0^h \times \mathcal{W}_f^h$:
\begin{equation}
    \begin{split}
        R ( \boldsymbol{s}^{n+1}, \boldsymbol{s}^n, \boldsymbol{d} )
        =~&\int_{\Omega_f^h} \rho_f \frac{\boldsymbol{v}_f^{n+1} - \boldsymbol{v}_f^n}{\delta_0} \cdot \boldsymbol{w}_f
        + \left[
            \mu_f D \boldsymbol{v}_f^{n+1}
            - p_f^{n+1} \boldsymbol{I}
        \right] : \nabla \boldsymbol{w}_f
        + q_f \nabla \cdot \boldsymbol{v}_f^{n+1}~d\Omega_f^h
        = 0,
    \end{split}
    \label{stenosed-valve-stokes-weak-form-eqn}
\end{equation}
where $D \boldsymbol{v} = \text{sym} \left( \boldsymbol{v} \right)$
and $\Omega_f^h$ refers to the discretized domain
and the definition of the function spaces are given as,
\begin{equation}
    S^p (\Omega_f^h)
    = \{ f : \Omega_f^h \rightarrow \mathbb{R}
    ~|~f \in \mathcal{C}^0 (\bar{\Omega}_f^h),
    f|_{\tau_e} \in \mathbb{P}^p (\tau_e), \forall~\tau_e \in T_f^h
    \},
    \label{stenosed-valve-stokes-function-spaces-eqn}
\end{equation}
which represent the general continuous $p^{th}$-order piecewise polynomial spaces
defined on $\Omega_f^h$,
and $T_f^h$ is the set of triangles composing $\Omega_f^h$.
Consequently,
\begin{equation}
    \boldsymbol{\mathcal{V}}^h = [ S^2 (\Omega_f^h) ]^2
    \qquad \text{and} \qquad
    \mathcal{W}_f = S^1 (\Omega_f^h),
    \label{stenosed-valve-stokes-test-spaces-eqn}
\end{equation}
and incorporating the boundary conditions yields,
\begin{align}
    \boldsymbol{\mathcal{V}}_0^h
    &= \{ \boldsymbol{v} \in \boldsymbol{\mathcal{V}}^h
    ~|~\boldsymbol{v} = \boldsymbol{0} \text{ on } \Gamma_f^W \}, \\
    \boldsymbol{\mathcal{V}}_D^h
    &= \{ \boldsymbol{v} \in \boldsymbol{\mathcal{V}}^h
    ~|~\boldsymbol{v} = \boldsymbol{v}_f
    \text{ according to \eqref{stenosed-valve-inflow-eqn}
    and \eqref{stenosed-valve-outflow-eqn}} \}.
    \label{stenosed-valve-stokes-test-spaces-bc-eqn}
\end{align}
\subsubsection{Convergence to a time-periodic steady state}
\label{stenosed-valve-steady-state-sec}
For the linear flow problem, the periodicity is boundary-driven
due to the pulsatile inflow.
Using sequential time-stepping, the system can be driven to a periodic steady-state
by simulating multiple cardiac cycles, such that each cycle starts off
with an improved initial state.
In contrast, time-periodic MGRIT as an iterative method
achieves a time-periodic steady-state solution by continuously
updating the initial condition during each iteration for a single cardiac cycle,
see Section~\ref{mgrit-extension-for-time-periodic-problems-sec}.

We obtain a time-periodic steady-state solution
by simulating $10$ subsequent cardic cycles using sequential time-stepping.
The solution over the $10^{th}$ cycle is thus set as the steady-state reference solution.
The reference solution is then compared with the sequential time-stepping solution for each cycle,
as well as the time-periodic MGRIT approximation over one cycle for each MGRIT iteration.
For comparison, we employ the $L_2 ( \Omega_t^h; L_2 ( \Omega_f^h ) )$-norm
over the velocity field over a whole temporal cycle.
For a better understanding of the initial condition convergence criterion,
we further compute the time point-wise difference $L_2 ( \Omega_f^h )$-norover the velocity field
for each sequential time-stepping cycle as well as time-periodic MGRIT iteration.
\subsubsection{Numerical setup}
\label{stenosed-numerical-setup-sec}
To solve the spatial problem described above,
we employ the finite element solver CHeart~\cite{LeeEtAl2016,HessenthalerRoehrleNordsletten2017},
which is based on the matrix solver MUMPS~\cite{AmestoyDuffLexcellent2000}.
For sequential time-stepping runs, CHeart steers the simulation by itself
and it is set up to run $10$ subsequent cardiac cycles.
For the time-periodic MGRIT runs, wrapper code was written
that implements the time-periodic MGRIT algorithm
(see Section~\ref{mgrit-extension-for-time-periodic-problems-sec})
based on the open-source library XBraid,
to solve the space-time problem over one cardiac cycle iteratively.
At the lowest level, the wrapper code calls CHeart to solve one time step
according to the step function defined in Algorithm~\ref{mgrit-step-function-time-periodic-alg},
using an initial condition, the current time and the current time step size
(i.e.\ fine or coarse grid time step size).

Measures that indicate the distance to the time-periodic steady-state reference solution
(as defined in Section~\ref{stenosed-valve-steady-state-sec})
are computed \emph{a posteriori} and are thus not part of the observed runtimes
of the respective algorithms.
\subsection{Fluid-structure interaction problem with analytic solution}
\label{analytic-fsi-sec}
In this section, we consider two cases of a class of analytic solutions
for fluid-structure interaction~\cite{HessenthalerBalmusRoehrleNordsletten2020}:
the transient interaction of flow with a linear solid
in a 2D channel (Section~\ref{methods-2D-linear-sec})
and the transient interaction of flow with a nonlinear solid
in a 3D vessel (Section~\ref{methods-3D-nonlinear-sec}).
\subsubsection{Transient interaction between flow and linear solid}
\label{methods-2D-linear-sec}
\begin{figure}[ht!]
    \centering
    \resizebox{0.35\textwidth}{!}{
        \begin{tikzpicture}                \def\L{4.0};
        \def\axl{0.2*\L};
        \def\ri{4.0};
        \def\ro{4.8};
                \draw[ultra thick, black] (0,0) -- (\L,0);
        \draw[ultra thick, black, below] (0.5*\L,0) node {$\Gamma_f^W$};
                \draw[ultra thick, myred] (0,0) -- (0,\ri);
        \draw[ultra thick, myred, left] (0,0.5*\ri) node {$\Gamma_f^I$};
                \draw[ultra thick, myred] (\L,0) -- (\L,\ri);
        \draw[ultra thick, myred, right] (\L,0.5*\ri) node {$\Gamma_f^O$};
                \draw[ultra thick, black, dashed] (0,\ri) -- (\L,\ri);
                        \draw[ultra thick, black, below] (0.5*\L,\ri) node {$\Gamma^\lambda$};
                \draw[ultra thick, black] (0,\ro) -- (\L,\ro);
        \draw[ultra thick, black, above] (0.5*\L,\ro) node {$\Gamma_s^W$};
                \draw[ultra thick, myblue] (0,\ri) -- (0,\ro);
        \draw[ultra thick, myblue, left] (0,0.5*\ro+0.5*\ri) node {$\Gamma_s^I$};
                \draw[ultra thick, myblue] (\L,\ri) -- (\L,\ro);
        \draw[ultra thick, myblue, right] (\L,0.5*\ro+0.5*\ri) node {$\Gamma_s^O$};
                \draw[ultra thick, black, left] (0.75*\L,0.5*\ri+0.5*\ro) node { \large $\Omega_s^0$};
        \draw[ultra thick, black, left] (0.75*\L,0.5*\ri) node { \large $\Omega_f^0$};
                \draw[->, line width=2pt,black] (0,0) -- (\axl,0);
        \draw[black, below] (\axl,0) node {$x$};
        \draw[->, line width=2pt,black] (0,0) -- (0,\axl);
        \draw[black, left] (0,\axl) node {$y$};
                \draw[<->, thick, black!60] (0.0,-1*\axl) -- (\L,-1*\axl);
        \draw[black!60, below] (0.5*\L,-1*\axl) node {\small $L = 1$};
        \draw[<->, thick, black!60] (-2.5*\axl,0) -- (-2.5*\axl,\ro);
        \draw[black!60, above] (-2.5*\axl,0.5*\ro) node[rotate=90] {\small $H_o = 1.2$};
        \draw[<->, thick, black!60] (-1.5*\axl,0) -- (-1.5*\axl,\ri);
        \draw[black!60, above] (-1.5*\axl,0.5*\ri) node[rotate=90] {\small $H_i = 1$};
                \draw[ultra thick, white, below] (0.5*\L,-1.5*\axl) node {~};
    \end{tikzpicture}}~~~\qquad\qquad~\resizebox{0.45\textwidth}{!}{
                \begin{tikzpicture}
                        \node [anchor=south west] at (0,0) { \includegraphics[width=0.6\linewidth]{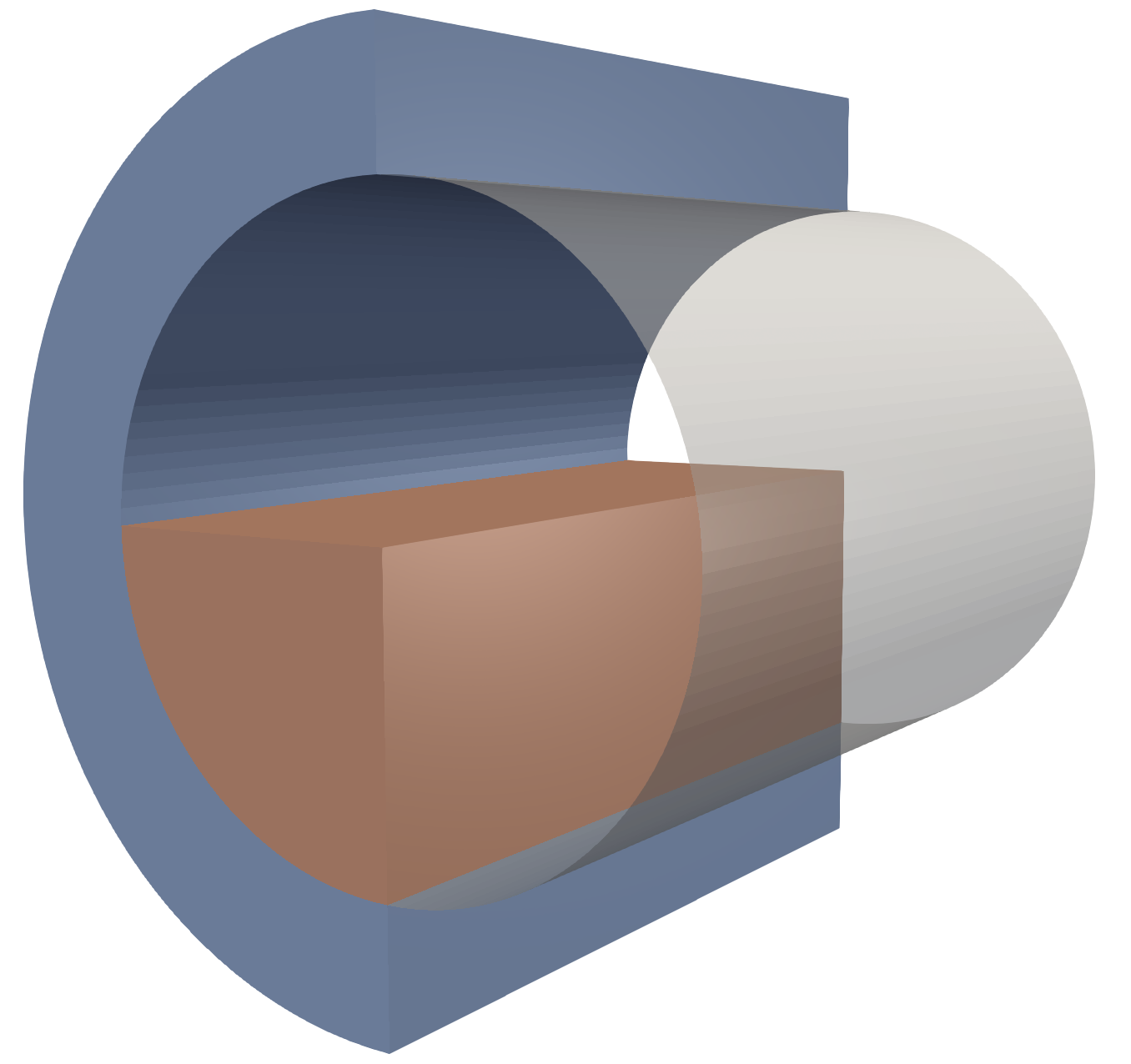} };
                        \draw[->, line width=2pt,black] (3.475,4.625) -- (4.475,4.775);
            \draw[black, below] (4.475,4.775) node {$z$};
            \draw[->, line width=2pt,black] (3.475,4.625) -- (2.475,4.69);
        \draw[black, below] (2.475,4.675) node {$y$};
            \draw[->, line width=2pt,black] (3.475,4.625) -- (3.475,3.625);
            \draw[black, right] (3.4755,3.625) node {$x$};
                        \draw[ultra thick, black, right] (4.5,3.5) node { \LARGE $\Omega_f^0$};
            \draw[ultra thick, black, right] (4.5,8.2) node { \LARGE $\Omega_s^0$};
                                            \draw[black, right] (8,6) node {\Large $\Gamma^\lambda$};
                        \draw[black, right] (2,3.25) node {\Large $\Gamma_f^I$};
                        \draw[<-, line width=1pt,black] (7.5,4) -- (10,4);
            \draw[<-, line width=1pt,black!60] (7.5,4) -- (9.31,4);
            \draw[black, right] (10,4) node {\Large $\Gamma_f^O$};
                        \draw[black, right] (0.85,7) node {\Large $\Gamma_s^I$};
                        \draw[<-, line width=1pt,black] (7.55,8) -- (10,8);
            \draw[black, right] (10,8) node {\Large $\Gamma_s^O$};
                        \draw[<-, line width=1pt,black] (5.2,1) -- (10,1);
            \draw[black, right] (10,1) node {\Large $\Gamma_s^W$};
                        \draw[dashed, line width=1pt, black] (-0.4,1.5) -- (3.2,1.5);
            \draw[dashed, line width=1pt, black] (-0.4,7.7) -- (3,7.7);
            \draw[<->, line width=1pt, black] (-0.5,1.5) -- (-0.5,7.7);
            \draw[black, above] (-0.5,4.7) node[rotate=90] {$2 H_i = 1.4$};
                        \draw[dashed, line width=1pt, black] (-1.4,0.25) -- (3.2,0.25);
            \draw[dashed, line width=1pt, black] (-1.4,9.1) -- (3,9.1);
            \draw[<->, line width=1pt, black] (-1.5,0.25) -- (-1.5,9.1);
            \draw[black, above] (-1.5,4.7) node[rotate=90] {$2 H_o = 2$};
                        \draw[dashed, line width=1pt, black] (3.5,0) -- (3.5,-0.4);
            \draw[dashed, line width=1pt, black] (7.3,2) -- (7.3,-0.4);
            \draw[<->, line width=1pt, black] (3.5,-0.5) -- (7.3,-0.5);
            \draw[black, below] (5.5,-0.5) node {$L = 1$};
        \end{tikzpicture}
    }
        \caption{Fluid and solid reference domains, $\Omega_f^0$ and $\Omega_s^0$, in two dimensions
            with respective boundaries at the inlet ($\Gamma_f^I$ and $\Gamma_s^I$),
            the outlet ($\Gamma_f^O$ and $\Gamma_s^O$) and the wall ($\Gamma_f^W$ and $\Gamma_s^W$).
            The common interface boundary is denoted as $\Gamma^\lambda$.
            Further, the domain length is denoted as $L$, the fluid domain height as $H_i$
            and the fluid / solid domain height as $H_o$. Modified from: \cite[Figure 2.1]{Hessenthaler2020_PhD}.}
    \label{domain-2D-3D-fig}
\end{figure}
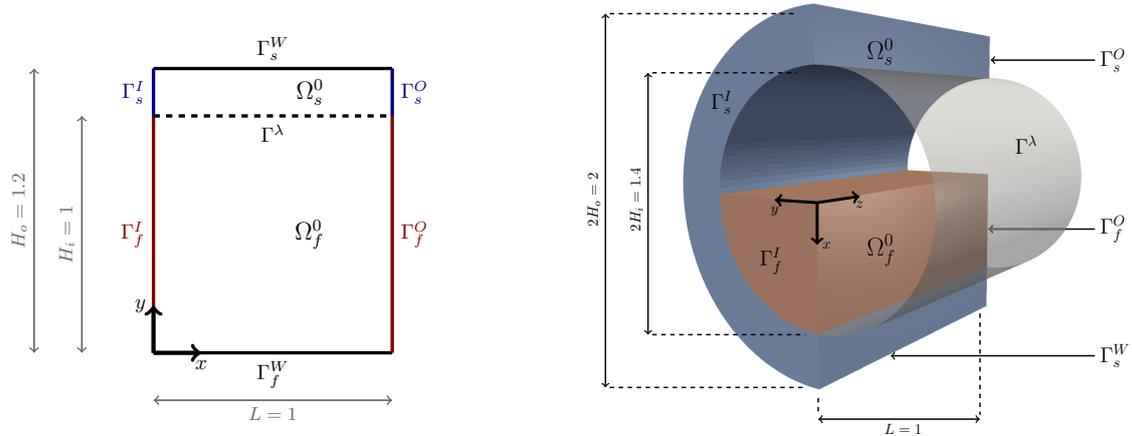

First, consider the time-periodic interaction of flow with a linear-elastic solid.
The space-time domain corresponds to a 2D channel (see Figure~\ref{domain-2D-3D-fig})
with extents $1 \times 1.2$ and cycle time of $T = 1.024$.
Fluid density and viscosity are selected as $\rho_f = 1$ and $\mu_f = 10^{-2}$.
Solid density and stiffness are selected as $\rho_s = 1$ and $\mu_s = 10^{-1}$.\footnote{In line
with \cite[Section 5.1]{HessenthalerBalmusRoehrleNordsletten2020},
parameters ``\emph{were selected to highlight key solution features
and for conducting a space-time convergence analysis.
But they are arbitrary in the sense that they were chosen
without a particular application area in mind. Thus, they are given in a unitless format.}''
One key solution feature is that the parameter set yields significant solid deformations,
whereas physiologically relevant parameters may not result in significant solid deformations,
see \cite[Section 5.3]{HessenthalerBalmusRoehrleNordsletten2020}.
Furthermore, results in the corresponding results section are likewise given in unitless format.}

The incompressible flow is modeled using the Stokes equations
(see Equation~\eqref{se-model-stenosed-valve-momentum-eqn}
and Equation~\eqref{se-model-stenosed-valve-divu-eqn}).
On the other hand, the incompressible, isotropic and homogeneous solid deformations are governed
by the transient linear elasticity equations in first-order form:
\begin{align}
	\partial_t \boldsymbol{u}_s &= \boldsymbol{v}_s && \text{in } \Omega_s, \label{linelas-dtu-v-eqn} \\
	\rho_s \partial_t \boldsymbol{v}_s
    &= \nabla \cdot \left[ \mu_s D \boldsymbol{u}_s - p_s \boldsymbol{I} \right] && \text{in } \Omega_s, \label{linelas-continuous-dtv-divsigma-eqn} \\
	\nabla \cdot \boldsymbol{v}_s &= 0 && \text{in } \Omega_s,
\end{align}
with displacement $\boldsymbol{u}_s$, velocity $\boldsymbol{v}_s$ and hydrostatic pressure $p_s$.
The two equation sets are coupled using the following interface conditions:
\begin{align}
    \boldsymbol{v}_f - \boldsymbol{v}_s &= \boldsymbol{0}, \label{fsi-coupling-1-eqn} \\
    \left[ \mu_f D \boldsymbol{v}_f - p_f \boldsymbol{I} \right] \cdot \boldsymbol{n}_f
    + \left[ \mu_s D \boldsymbol{u}_s - p_s \boldsymbol{I} \right] \cdot \boldsymbol{n}_s
    &= \boldsymbol{0}, \label{fsi-coupling-2-eqn}
\end{align}
with outer boundary normal vectors for the fluid and solid subdomains
$\boldsymbol{n}_f$ and $\boldsymbol{n}_s$, respectively.
\paragraph{Space-time discretization and weak formulation}
Similar to Section~\ref{stenosed-valve-sec},
an inf-sup stable Taylor-Hood finite element discretization is employed in space
with weak enforcement of the coupling constraints
\eqref{fsi-coupling-1-eqn} / \eqref{fsi-coupling-2-eqn}
through the introduction of a Lagrange multiplier variable,
$\boldsymbol{\lambda} = \boldsymbol{t}_f = - \boldsymbol{t}_s$,
see \cite{NordslettenKaySmith2010}.
Quadratic quadrilateral elements are used for the fluid velocity, solid velocity and solid displacement variables,
linear quadrilateral elements for the fluid pressure and solid pressure variables
and quadratic line elements for the Lagrange multiplier variable.
The spatial step size is selected as $0.025$.
For the time discretization, a backward Euler time discretization
scheme~\cite{HessenthalerNordslettenRoehrleSchroderFalgout2018}
is selected with time step size $0.002$.
Overall, the space-time discretization amounts to $5120$ time steps per cycle
and $18088$ spatial degrees-of-freedom.

The derivation of the weak form equations
is analogeous to Section~\ref{stenosed-valve-space-time-discretization-sec}
and is therefore omitted.
\subsubsection{Transient interaction between flow and nonlinear solid}
\label{methods-3D-nonlinear-sec}
Now adding complexity, we consider the time-periodic interaction
of flow with a hyperelastic solid in a 3D vessel geometry.
The spatial domain is derived by rotating the domain from Section~\ref{methods-2D-linear-sec}
around the $x$-axis (see Figure~\ref{domain-2D-3D-fig})
with extents $1 \times 1$ (solid thickness of $0.3$). The cycle length is $1.024$.
Fluid density and viscosity are selected as $\rho_f = 2.1$ and $\mu_f = 3 \cdot 10^{-2}$.
Solid density and stiffness are selected as $\rho_s = 1$ and $\mu_s = 10^{-1}$.\footnote{Similar
to Section~\ref{methods-2D-linear-sec}, parameters were selected
according to \cite[Section 5.2]{HessenthalerBalmusRoehrleNordsletten2020}.}

The incompressible flow is modeled using the Navier-Stokes equations in Arbitrary Lagrangian-Eulerian (ALE) form.
On the other hand, the incompressible, isotropic and homogeneous solid deformations are governed
by the transient finite elasticity equations using a Neo-Hookean material law.
Then, the FSI problem takes the following general form in accordance to \cite{HessenthalerBalmusRoehrleNordsletten2020}:
\begin{align}
                \rho_f \partial_t \boldsymbol{v}_f +  \rho_f (\boldsymbol{v}_f - \boldsymbol{w}_f) \cdot \nabla_{\boldsymbol{x}}
     \boldsymbol{v}_f
    +\nabla_{\boldsymbol{x}} \cdot \boldsymbol{\sigma}_f &= \boldsymbol{0} && \text{in } \Omega_f,
    \label{eq:strong_form_nonlinear_fluid_momentum_balance}
    \\
    \nabla_{\boldsymbol{x}} \cdot \boldsymbol{v}_f &= 0 && \text{in } \Omega_f,
    \\
    [\boldsymbol{v}_f]_k & = 0 && \text{on } \Gamma_f^I \cup \Gamma_f^O,~\text{for } k\in\{ x, y \},
    \\
    (\boldsymbol{\sigma}_f \cdot \boldsymbol{n}_f ) \cdot \boldsymbol{e}_z
    & = [\boldsymbol{t}_f]_z && \text{on } \Gamma_f^I \cup \Gamma_f^O,
    \\
    \boldsymbol{v}_f (\cdot, 0) &= \boldsymbol{v}_f^0 && \text{in } \Omega_f (0),
    \\
        \cline{1-4}
                    \rho_s \partial_{tt} \boldsymbol{u}_s -\nabla_{\boldsymbol{X}}
    \cdot \boldsymbol{P}_s &= \boldsymbol{0} && \text{in } \Omega_s(0)\times[0,T],
    \label{eq:strong_form_nonlinear_solid_momentum_balance}
    \\
    \text{det}\boldsymbol{F} - 1 &= 0 && \text{in } \Omega_s(0) \times [0,T],
    \\
    \boldsymbol{u}_s (\cdot, t) &= \boldsymbol{0} && \text{on } \Gamma_s^W,
    \\
    [\boldsymbol{u}_s]_k & = 0 && \text{on }
    \Gamma_s^I \cup \Gamma_s^O,~\text{for } k\in \{ x, y \}
    \\
    (\boldsymbol{P}_s \cdot \boldsymbol{N}_s) \cdot \boldsymbol{e}_z
    &= \boldsymbol{t}_s && \text{on }
    \Gamma_s^I \cup \Gamma_s^O,
    \\
    \boldsymbol{u}_s (\cdot, 0) &= \boldsymbol{u}_s^0 && \text{in } \Omega_s (0),
    \\
    \boldsymbol{v}_s (\cdot, 0) & = \boldsymbol{v}_s^0 && \text{in } \Omega_s (0),
    \\
        \cline{1-4}
                    \boldsymbol{\sigma}_f \cdot \boldsymbol{n}_f + \boldsymbol{\sigma}_s \cdot \boldsymbol{n}_s &= \boldsymbol{0} && \text{on } \Gamma^\lambda, \label{eq:strong_form_nonlinear_dynamic_constraint}\\
    \boldsymbol{v}_f - \boldsymbol{v}_s &= \boldsymbol{0} && \text{on } \Gamma^\lambda,
    \label{eq:strong_form_nonlinear_kinematic_constraint}
\end{align}
with fluid velocity $\boldsymbol{v}_f$,
fluid domain velocity $\boldsymbol{w}_f$,
fluid Cauchy stress tensor
$\boldsymbol{\sigma}_f = \mu_f D \boldsymbol{v}_f - p_f \boldsymbol{I}$
with fluid pressure $p_f$,
solid displacement $\boldsymbol{u}_s$,
the first Piola-Kirchhoff stress tensor,
\begin{equation*}
	\boldsymbol{P}_s = \frac{\mu_s}{(\text{det}\boldsymbol{F})^{2/3}}
	 \left[\boldsymbol{F} - \frac{\boldsymbol{F} : \boldsymbol{F}}{3} \boldsymbol{F}^{-T}\right] - p_s \boldsymbol{F}^{-T},
\end{equation*}
the deformation gradient tensor
$\boldsymbol{F} = \nabla_{\boldsymbol{X}} \boldsymbol{u}_s + \boldsymbol{I}$,
the solid Cauchy stress tensor
$\boldsymbol{\sigma}_s = \frac{1}{\det{\boldsymbol{F}}} \boldsymbol{P}_s \boldsymbol{F}^T$,
the solid pressure $p_s$
and the outer boundary normals $\boldsymbol{n}_f$ on the fluid domain
and $\boldsymbol{n}_s$ and $\boldsymbol{N}_s$
on the solid and solid reference domain, respectively.

In the discrete setting, care must be taken in order to properly manage
coupling between (fluid and solid) meshes.
This is achieved by mapping both coupling surfaces to a common interface surface $\Gamma^\lambda$.
Further, note that using Nanson's formula
(i.e.\ $\boldsymbol{n}_s = (\det \boldsymbol{F}) \boldsymbol{F}^{-T} \boldsymbol{N}_s$)
in Equation~\ref{eq:strong_form_nonlinear_dynamic_constraint},
the solid traction can be written using
the Cauchy and first Piola-Kirchhoff stress tensor as
$\boldsymbol{\sigma}_s \cdot \boldsymbol{n}_s = \boldsymbol{P}_s \cdot \boldsymbol{N}_s$.

Here, we omit the expansive definition of initial, Dirichlet and Neumann boundary conditions
(i.e.\ $\boldsymbol{v}_f^0$, $\boldsymbol{t}_f$, $\boldsymbol{t}_s$, $\boldsymbol{u}_s^0$, $\boldsymbol{v}_s^0$),
as those are adopted unmodified from the original work~\cite{HessenthalerBalmusRoehrleNordsletten2020}.\footnote{For the time-periodic MGRIT run,
the initial condition is considered to be the analytic solution for all variables except
for the fluid velocity, which is initially zero everywhere.
This is to achieve a good initial condition and initial guess with a pertubation
to simulate a practical case where some (experimental) data are available
but convergence to a periodic steady-state is required.}
\paragraph{Space-time discretization and weak formulation}
Again, using inf-sup quadratic-linear Taylor-Hood finite elements
(fluid: tetrahedral mesh; solid: hexahedral mesh; Lagrange multiplier: triangular mesh)
and backward Euler time integration,
we consider two mesh \emph{refinement levels}: \emph{coarse} and \emph{fine}.
The domains are discretized using approximate space-time step sizes of $(0.314, 0.002)$
and $(0.157, 0.0005)$ for the \emph{coarse} and \emph{fine} meshes,
which results in $32793 \times 512$ and $245009 \times 2048$ space-time degrees-of-freedom,
respectively.

For conciseness, we omit the weak form of
\eqref{eq:strong_form_nonlinear_fluid_momentum_balance} -- \eqref{eq:strong_form_nonlinear_kinematic_constraint}
as it is derived similarly to Section~\ref{stenosed-valve-sec},
however, note that due to the ALE term in Equation~\eqref{eq:strong_form_nonlinear_kinematic_constraint}
the weak form is expanded by an ALE diffusion problem to account for the fluid domain motion
according to~\cite[Equation~(85)]{HessenthalerBalmusRoehrleNordsletten2020}.

The fluid-solid coupling across these discrete interfaces requires care,
as the discretized form consists of triangles (fluid side)
and curvilinear quadrilaterals (solid side).
This is achieved by developing an interface discretization with a bijective map
formed between both fluid and solid discretizations
(see~\cite{NordslettenKaySmith2010} for details).
The fluid-solid systems are coupled through the use of a Lagrange multiplier
(representing the surface traction) and solved
as a monolithic system~\cite{NordslettenKaySmith2010,HessenthalerBalmusRoehrleNordsletten2020}.
\subsubsection{Convergence to a time-periodic steady state}
\label{analytic-steady-state-sec}
Due to the presence of an analytic solution \cite{HessenthalerBalmusRoehrleNordsletten2020},
it is possible to compute the space-time error for each algorithm
at each cycle and iteration, respectively.
Thus, the solution progress can be tracked
and the occurence of a time-periodic steady-state can be detected.
This is done by computing the spatial $\ell_2$ error
at each time point (over the length of a cycle),
as well as by numerically integrating the (time) point-wise spatial errors
over the temporal cycle length.
Thus, it is possible to track where errors occur in time (depending on the distance to the initial condition)
and how the space-time error is reduced \emph{cycle-to-cycle}
and \emph{iteration-to-iteration}.
\subsubsection{Numerical setup}
\label{analytic-numerical-setup-sec}
Similar to Section~\ref{stenosed-valve-sec}, CHeart and XBraid are employed
to simulate the space-time system using sequential time-stepping
($10$ and $7$ cardiac cycles for the linear and nonlinear case, respectively)
and time-periodic MGRIT (up to $12$ iterations) with FCF-relaxation.
In the nonlinear PDE case, a Newton-Raphson-Shimanskii solver~\cite{Shamanskii1967}
is employed to effectively reduce the computational cost
through reuse of the Jacobian matrix~\cite{HessenthalerRoehrleNordsletten2017}.
In the MGRIT case, this is handled by storing separate matrix objects~\cite{Hessenthaler2020_PhD},
such that matrix reuse can be applied on each time grid independently.

By adding a solver step to evaluate the analytic solution in CHeart,
the space-time error is computed on-the-fly.
Therefore, reported runtimes in Section~\ref{results-sec}
include the time it takes to compute the current space-time error.
\subsection{Flow in a left atrium / left ventricle geometry}
\label{lv-flow-application-model-problem-sec}
\begin{figure}[ht!]
    \centering
    \begin{minipage}{0.62\linewidth}
        \includegraphics[height=0.5\linewidth]{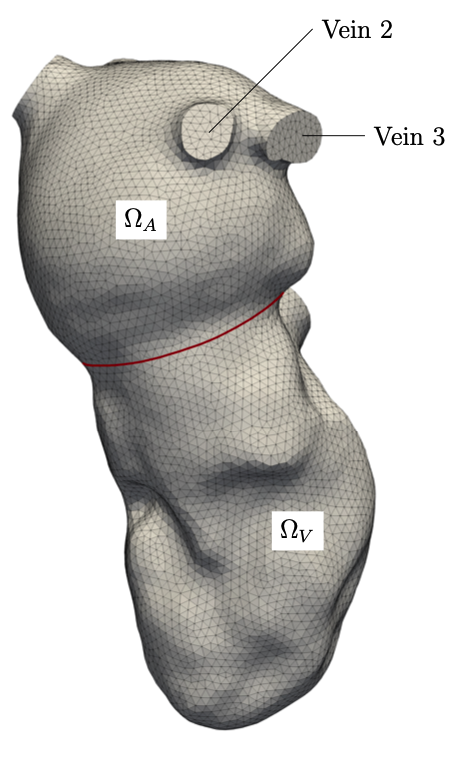}
        \includegraphics[height=0.5\linewidth]{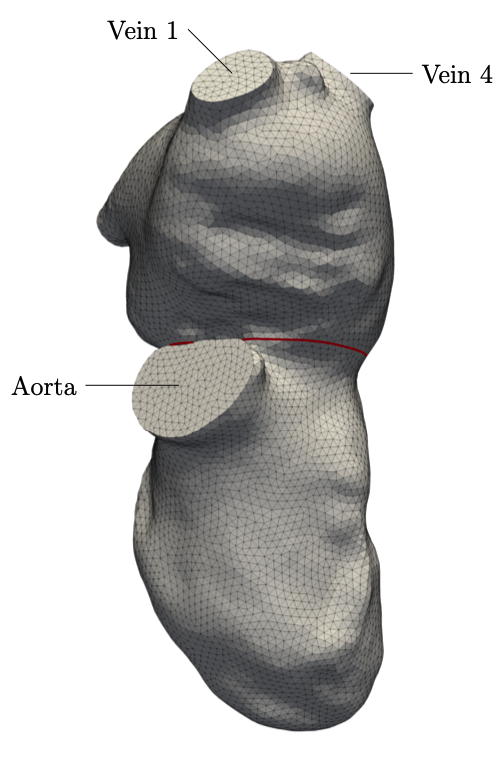}
    \end{minipage}\begin{minipage}{0.37\linewidth}
        \hspace{-3cm}
        \footnotesize
        \begin{tabular}{ l  c  l }
                                    $\Omega_A$      &:& LA domain                                                           \\
            $\Gamma_{V1}$   &:& Vein $1$ boundary                                                   \\
            $\Gamma_{V2}$   &:& Vein $2$ boundary                                                   \\
            $\Gamma_{V3}$   &:& Vein $3$ boundary                                                   \\
            $\Gamma_{V4}$   &:& Vein $4$ boundary                                                   \\
            $\Gamma_{Vn}$   &:& $\Gamma_{V1} \cup \Gamma_{V2} \cup \Gamma_{V3} \cup \Gamma_{V4}$    \\
            $\Gamma_{CA}$   &:& Common surface with LV                                              \\
            $\Gamma_{CAo}$  &:& Coupling surface of LA that is open during diastole  \\
            $\Gamma_{WA}$   &:& Endocardial wall of LA domain                                       \\
                        \hline
            $\Omega_{V}$    &:& LV domain                                                           \\
            $\Gamma_{AV}$   &:& Aortic valve boundary                                               \\
            $\Gamma_{CV}$   &:& Common surface with LA                                              \\
            $\Gamma_{CVo}$  &:& Coupling surface of LV that is open during diastole  \\
            $\Gamma_{WV}$   &:& Endocardial wall of LV domain                                       \\
            \hline
            $\Omega_{LM}$   &:& $\Omega_A \cap \Omega_V$, Lagrange multiplier (LM) domain         \\
            $\Gamma_{LM}$   &:& LM boundary                             \\
        \end{tabular}
    \end{minipage}~\\
    \caption{Patient-specific model of the left atrium ($\Omega_A$) and left ventricle ($\Omega_V$),
        showing initial mesh and the different inflow and outflow boundaries at pulmonary veins and the aorta.
        The red line indicates the coupling surface between atrium and ventricle.
        Figure modified from~\cite[Figure 3]{MarleviEtAl2021}.}
    \label{lv-flow-application-coarse-mesh-views-fig}
\end{figure}

Lastly, we consider nonlinear flow in a patient-specific left atrium (LA) / left ventricle (LV) geometry.
The geometry (see Figure~\ref{lv-flow-application-coarse-mesh-views-fig}),
as well as the time-dependent deformation of the LA and LV were extracted
from computed tomography (CT) data.
The CT data set stems from a Cardiac Resynchronization Therapy (CRT) patient
that suffers from an abnormal wall motion and a reduced ejection fraction (see Section~\ref{lv-results-sec}).
To reduce the exposure of the patient to ionizing radiation during a CT scan,
the acquisition time for such patients is kept as short as possible.
Thus, no (additional) flow data
(e.g., using Doppler echocardiography~\cite{NishimuraMillerjrCallahanBenassiSewardTajik1985})
could be acquired, such that the data from numerical models are the only data available to assess cardiac function
and, in particular, how blood flow through the LA and LV are affected.

For the computational model, flow over the LA and LV subdomains
are modeled separately and coupling is enforced at the common interface
using appropriate coupling constraints defined in the following.
The common interface is selected as the planar section,
where the mitral valve (MV) would be located.
The MV geometry and deformation were not available from the data set
provided, and thus, the MV opening at $t = 0~s$ is approximated as an ellipse
with minor and major axes of $l_{MV}^{min} \approx 2.66~cm$ and $l_{MV}^{max} \approx 4.64~cm$.
The MV is open for $t \in (0, 0.37]~s$ (diastole) and closed for $t \in (0.37, 0.8]~s$ (systole).
To switch between diastole and systole, we introduce a scalar function,
\begin{alignat}{4}
    \alpha (t)
    &= \begin{cases}
        0 & \text{for } t \in (0, 0.37], \\
        1 & \text{for } t \in (0.37, 0.8].
    \end{cases}\label{alpha-switch-eqn}
\end{alignat}

The domains and domain boundaries are denoted as illustrated in Figure~\ref{lv-flow-application-coarse-mesh-views-fig}.
Blood is modeled as an incompressible Newtonian fluid
with density $\rho = 1.025~g / cm^3$
and viscosity $\mu = 0.004~Pa \cdot s$.
Flow in the atrium and ventricle is governed
by the nonconservative ALE Navier-Stokes equations for a Newtonian fluid.
This enables the solution of fluid flow in the moving domains $\Omega_A$
and $\Omega_V$, with the motion given by domain velocities $w_A$ and $w_V$.
The surface motion of both domains were extracted from CT data,
with the internal motion defined using a pseudo-elastic problem
with element based stiffening (see~\cite{BalmusMassingHoffmanRazaviNordsletten2020}).

The strong form equations, coupling, boundary and initial conditions are given as,
\begin{alignat}{4}
    \rho \partial_t \boldsymbol{v}_A
    + \rho \left( \boldsymbol{v}_A - \boldsymbol{w}_A \right) \cdot \nabla_{\boldsymbol{x}} \boldsymbol{v}_A
    - \nabla_{\boldsymbol{x}} \cdot \boldsymbol{\sigma}_A
    &= \boldsymbol{0}
    &&\quad \text{in } \Omega_A, \label{lv-atrium-momentum-eqn} \\
    \nabla_{\boldsymbol{x}} \cdot \boldsymbol{v}_A
    &= 0
    &&\quad \text{in } \Omega_A, \label{lv-atrium-div-eqn} \\
    \boldsymbol{v}_A
    &= \boldsymbol{w}_A
    &&\quad \text{on } \Gamma_{WA}, \\
    \boldsymbol{\sigma}_A \cdot \boldsymbol{n}_A
    &= \boldsymbol{t}_A
    &&\quad \text{on } \Gamma_{Vn}, \label{lv-atrium-veins-BC-eqn} \\
    \boldsymbol{v}_A (\cdot, 0)
    &= \boldsymbol{w}_A
    &&\quad \text{on } \Omega_A (0), \\
        \cline{1-4}
        \rho \partial_t \boldsymbol{v}_V
    + \rho \left( \boldsymbol{v}_V - \boldsymbol{w}_V \right) \cdot \nabla_{\boldsymbol{x}} \boldsymbol{v}_V
    - \nabla_{\boldsymbol{x}} \cdot \boldsymbol{\sigma}_V
    &= \boldsymbol{0}
    &&\quad \text{in } \Omega_V, \label{lv-ventricle-momentum-eqn} \\
    \nabla_{\boldsymbol{x}} \cdot \boldsymbol{v}_V
    &= 0
    &&\quad \text{in } \Omega_V, \label{lv-ventricle-div-eqn} \\
    \boldsymbol{v}_V
    &= \boldsymbol{w}_V
    &&\quad \text{on } \Gamma_{WV}, \\
    \boldsymbol{v}_V
    &= \boldsymbol{w}_V
    &&\quad \text{on } \Gamma_{AV} \nonumber \\
    &&&\quad \text{for } t \in (0, 0.37], \label{lv-ventricle-aortic-valve-BC-diastole-eqn} \\
    \boldsymbol{\sigma}_V \cdot \boldsymbol{n}_V
    &= \boldsymbol{t}_V
    &&\quad \text{on } \Gamma_{AV} \nonumber \\
    &&&\quad \text{for } t \in (0.37, 0.8], \label{lv-ventricle-aortic-valve-BC-systole-eqn} \\
    \boldsymbol{v}_V
    &= \boldsymbol{w}_V
    &&\quad \text{on } \Gamma_{CV} \setminus \Gamma_{CVo} \nonumber \\
    &&&\quad \text{for } t \in (0, 0.37], \\
    \boldsymbol{v}_V (\cdot, 0)
    &= \boldsymbol{w}_V
    &&\quad \text{on } \Omega_V (0), \\
        \cline{1-4}
        \boldsymbol{v}_A
    - \alpha \boldsymbol{w}_A
    - (1 - \alpha) \boldsymbol{v}_V
    &= \boldsymbol{0}
    &&\quad \text{on } \Omega_{LM}, \label{lv-atrium-coupling-eqn} \\
                \alpha \left( \boldsymbol{v}_V - \boldsymbol{w}_V \right)
    + (1 - \alpha) \left( \boldsymbol{\sigma}_V \cdot \boldsymbol{n}_V + \boldsymbol{\sigma}_A \cdot \boldsymbol{n}_A \right)
    &= \boldsymbol{0}
    &&\quad \text{on } \Omega_{LM}, \label{lv-ventricle-coupling-eqn}
\end{alignat}
with velocity $\boldsymbol{v}_A$ (atrium) and $\boldsymbol{v}_V$ (ventricle),
domain velocity $\boldsymbol{w}_A$ (atrium) and $\boldsymbol{w}_V$ (ventricle)
obtained from CT data (see Section~\ref{lv-flow-application-model-problem-sec}),
outward boundary normal $\boldsymbol{n}_A$ (atrium) and $\boldsymbol{n}_V$ (ventricle),
and Cauchy stress tensor,
\begin{alignat}{4}
    \boldsymbol{\sigma}_A
    &= \mu D \boldsymbol{v}_A - p_A \boldsymbol{I}, \\
    \boldsymbol{\sigma}_V
    &= \mu D \boldsymbol{v}_V - p_V \boldsymbol{I},
\end{alignat}
with pressure $p_A$ (atrium) and $p_V$ (ventricle).
Further, the scalar function $\alpha = \alpha (t)$ in Equation~\eqref{lv-atrium-coupling-eqn}
and Equation~\eqref{lv-ventricle-coupling-eqn}
is given by Equation~\eqref{alpha-switch-eqn}
and enables switching the coupling constraints between diastole and systole:
(i) during diastole, continuity of velocity and traction is enforced across the coupling boundary
and
(ii) during systole, a no-slip condition is prescribed on the coupling boundary.

Further, outflow stabilization is employed at the veins
(see Equation~\ref{lv-atrium-veins-BC-eqn}) to deal with potential reflow
on~$\Gamma_{Vn}$,
and at the aortic valve boundary during systole
(see Equation~\eqref{lv-ventricle-aortic-valve-BC-systole-eqn}),
\begin{alignat}{4}
    \boldsymbol{t}_A
    &= \frac{\rho \beta}{2}
        \frac{ \left( \boldsymbol{v}_A \cdot \boldsymbol{n}_A \right)^2 }{ \left( \boldsymbol{v}_A \cdot \boldsymbol{n}_A \right)^2 + 0.01}
        \min \left\{ \boldsymbol{v}_A \cdot \boldsymbol{n}_A, 0 \right\}
        \boldsymbol{v}_A, \label{lv-atrium-outflow-stabilization-eqn} \\
    \boldsymbol{t}_V
    &= \frac{\rho \beta}{2}
        \frac{ \left( \boldsymbol{v}_V \cdot \boldsymbol{n}_V \right)^2 }{ \left( \boldsymbol{v}_V \cdot \boldsymbol{n}_V \right)^2 + 0.01}
        \min \left\{ \boldsymbol{v}_V \cdot \boldsymbol{n}_V, 0 \right\}
        \boldsymbol{v}_V. \label{lv-ventricle-outflow-stabilization-eqn}
\end{alignat}

Note, that Equation~\eqref{lv-atrium-outflow-stabilization-eqn}
and Equation~\eqref{lv-ventricle-outflow-stabilization-eqn}
provide a modified version of the outflow stabilization proposed
in~\cite{BazilevsGoheanHughesMoserZhang2008}
with a scaling of~$\beta = 0.2$,
as suggested in~\cite{MoghadamBazilevsHsiaVignonclementelMarsden2011}.

On the coupling domain, Lagrange multiplier variables are defined
as $\boldsymbol{\lambda}_A = \boldsymbol{\sigma}_A \cdot \boldsymbol{n}_A$
and $\boldsymbol{\lambda}_V = - \boldsymbol{\sigma}_V \cdot \boldsymbol{n}_V$
to enforce the coupling constraints,
such that during dyastole: $\boldsymbol{\lambda}_A - \boldsymbol{\lambda}_V = \boldsymbol{0}$.
The flow is modeled by a stabilized general Galerkin scheme
(instead of using, e.g., an inf-sup stable Taylor-Hood finite element discretization scheme)
for the incompressible Navier-Stokes equations;
namely the cG(1)cG(1) scheme as given in the study of
Hoffman et al.~\cite{HoffmanJanssonDeabreu2011}.
The scheme was implemented in CHeart~\cite{CHeart,LeeEtAl2016},
and validated in a previous work~\cite{HessenthalerRoehrleNordsletten2017}.
\subsubsection{Spatiotemporal discretization}
\label{lv-flow-application-spatiotemporal-discretization-sec}
The temporal domain is discretized using $800$~equidistant time steps
$0~s = t_0 < t_1 < \ldots < t_{N_0 - 1} = 0.8~s$
with constant time step size~$\delta_0 = t_{n+1} - t_n = 0.001~s$ for~$n = 0, \ldots, N_0 - 2$.
A backward Euler time discretization scheme is employed
to discretize the flow problem in time.\footnote{Preliminary
numerical tests showed that using a backward Euler-like scheme
with midpoint rule in combination with parallel-in-time integration
can result in unphysical oscillations at inflow and outflow boundaries,
similar to those seen in literature for other problems (see \cite{HessenthalerNordslettenRoehrleSchroderFalgout2018}),
causing the scheme to be nonconvergent.
Several works have found the benefit of time-discretization schemes with the property of L-stability
for parallel-in-time discretization schemes
\cite{DobrevKolevPeterssonSchroder2017,Southworth2019,HessenthalerSouthworthNordslettenRoehrleFalgoutSchroder2020}.
In this work, however,
a pure backward Euler and thus, L-stable time discretization scheme was employed.}
The atrial and ventricular domains, $\Omega_A$ and $\Omega_V$,
are discretized using $66484$ and $69716$~tetrahedral elements
with spatial step sizes in the range of
$0.163 - 0.290~cm$ and $0.194 - 0.287~cm$.
The coupling domain $\Omega_{LM}$ is discretized using $945$~triangular elements
that conform with the tetrahedral elements at the coupling boundary
of the atrium and ventricle.
In the following, we use~$\Omega_A^0$ and~$\Omega_V^0$ to refer to the initial meshes
discretizing atrium and ventricle; further,~$\Omega_A^n$ and~$\Omega_V^n$
refer to the respective current meshes.

Finite element discretizations were constructed using $\mathbb{P}^1 - \mathbb{P}^1$~elements
for fluid velocity and pressure and $\mathbb{P}^1$~elements for the Lagrange multipliers
on the coupling domain,
resulting in $55842$~degrees-of-freedom (DOFs).
Here, we omit the weak form as it is quite expansive,
however, refer the interested reader
to the Supplementary Materials~\ref{lv-flow-application-weak-formulation-suppsec}.
\subsubsection{Convergence to a time-periodic steady state}
\label{lv-steady-state-sec}
Similar to the problems described in the previous sections,
the system has to be driven to a periodic steady-state
because no data is available to prescribe a meaningful initial flow condition.
Thus, zero initial flow is assumed and multiple cycles are simulated
(sequential time-stepping) or multiple time-periodic MGRIT iterations
over the length of one cycle are performed.
To detect a time-periodic steady-state,
the rate of inflow / outflow at the pulmonary veins\footnote{Note,
that the flow rates at the MV and AV surfaces can be captured with machine precision
since the fluxes reflect the prescribed volume change in the ventricle
during diastole (flow between LA and LV) and systole (flow across AV surface).}
is measured over the entire cycle length at~$10~ms$ increments.
\subsubsection{Numerical setup}
\label{lv-numerical-setup-sec}
Similar to Section~\ref{stenosed-valve-sec}
and Section~\ref{analytic-fsi-sec}, CHeart and XBraid are employed
to simulate the space-time system using sequential time-stepping ($10$ cardiac cycles)
and time-periodic MGRIT ($10$ iterations) with FCF-relaxation.
Again, a Newton-Raphson-Shimanskii solver~\cite{Shamanskii1967}
is employed to efficiently solve the coupled nonlinear system.
Here, we add a solver step to evaluate the flow rate at all pulmonary veins on-the-fly.
Thus, reported runtimes in Section~\ref{results-sec} include the time it takes to compute the flow rates.
\section{Results}\label{results-sec}
In this section, we present results for each of the application classes
described in Section~\ref{numerical-experiments-sec},
using the methods described in Section~\ref{methodology-sec}.
For each application, we start by describing the numerical solution itself.
We continue by highlighting how the sequential time-stepping solution
converges to a time-periodic steady-state
(by running multiple cycles consecutively) and how this compares
to applying the time-periodic MGRIT algorithm to only a single time period.
Finally, we present timing and speedup results for both
the sequential time-stepping and the parallel-in-time algorithm.
\subsection{Simplified flow in the ascending aorta}
\label{stenosed-valve-results}
\begin{figure}[ht!]
    \centering
    \begin{subfigure}[b]{0.49\textwidth}
        \centering
        \includegraphics[angle=180,width=\linewidth]{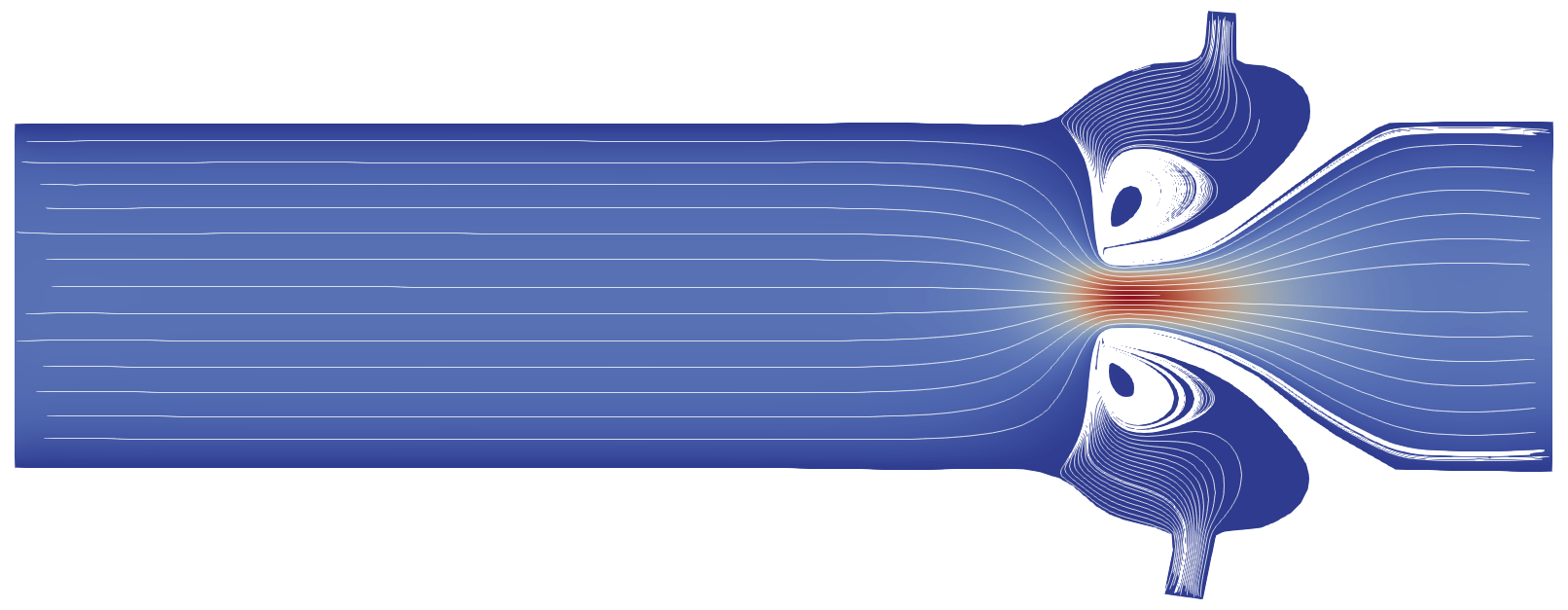}        \vspace{0.35cm}
    \end{subfigure}~~~~~~~~~~\begin{subfigure}[b]{0.15\textwidth}
                \centering
        \begin{tikzpicture}[scale=1]
                        \node [anchor=south west, rotate=90, inner sep=0pt] at (0,0) {
                \includegraphics[width=1.3\linewidth, height=0.125\linewidth]{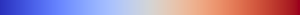}
                };
                        \draw[anchor=west, align=left, black, right] (0.125,0) node {$0$};
            \draw[anchor=west, align=left, black, right] (0.125,1.6) node {$440$};
            \draw[anchor=west, align=left, black, right] (0.125,3.2) node {$880$};
            \draw[anchor=west, align=left, black, right] (1.2,1.6) node {\small\rotatebox{90}{\parbox{1.3\linewidth}{\centering Velocity Magnitude\\$[mm/s]$}}};
        \end{tikzpicture}
            \end{subfigure}
    \caption{Stokes flow:
        Velocity magnitude at $3/8$ of the $10^{th}$ cycle.
        The parabolic inflow becomes a jet between the valves.
        The jet quickly widens to an approximately parabolic flow profile,
        similar to Poiseuille flow in a channel.
        }
    \label{stenosed-valve-stokes-solution-fig}
\end{figure}
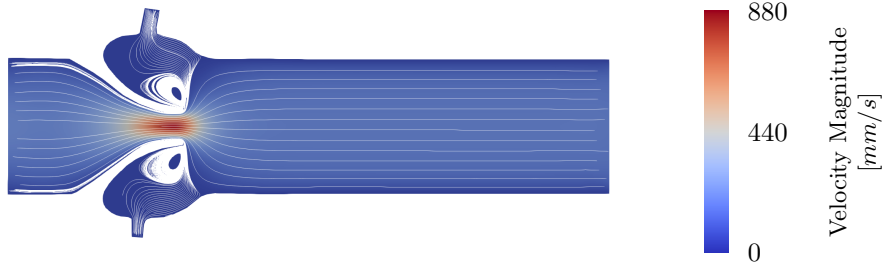

Figure~\ref{stenosed-valve-stokes-solution-fig} shows
a snapshot of the flow field at $3/8$ of the $10^{th}$ cycle due to the pulsatile flow.
The parabolic inflow profile (see Equation~\eqref{stenosed-valve-inflow-eqn})
is constricted between the rigid valves to form a jet.
The jet quickly widens to an approximately parabolic flow profile,
similar to Poiseuille flow in a channel; for example, note the (close to) parallel
streamlines. In the linear flow case considered here, the pulsatile inflow effectively scales
the observed dynamic velocity field across the flow domain.
Further note the small reflow regions between the valves and the coronary arteries.
\FloatBarrier
\FloatBarrier
\subsubsection{Time-periodic steady-state}
\begin{figure}[ht!]
    \centering
    \setlength{\figurewidth}{0.85\textwidth}
    \setlength{\figureheight}{0.25\textwidth}
            \begin{tikzpicture}

\definecolor{myred}{RGB}{128,0,0}
\definecolor{myblue}{RGB}{0,0,128}
\definecolor{mygreen}{RGB}{0,128,128}

\begin{axis}[width=0.951\figurewidth,
height=\figureheight,
at={(0\figurewidth,0\figureheight)},
scale only axis,
clip=false,
xmin=1,
xmax=10,
xtick={1, 2, 3, 4, 5, 6, 7, 8, 9, 10},
xminorticks=true,
xlabel style={font=\color{white!15!black}},
xlabel={Cycle / Iteration $q$},
ymin=1e-9,
ymax=1e3,
ymode=log,
yminorticks=true,
ylabel style={font=\color{white!15!black}},
ylabel={Velocity error},
axis background/.style={fill=white},
legend cell align=left,
legend style={draw=none}
]

\addplot [color=myred, line width=1.0pt]
  table[row sep=crcr]{1 4.77376395e+02\\
2 6.97369734e+00\\
3 1.40623159e-01\\
4 7.34252496e-03\\
5 1.06602139e-03\\
6 1.72250524e-04\\
7 2.83289503e-05\\
8 4.68632243e-06\\
9 7.64142089e-07\\
10 1.09653778e-07\\
};
\addlegendentry{~Sequential time-stepping}

\addplot [color=mygray, dashed, line width=1.0pt]
  table[row sep=crcr]{1 4.77376395e+02\\
2 4.77376395e+01\\
3 4.77376395e+00\\
4 4.77376395e-01\\
5 4.77376395e-02\\
6 4.77376395e-03\\
7 4.77376395e-04\\
8 4.77376395e-05\\
9 4.77376395e-06\\
10 4.77376395e-07\\
};
\addlegendentry{~First-order reduction}

\addplot [color=myred, line width=1.5pt, mark size=3pt, mark=square, mark options={myred}]
  table[row sep=crcr]{1  7.30194907e+00\\
2  1.00185973e-01\\
3  6.81201152e-03\\
4  1.04162638e-03\\
5  1.67481518e-04\\
6  2.73504777e-05\\
7  4.49136861e-06\\
8  7.26038102e-07\\
9  1.02320089e-07\\
10 1.55647189e-09\\
};
\addlegendentry{MGRIT, F-relaxation}

\addplot [color=myblue, line width=1.5pt, mark size=1.5pt, mark=*, mark options={solid, myblue}]
  table[row sep=crcr]{1  7.30563122e+00\\
2  1.00009870e-01\\
3  6.78727708e-03\\
4  1.04096447e-03\\
5  1.67456527e-04\\
6  2.73533032e-05\\
7  4.49207704e-06\\
8  7.26206128e-07\\
9  1.02351212e-07\\
10 1.54939154e-09\\
};
\addlegendentry{MGRIT, FCF-relaxation}

\end{axis}
\end{tikzpicture}    \caption{Convergence of the velocity field at the beginning of each cycle
        at $t = (q - 1) T$ for $q = 1, 2, \ldots$ sequential time-stepping cycles
        and at $t = 0$ for $q = 0, 1, 2 \ldots$ time-periodic MGRIT iterations (note,
        that the zeroth cycle initializes the space-time solution and is omitted here):
        Velocity error with respect to the solution
        at the end of the $10^{th}$ cycle (at $t = 10 T$) from sequential time-stepping.}
    \label{stenosed-valve-error-ic-stokes-seq-vs-pint-fig}
\end{figure}
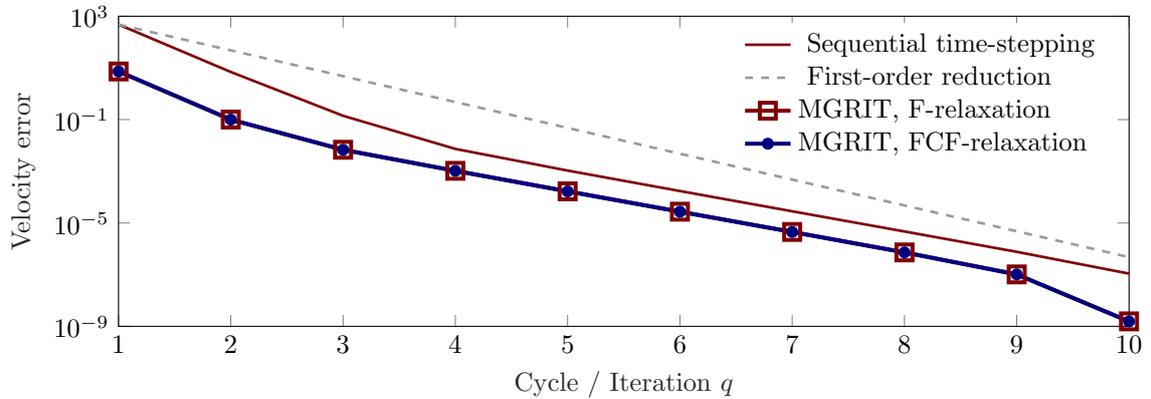

Figure~\ref{stenosed-valve-error-ic-stokes-seq-vs-pint-fig} illustrates convergence
of the solution at the beginning of each cycle
to the solution at the end of the $10^{th}$ cycle for the Stokes flow model.
Here, the average error reduction is approximately one order of magnitude
for each simulated cycle.
This highlights the fact that the accuracy of the periodic steady-state solution
is directly linked to the number of simulated cycles.
As we will see later, each reduction of the cycle-to-cycle error
by approximately one order of magnitude
increases the total computational cost by the computational cost of adding one cycle length
to the temporal domain.
In the context of using parallel-in-time integration with MGRIT,
this motivated reducing the computational cost
by simulating only one cardiac cycle but making the time grid periodic,
see Section~\ref{mgrit-extension-for-time-periodic-problems-sec}.
\FloatBarrier
\FloatBarrier

On the other hand, two-level MGRIT
converges at a similar rate, see Figure~\ref{stenosed-valve-error-ic-stokes-seq-vs-pint-fig}.
That is, one time-periodic MGRIT iteration reduces the error at the beginning of the cycle
similarly quickly compared to one cycle using sequential time-stepping.
Here, F- and FCF-relaxation don't show any significant difference in periodic steady-state
convergence.

\begin{figure}[ht!]
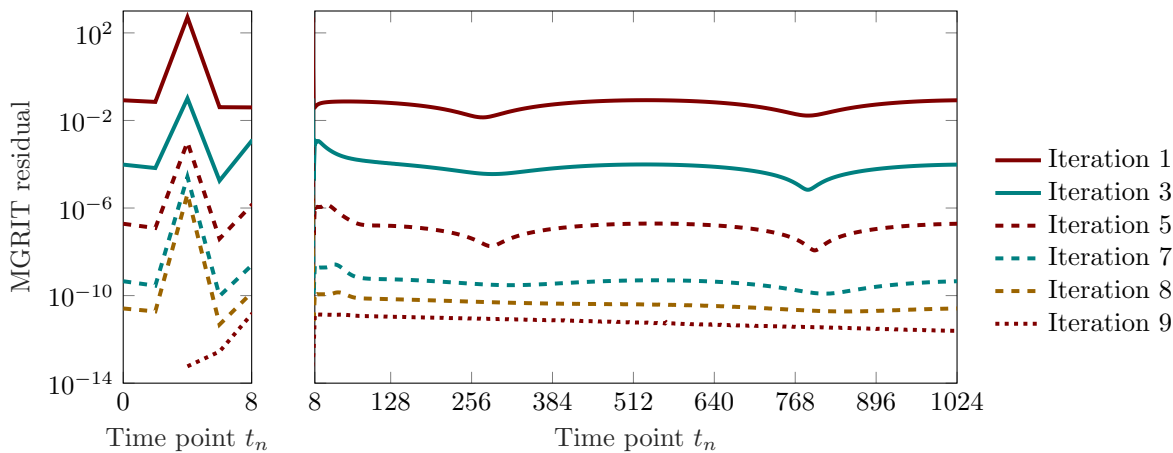

    \centering
    \setlength{\figurewidth}{0.25\textwidth}
    \setlength{\figureheight}{0.3\textwidth}
    \begin{subfigure}[b]{0.2\linewidth}
                        \definecolor{mycolor1}{rgb}{0.00000,0.44700,0.74100}\definecolor{mycolor2}{rgb}{0.85000,0.32500,0.09800}\definecolor{mycolor3}{rgb}{0.92900,0.69400,0.12500}\definecolor{mycolor4}{rgb}{0.49400,0.18400,0.55600}\definecolor{mycolor5}{rgb}{0.46600,0.67400,0.18800}\definecolor{mycolor6}{rgb}{0.30100,0.74500,0.93300}

    \end{subfigure}
    \caption{Two-level convergence of MGRIT with FCF-relaxation and temporal coarsening factor $2$
        for the time-periodic Stokes problem:
        MGRIT residual at C-points for multiple iterations if IC tolerance $10^{-10}$ is used.}
    \label{stenosed-valve-stokes-restime-convergence-FCF-relaxations-with-ic-tolerance-fig}
\end{figure}

Now, considering how the time point-wise MGRIT residual is distributed,
we can connect updating the initial condition during each MGRIT iteration
with the largest observed time-point wise residual:
Figure~\ref{stenosed-valve-stokes-restime-convergence-FCF-relaxations-with-ic-tolerance-fig}
illustrates how large residuals occur at the beginning of the temporal cycle for each MGRIT iteration.
This makes sense as the largest mismatch / discontinuity in the space-time solution will occur
at $t = 0$ due to the continuously updated initial condition; it is this mismatch that is then propagated
in time and appears as a relatively larger residual norm, which we refer to as a \emph{spike}.
Looking closer at Figure~\ref{stenosed-valve-stokes-restime-convergence-FCF-relaxations-with-ic-tolerance-fig},
it is seen that the spikes occur at the fourth time point for FCF-relaxation
(and at second time point for F-relaxation; plot omitted),
which is due to the fact that MGRIT propagates the exact solution across two coarse-grid intervals
for FCF-relaxation (and one coarse-grid interval for F-relaxation).
This is in contrast to traditional (non-periodic) MGRIT,
where the largest residual norms tend to occur towards the end of the temporal domain,
i.e.\ with greater distance to the initial condition.

Since the spike only occurs very locally and right at the beginning of the temporal domain,
skipping the updates of the initial condition once it is considered converged
(see Section~\ref{mgrit-extension-for-time-periodic-problems-sec})
very effectively zeroes the residual
due to the exactness property of traditional MGRIT~\cite{DobrevKolevPeterssonSchroder2017}.
This leaves fairly balanced MGRIT residuals across the temporal domain.

\subsubsection{Parallel performance}
Let's now consider parallel performance with respect to parallelism
applied to the spatial domain, the temporal domain and the spatiotemporal domain.
Runtimes reported here were obtained using two small-scale clusters:
LEAD and TheoSim (see Section~\ref{HPC-machines-sec}).
The reported runtime of the serial algorithm is the best of five runs
and of the parallel algorithms it is the best of one run.
Reported runtimes were further rounded to the nearest integer number in seconds.

If no parallelism is employed, the sequential time-stepping algorithm
takes $523~s$ on LEAD and $332~s$ on TheoSim (referred to as \emph{baseline}).
Employing spatial parallelism, the time-to-solution can be reduced
(see Figure~\ref{stenosed-valve-periodic-stokes-runtimes-LEAD-ASES-nl2-F-FCF-relaxation-fig}) to
$56~s$ on LEAD (speedup: $9.34$x) and $43~s$ on TheoSim (speedup: $7.72$x),
using $16$ processors.

\begin{figure}[ht!]
    \centering
    \begin{subfigure}[b]{0.475\linewidth}
        \setlength{\figurewidth}{0.9\textwidth}
        \setlength{\figureheight}{0.4\textwidth}
                      \begin{tikzpicture}

\begin{axis}[width=0.85\figurewidth,
height=\figureheight,
at={(0\figurewidth,0\figureheight)},
scale only axis,
xmode=log,
xmin=1,
xmax=16,
xlabel style={font=\color{white!15!black}},
xlabel={Number of processors},
xtick={1, 2, 4, 8, 16},
xticklabels={1, 2, 4, 8, 16},
ymode=log,
ymin=3e1,
ymax=2e3,
yminorticks=true,
ylabel style={font=\color{white!15!black}},
ylabel={Wall-clock time $[s]$},
ytick={1e2, 1e3},
axis background/.style={fill=white},
title style={font=\bfseries},
legend style={at={(1.2,0.2)}, anchor=south, legend cell align=left, align=left, fill=none, draw=none}
]

\addplot [color=mygray, dashed, line width=1.5pt]
table[row sep=crcr]{1   523.05\\
16  523.05\\
};

\addplot [color=mygray, dotted, line width=1.5pt]
table[row sep=crcr]{1   523.05\\
2   261.63\\
4   130.76\\
8   65.38\\
16  32.69\\
};

\addplot [color=mygray, dashed, line width=1.5pt, mark size=3pt, mark=square, mark options={solid, mygray}]
table[row sep=crcr]{1   523.05\\
2   264.00\\
4   142.41\\
8   82.761\\
16  55.887\\
};

\addplot [color=myred, line width=1.5pt, mark size=3pt, mark=square, mark options={myred}]
table[row sep=crcr]{1   1490\\
2   919\\
4   652\\
8   504\\
16  446\\
};
\addplot [color=myblue, line width=1.5pt, mark size=3pt, mark=square, mark options={myblue}]
table[row sep=crcr]{1   1170\\
2   688\\
4   445\\
8   318\\
16  265\\
};
\addplot [color=mygreen, line width=1.5pt, mark size=3pt, mark=square, mark options={mygreen}]
table[row sep=crcr]{1   1120\\
2   629\\
4   368\\
8   239\\
16  177\\
};
\addplot [color=myorange, line width=1.5pt, mark size=3pt, mark=square, mark options={myorange}]
table[row sep=crcr]{1   1180\\
2   631\\
4   346\\
8   206\\
16  141\\
};
\addplot [color=myred, dashed, line width=1.5pt, mark size=3pt, mark=square, mark options={solid, myred}]
table[row sep=crcr]{1   1130\\
2   582\\
4   306\\
8   169\\
16  104\\
};

\end{axis}

\end{tikzpicture}
        \caption{LEAD: MGRIT with F-relaxation.
            Maximum observed speedup with time-only parallelism is $5.03$x with $16$ processors.}
        \label{stenosed-valve-periodic-stokes-runtimes-LEAD-nl2-F-relaxation-fig}
    \end{subfigure}\hfill    \begin{subfigure}[b]{0.475\linewidth}
        \setlength{\figurewidth}{0.9\textwidth}
        \setlength{\figureheight}{0.4\textwidth}
                      \begin{tikzpicture}

\begin{axis}[width=0.85\figurewidth,
height=\figureheight,
at={(0\figurewidth,0\figureheight)},
scale only axis,
xmode=log,
xmin=1,
xmax=16,
xlabel style={font=\color{white!15!black}},
xlabel={Number of processors},
xtick={1, 2, 4, 8, 16},
xticklabels={1, 2, 4, 8, 16},
ymode=log,
ymin=3e1,
ymax=2e3,
yminorticks=true,
ylabel style={font=\color{white!15!black}},
ylabel={Wall-clock time $[s]$},
ytick={1e2, 1e3},
axis background/.style={fill=white},
title style={font=\bfseries},
legend style={at={(1.2,0.2)}, anchor=south, legend cell align=left, align=left, fill=none, draw=none}
]

\addplot [color=mygray, dashed, line width=1.5pt]
table[row sep=crcr]{1   523.05\\
16  523.05\\
};

\addplot [color=mygray, dotted, line width=1.5pt]
table[row sep=crcr]{1   523.05\\
2   261.63\\
4   130.76\\
8   65.38\\
16  32.69\\
};

\addplot [color=mygray, dashed, line width=1.5pt, mark size=3pt, mark=square, mark options={solid, mygray}]
table[row sep=crcr]{1   523.05\\
2   264.00\\
4   142.41\\
8   82.761\\
16  55.887\\
};

\addplot [color=myred, line width=1.5pt, mark size=3pt, mark=square, mark options={myred}]
table[row sep=crcr]{1   1430\\
2   826\\
4   538\\
8   384\\
16  315\\
};
\addplot [color=myblue, line width=1.5pt, mark size=3pt, mark=square, mark options={myblue}]
table[row sep=crcr]{1   1400\\
2   800\\
4   472\\
8   313\\
16  234\\
};
\addplot [color=mygreen, line width=1.5pt, mark size=3pt, mark=square, mark options={mygreen}]
table[row sep=crcr]{1   1300\\
2   689\\
4   383\\
8   226\\
16  150\\
};
\addplot [color=myorange, line width=1.5pt, mark size=3pt, mark=square, mark options={myorange}]
table[row sep=crcr]{1   1350\\
2   687\\
4   371\\
8   202\\
16  130\\
};
\addplot [color=myred, dashed, line width=1.5pt, mark size=3pt, mark=square, mark options={solid, myred}]
table[row sep=crcr]{1   1300\\
2   677\\
4   339\\
8   178\\
16  101\\
};

\end{axis}

\end{tikzpicture}
        \caption{LEAD: MGRIT with FCF-relaxation.
            Maximum observed speedup with time-only parallelism is $5.18$x with $16$ processors.}
        \label{stenosed-valve-periodic-stokes-runtimes-LEAD-nl2-FCF-relaxation-fig}
    \end{subfigure}
    {
        \begin{subfigure}[b]{0.9\linewidth}
            \centering
            \setlength{\figurewidth}{0.4\textwidth}
            \setlength{\figureheight}{0.2\textwidth}
                                    \begin{tikzpicture}

\begin{axis}[width=\figurewidth,
height=0.667\figureheight,
at={(0\figurewidth,0\figureheight)},
hide axis,
unbounded coords=jump,
xmin=-0.5,
xmax=0.5,
ymin=0,
ymax=1,
axis background/.style={fill=white},
title style={font=\bfseries},
legend style={at={(0.95,-0.03)}, anchor=north, legend columns=3, legend cell align=left, align=left, fill=none, draw=none}
]

\addlegendimage{color=mygray, dashed, line width=1.5pt}
\addlegendentry{Baseline}

\addlegendimage{color=mygray, dotted, line width=1.5pt}
\addlegendentry{Optimal}

\addlegendimage{color=mygray, dashed, line width=1.5pt, mark size=3pt, mark=square, mark options={solid, mygray}}
\addlegendentry{Space-only}

\addlegendimage{color=myred, line width=1.5pt, mark size=3pt, mark=square, mark options={myred}}
\addlegendentry{Time-only, $m_0 = 2$}

\addlegendimage{color=myblue, line width=1.5pt, mark size=3pt, mark=square, mark options={myblue}}
\addlegendentry{Time-only, $m_0 = 4$}

\addlegendimage{color=mygreen, line width=1.5pt, mark size=3pt, mark=square, mark options={mygreen}}
\addlegendentry{Time-only, $m_0 = 8$}

\addlegendimage{color=myorange, line width=1.5pt, mark size=3pt, mark=square, mark options={myorange}}
\addlegendentry{Time-only, $m_0 = 16~~$}

\addlegendimage{color=myred, dashed, line width=1.5pt, mark size=3pt, mark=square, mark options={solid, myred}}
\addlegendentry{Time-only, $m_0 = 32~~$}

\addlegendimage{color=myblue, dashed, line width=1.5pt, mark size=3pt, mark=square, mark options={solid, myblue}}
\addlegendentry{Space-time, $m_0 = 32$}

\end{axis}
\end{tikzpicture}        \end{subfigure}
    }~\\[3ex]
    \begin{subfigure}[b]{0.475\linewidth}
        \setlength{\figurewidth}{0.9\textwidth}
        \setlength{\figureheight}{0.4\textwidth}
                        \begin{tikzpicture}

\begin{axis}[width=0.85\figurewidth,
height=\figureheight,
at={(0\figurewidth,0\figureheight)},
scale only axis,
xmode=log,
xmin=1,
xmax=256,
xlabel style={font=\color{white!15!black}},
xlabel={Number of processors},
xtick={1, 2, 4, 8, 16, 32, 64, 128, 256},
xticklabels={1, 2, 4, 8, 16, 32, 64, 128, 256},
ymode=log,
ymin=1e0,
ymax=2e3,
yminorticks=true,
ylabel style={font=\color{white!15!black}},
ylabel={Wall-clock time $[s]$},
ytick={1e0, 1e1, 1e2, 1e3},
axis background/.style={fill=white},
title style={font=\bfseries},
legend style={at={(1.25,0.2)}, anchor=south, legend cell align=left, align=left, fill=none, draw=none}
]

\addplot [color=mygray, dashed, line width=1.5pt]
table[row sep=crcr]{1   332.16\\
256  332.16\\
};

\addplot [color=mygray, dotted, line width=1.5pt]
table[row sep=crcr]{1   332.16\\
2   166.08\\
4   83.04\\
8   41.52\\
16  20.76\\
32  10.38\\
64  5.19\\
128 2.60\\
256 1.30\\
};

\addplot [color=mygray, dashed, line width=1.5pt, mark size=3pt, mark=square, mark options={solid, mygray}]
table[row sep=crcr]{1   332.16\\
2   181.66\\
4   99.00\\
8   63.75\\
16  42.82\\
32  260.92\\
};

\addplot [color=myred, line width=1.5pt, mark size=3pt, mark=square, mark options={myred}]
table[row sep=crcr]{1   946\\
2   587\\
4   423\\
8   354\\
16  323\\
32  299\\
};
\addplot [color=myblue, line width=1.5pt, mark size=3pt, mark=square, mark options={myblue}]
table[row sep=crcr]{1   753\\
2   445\\
4   294\\
8   224\\
16  192\\
32  170\\
};
\addplot [color=mygreen, line width=1.5pt, mark size=3pt, mark=square, mark options={mygreen}]
table[row sep=crcr]{1   723\\
2   399\\
4   242\\
8   168\\
16  131\\
32  111\\
};
\addplot [color=myorange, line width=1.5pt, mark size=3pt, mark=square, mark options={myorange}]
table[row sep=crcr]{1   753\\
2   403\\
4   230\\
8   148\\
16  103\\
32  76.6\\
};
\addplot [color=myred, dashed, line width=1.5pt, mark size=3pt, mark=square, mark options={solid, myred}]
table[row sep=crcr]{1   713\\
2   373\\
4   202\\
8   120\\
16  77.6\\
32  75.7\\
};
\addplot [color=myblue, dashed, line width=1.5pt, mark size=3pt, mark=square, mark options={solid, myblue}]
table[row sep=crcr]{4   210\\
16  55.0\\
64  18.3\\
144 12.8\\
256 9.47\\
};

\end{axis}

\end{tikzpicture}
        \caption{TheoSim: MGRIT with F-relaxation.
            Maximum observed speedup with time-only parallelism is $4.37$x with $32$ processors.
            Speedup with space-time parallelism is $36.89$x with $256$ processors.}
        \label{stenosed-valve-periodic-stokes-runtimes-ASES-nl2-F-relaxation-fig}
    \end{subfigure}\hfill    \begin{subfigure}[b]{0.475\linewidth}
        \setlength{\figurewidth}{0.9\textwidth}
        \setlength{\figureheight}{0.4\textwidth}
                        \begin{tikzpicture}

\begin{axis}[width=0.85\figurewidth,
height=\figureheight,
at={(0\figurewidth,0\figureheight)},
scale only axis,
xmode=log,
xmin=1,
xmax=256,
xlabel style={font=\color{white!15!black}},
xlabel={Number of processors},
xtick={1, 2, 4, 8, 16, 32, 64, 128, 256},
xticklabels={1, 2, 4, 8, 16, 32, 64, 128, 256},
ymode=log,
ymin=1e0,
ymax=2e3,
yminorticks=true,
ylabel style={font=\color{white!15!black}},
ylabel={Wall-clock time $[s]$},
ytick={1e0, 1e1, 1e2, 1e3},
axis background/.style={fill=white},
title style={font=\bfseries},
legend style={at={(1.25,0.2)}, anchor=south, legend cell align=left, align=left, fill=none, draw=none}
]

\addplot [color=mygray, dashed, line width=1.5pt]
table[row sep=crcr]{1   332.16\\
256 332.16\\
};

\addplot [color=mygray, dotted, line width=1.5pt]
table[row sep=crcr]{1   332.16\\
2   166.08\\
4   83.04\\
8   41.52\\
16  20.76\\
32  10.38\\
64  5.19\\
128 2.60\\
256 1.30\\
};

\addplot [color=mygray, dashed, line width=1.5pt, mark size=3pt, mark=square, mark options={solid, mygray}]
table[row sep=crcr]{1   332.16\\
2   181.66\\
4   99.00\\
8   63.75\\
16  42.82\\
32  260.92\\
};

\addplot [color=myred, line width=1.5pt, mark size=3pt, mark=square, mark options={myred}]
table[row sep=crcr]{1   917\\
2   533\\
4   356\\
8   268\\
16  233\\
32  204\\
};
\addplot [color=myblue, line width=1.5pt, mark size=3pt, mark=square, mark options={myblue}]
table[row sep=crcr]{1   909\\
2   513\\
4   312\\
8   217\\
16  168\\
32  141\\
};
\addplot [color=mygreen, line width=1.5pt, mark size=3pt, mark=square, mark options={mygreen}]
table[row sep=crcr]{1   829\\
2   442\\
4   253\\
8   162\\
16  113\\
32  83.9\\
};
\addplot [color=myorange, line width=1.5pt, mark size=3pt, mark=square, mark options={myorange}]
table[row sep=crcr]{1   857\\
2   444\\
4   247\\
8   143\\
16  91.1\\
32  61.9\\
};
\addplot [color=myred, dashed, line width=1.5pt, mark size=3pt, mark=square, mark options={solid, myred}]
table[row sep=crcr]{1   831\\
2   432\\
4   225\\
8   129\\
16  80.8\\
32  48.2\\
};
\addplot [color=myblue, dashed, line width=1.5pt, mark size=3pt, mark=square, mark options={solid, myblue}]
table[row sep=crcr]{4   243\\
16  65.0\\
64  20.8\\
144 13.5\\
256 9.71\\};

\end{axis}

\end{tikzpicture}
        \caption{TheoSim: MGRIT with FCF-relaxation.
            Maximum observed speedup with time-only parallelism is $6.92$x with $32$ processors.
            Speedup with space-time parallelism is $33.20$x with $256$ processors.}
        \label{stenosed-valve-periodic-stokes-runtimes-ASES-nl2-FCF-relaxation-fig}
    \end{subfigure}
    \caption{Runtimes on LEAD (top row) and TheoSim (bottom row) for sequential time-stepping ($10$ cycles)
        and time-periodic two-level MGRIT ($q$ iterations, such that the initial condition tolerance $10^{-10}$ was satisfied)
        with F-relaxation (left column) and FCF-relaxation (right column).}
    \label{stenosed-valve-periodic-stokes-runtimes-LEAD-ASES-nl2-F-FCF-relaxation-fig}
\end{figure}

Using two-level MGRIT without parallelism yields a more expensive algorithm,
e.g.\ see Figure~\ref{stenosed-valve-periodic-stokes-runtimes-LEAD-nl2-F-relaxation-fig}.
The benefit of MGRIT is exploited by using more processors and by coarsening more aggressively.
For time-only parallelism, the largest speedup of $6.92$x can be observed
for MGRIT with FCF-relaxation and coarsening factor~$m = 32$ when using $32$~processors on TheoSim.
For space-time parallelism, the largest speedup of $36.89$x can be observed
for MGRIT with F-relaxation and coarsening factor~$m = 32$ when using $256$~processors
(i.e.\ $16$~processors are used for parallelizing in the temporal domain,
while each spatial problem is solved in parallel using $16$~processors) on TheoSim.

More detailed runtime data are reported in Table~\ref{stenosed-valve-runtimes-speedup-LEAD-TheoSim-tab}.
From the data, we can conclude that for time-only parallelism,
FCF-relaxation gives a slightly better speedup than F-relaxation, however,
F-relaxation gives a slightly better speedup if space-time parallelism is employed.\footnote{
F- and FCF-relaxation yield a similar convergence behavior,
see Figure~\ref{stenosed-valve-error-ic-stokes-seq-vs-pint-fig}.
Thus, it is likely that the additional cost associated with FCF-relaxation impacts the runtime
more negatively if more processors are assigned to the spatial component as opposed to the temporal component.}
Overall, the achieved speedups for two-level MGRIT are quite large compared to results reported in literature.
For example, for a two-dimensional compressible fluid dynamics application
Falgout et al.~\cite{FalgoutKatzKolevSchroderWissinkYang2015} reported a speedup of $7.53$x
using time-only parallelism on $4096$~processors, i.e.\ on a considerably larger number of processors.
While the setting is slightly different (e.g., different $Re$ number, different space-time resolution, etc.),
the increased parallel efficiency is arguably achieved through exploiting the periodicity of the solution in time,
and thus, to the development of the new time-periodic MGRIT variant
in Section~\ref{mgrit-extension-for-time-periodic-problems-sec}.
\subsection{Fluid-structure interaction problem with analytic solution}
\label{results-fsi}
In this section, we report results for the linear and nonlinear FSI problems with analytic solutions
in Sections~\ref{results-linear-fsi} and \ref{results-nonlinear-fsi},
respectively.
\subsubsection{Transient interaction between flow and linear solid}
\label{results-linear-fsi}
\begin{figure}[ht!]
    \centering
    \begin{minipage}{\textwidth}
        \hspace{-10pt}
        \vfill
        \begin{minipage}{0.5\textwidth}
            \centering
            \setlength{\figurewidth}{0.8\textwidth}
            \setlength{\figureheight}{0.4\textwidth}
                        \begin{tikzpicture}

\begin{axis}[width=\figurewidth,
height=0.979\figureheight,
at={(0\figurewidth,0\figureheight)},
scale only axis,
xmin=-0.5,
xmax=0.5,
xtick={-0.4, -0.2,    0,  0.2,  0.4},
xlabel style={font=\color{white!15!black}},
xlabel={$v_f$},
ymin=-0,
ymax=1,
ytick={  0, 0.2, 0.4, 0.6, 0.8,   1},
ylabel style={font=\color{white!15!black}},
ylabel={$y$},
axis background/.style={fill=white},
legend style={at={(0.97,0.03)}, anchor=south east, legend cell align=left, align=left, fill=none, draw=none}
]
\addplot [color=black, line width=1.0pt]
  table[row sep=crcr]{-0.000999985894381663	0\\
-0.000999985488143186	0.01\\
-0.000999984284423869	0.02\\
-0.000999982329732123	0.03\\
-0.000999979706591447	0.04\\
-0.000999976540848007	0.05\\
-0.000999973011467733	0.06\\
-0.000999969362342958	0.07\\
-0.000999965915427412	0.08\\
-0.000999963084273058	0.09\\
-0.000999961386748144	0.1\\
-0.000999961455371473	0.11\\
-0.000999964043306637	0.12\\
-0.000999970023631054	0.13\\
-0.000999980379044604	0.14\\
-0.000999996178737222	0.15\\
-0.00100001853873012	0.16\\
-0.00100004856169005	0.17\\
-0.00100008725205257	0.18\\
-0.00100013540235608	0.19\\
-0.00100019344707624	0.2\\
-0.00100026128106872	0.21\\
-0.00100033804109817	0.22\\
-0.00100042185098668	0.23\\
-0.00100050953379144	0.24\\
-0.00100059629825611	0.25\\
-0.00100067541169373	0.26\\
-0.00100073787754539	0.27\\
-0.00100077214316537	0.28\\
-0.00100076387188796	0.29\\
-0.00100069582301438	0.3\\
-0.00100054789377232	0.31\\
-0.00100029738813258	0.32\\
-0.000999919588000479	0.33\\
-0.000999388711873746	0.34\\
-0.00099867935342821	0.35\\
-0.000997768496195383	0.36\\
-0.000996638198721886	0.37\\
-0.000995279035182248	0.38\\
-0.00099369435683523	0.39\\
-0.000991905407137664	0.4\\
-0.000989957274688696	0.41\\
-0.000987925600287944	0.42\\
-0.000985923864141489	0.43\\
-0.000984110963857339	0.44\\
-0.000982698651225096	0.45\\
-0.000981958224867788	0.46\\
-0.000982225677326116	0.47\\
-0.000983904271912818	0.48\\
-0.000987463282716696	0.49\\
-0.000993431380270668	0.5\\
-0.00100238290022905	0.51\\
-0.00101491501323908	0.52\\
-0.00103161364797293	0.53\\
-0.00105300594037072	0.54\\
-0.00107949703292546	0.55\\
-0.00111128927901406	0.56\\
-0.00114828237762174	0.57\\
-0.00118995373926896	0.58\\
-0.00123521953584101	0.59\\
-0.00128227848904659	0.6\\
-0.00132844257611248	0.61\\
-0.00136996154071261	0.62\\
-0.00140185143958793	0.63\\
-0.00141774145206379	0.64\\
-0.00140975781301211	0.65\\
-0.00136846892806982	0.66\\
-0.00128292135008213	0.67\\
-0.00114080210391227	0.68\\
-0.000928768497215063	0.69\\
-0.00063299156826408	0.7\\
-0.000239963064726708	0.71\\
0.000262382487290156	0.72\\
0.000883180469780033	0.73\\
0.00162673912203117	0.74\\
0.00249019954837309	0.75\\
0.00346073803238681	0.76\\
0.00451232317677228	0.77\\
0.00560207818471059	0.78\\
0.0066663485729168	0.79\\
0.00761663921392559	0.8\\
0.00833566291461691	0.81\\
0.00867383619240402	0.82\\
0.00844666603130114	0.83\\
0.00743359243572717	0.84\\
0.00537898211992196	0.85\\
0.00199610311230862	0.86\\
-0.00302495976630241	0.87\\
-0.0100043739788609	0.88\\
-0.0192534541937213	0.89\\
-0.0310481896175634	0.9\\
-0.0455942482590174	0.91\\
-0.0629824409900489	0.92\\
-0.0831338990161047	0.93\\
-0.105734657255013	0.94\\
-0.130159981612019	0.95\\
-0.155389669705611	0.96\\
-0.179916730300405	0.97\\
-0.201653339786383	0.98\\
-0.217839808500255	0.99\\
-0.224964474615423	1\\
};

\addplot [color=black!40, line width=1.0pt]
  table[row sep=crcr]{0.134950376538523	0\\
0.134950376562731	0.01\\
0.134950376606365	0.02\\
0.134950376582369	0.03\\
0.13495037634552	0.04\\
0.134950375692607	0.05\\
0.134950374363448	0.06\\
0.134950372043516	0.07\\
0.134950368369136	0.08\\
0.134950362936451	0.09\\
0.134950355315507	0.1\\
0.134950345070978	0.11\\
0.134950331791132	0.12\\
0.134950315126661	0.13\\
0.134950294840893	0.14\\
0.134950270872688	0.15\\
0.134950243412851	0.16\\
0.134950212994248	0.17\\
0.13495018059486	0.18\\
0.134950147751696	0.19\\
0.134950116681806	0.2\\
0.134950090404552	0.21\\
0.134950072856682	0.22\\
0.134950068988798	0.23\\
0.134950084828318	0.24\\
0.13495012749025	0.25\\
0.1349502051131	0.26\\
0.134950326693137	0.27\\
0.134950501786504	0.28\\
0.134950740045475	0.29\\
0.134951050553176	0.3\\
0.134951440920846	0.31\\
0.134951916114007	0.32\\
0.134952476979624	0.33\\
0.134953118456486	0.34\\
0.13495382746673	0.35\\
0.13495458050891	0.36\\
0.134955341003334	0.37\\
0.134956056479835	0.38\\
0.134956655747451	0.39\\
0.134957046245346	0.4\\
0.134957111844663	0.41\\
0.134956711451183	0.42\\
0.134955678846874	0.43\\
0.134953824301669	0.44\\
0.134950938580244	0.45\\
0.134946800055633	0.46\\
0.134941185712961	0.47\\
0.134933886870828	0.48\\
0.134924730450115	0.49\\
0.134913606562545	0.5\\
0.134900503053336	0.51\\
0.134885547389981	0.52\\
0.134869055916652	0.53\\
0.134851589963392	0.54\\
0.134834017583875	0.55\\
0.134817578769579	0.56\\
0.134803950831279	0.57\\
0.134795309238269	0.58\\
0.134794377562126	0.59\\
0.134804458303067	0.6\\
0.134829434325403	0.61\\
0.134873728467339	0.62\\
0.134942206731576	0.63\\
0.135040008465609	0.64\\
0.135172285317878	0.65\\
0.135343829784069	0.66\\
0.135558574180832	0.67\\
0.135818942317661	0.68\\
0.136125039468693	0.69\\
0.136473672027131	0.7\\
0.136857197061091	0.71\\
0.137262214516607	0.72\\
0.137668131663897	0.73\\
0.138045651138531	0.74\\
0.138355261057359	0.75\\
0.138545838460498	0.76\\
0.138553515716934	0.77\\
0.138301003081743	0.78\\
0.137697608290042	0.79\\
0.136640244169336	0.8\\
0.135015765092099	0.81\\
0.132705018923109	0.82\\
0.129589037909442	0.83\\
0.125557813233194	0.84\\
0.120522095631588	0.85\\
0.114428628860143	0.86\\
0.107279142461268	0.87\\
0.0991532924289255	0.88\\
0.0902355288626038	0.89\\
0.0808455738485774	0.9\\
0.0714717959938282	0.91\\
0.0628062569247579	0.92\\
0.055779569063734	0.93\\
0.0515929372158165	0.94\\
0.0517438601273694	0.95\\
0.0580409533830163	0.96\\
0.072602246294396	0.97\\
0.097830144456978	0.98\\
0.13635509935835	0.99\\
0.19093897538784	1\\
};

\addplot [color=black, dashed, line width=1.0pt]
  table[row sep=crcr]{0.150948810173562	0\\
0.150948809794223	0.01\\
0.150948808638986	0.02\\
0.150948806657631	0.03\\
0.150948803771319	0.04\\
0.150948799880097	0.05\\
0.150948794873835	0.06\\
0.15094878864694	0.07\\
0.150948781117272	0.08\\
0.150948772249641	0.09\\
0.150948762084176	0.1\\
0.150948750769689	0.11\\
0.150948738601851	0.12\\
0.150948726065606	0.13\\
0.150948713880682	0.14\\
0.150948703048333	0.15\\
0.15094869489659	0.16\\
0.150948691120209	0.17\\
0.150948693810301	0.18\\
0.150948705467235	0.19\\
0.150948728988943	0.2\\
0.150948767625215	0.21\\
0.150948824887096	0.22\\
0.150948904399223	0.23\\
0.150949009681959	0.24\\
0.150949143849823	0.25\\
0.15094930921315	0.26\\
0.1509495067715	0.27\\
0.150949735590446	0.28\\
0.150949992058352	0.29\\
0.150950269027151	0.3\\
0.150950554851223	0.31\\
0.150950832351934	0.32\\
0.150951077752285	0.33\\
0.150951259647057	0.34\\
0.150951338098585	0.35\\
0.150951263976991	0.36\\
0.150950978695647	0.37\\
0.150950414527	0.38\\
0.150949495719151	0.39\\
0.150948140667466	0.4\\
0.150946265425073	0.41\\
0.150943788857274	0.42\\
0.150940639752694	0.43\\
0.150936766192186	0.44\\
0.150932147437706	0.45\\
0.150926808529183	0.46\\
0.150920837658308	0.47\\
0.150914406214046	0.48\\
0.150907791155271	0.49\\
0.150901399051207	0.5\\
0.150895790731836	0.51\\
0.150891705002096	0.52\\
0.150890079293469	0.53\\
0.150892064458422	0.54\\
0.15089903016901	0.55\\
0.150912556583321	0.56\\
0.150934407128215	0.57\\
0.150966476466178	0.58\\
0.151010707039686	0.59\\
0.151068967112079	0.6\\
0.151142883068244	0.61\\
0.151233619046356	0.62\\
0.15134159791567	0.63\\
0.151466159392479	0.64\\
0.151605153916703	0.65\\
0.151754475029871	0.66\\
0.151907538640982	0.67\\
0.15205472496771	0.68\\
0.152182808292063	0.69\\
0.152274411106572	0.7\\
0.152307532788006	0.71\\
0.152255218520265	0.72\\
0.15208545150599	0.73\\
0.151761370018722	0.74\\
0.151241929701877	0.75\\
0.150483149482403	0.76\\
0.149440094848312	0.77\\
0.148069762833226	0.78\\
0.146335036074965	0.79\\
0.144209865372102	0.8\\
0.141685817232457	0.81\\
0.13878008037956	0.82\\
0.135544957944969	0.83\\
0.132078774675068	0.84\\
0.128537995388182	0.85\\
0.125150176887578	0.86\\
0.122227156117002	0.87\\
0.120177609537659	0.88\\
0.119517801787409	0.89\\
0.120878978220879	0.9\\
0.125009452968259	0.91\\
0.132769014579602	0.92\\
0.145112835381048	0.93\\
0.163061657557327	0.94\\
0.187654678468612	0.95\\
0.219881321696976	0.96\\
0.260588024083454	0.97\\
0.310356372090992	0.98\\
0.369349477148114	0.99\\
0.437124496712916	1\\
};

\addplot [color=black!40, dashed, line width=1.0pt]
  table[row sep=crcr]{0.0327749547457922	0\\
0.0327749543000845	0.01\\
0.032774952972821	0.02\\
0.0327749507952537	0.03\\
0.0327749478250034	0.04\\
0.0327749441542226	0.05\\
0.0327749399207208	0.06\\
0.0327749353216976	0.07\\
0.032774930629559	0.08\\
0.0327749262090608	0.09\\
0.0327749225347462	0.1\\
0.0327749202072904	0.11\\
0.0327749199669581	0.12\\
0.0327749227019013	0.13\\
0.032774929448507	0.14\\
0.0327749413804499	0.15\\
0.0327749597825552	0.16\\
0.0327749860050687	0.17\\
0.0327750213935262	0.18\\
0.0327750671891821	0.19\\
0.032775124394994	0.2\\
0.0327751936025734	0.21\\
0.0327752747764365	0.22\\
0.0327753669934615	0.23\\
0.0327754681378502	0.24\\
0.0327755745552609	0.25\\
0.0327756806742954	0.26\\
0.0327757786093361	0.27\\
0.0327758577659587	0.28\\
0.0327759044788571	0.29\\
0.0327759017223954	0.3\\
0.0327758289454187	0.31\\
0.0327756620945266	0.32\\
0.0327753739031701	0.33\\
0.0327749345369504	0.34\\
0.0327743126973732	0.35\\
0.0327734772956907	0.36\\
0.0327723998136204	0.37\\
0.0327710574665062	0.38\\
0.0327694372743087	0.39\\
0.0327675411236534	0.4\\
0.0327653918666297	0.41\\
0.0327630404454115	0.42\\
0.0327605739521904	0.43\\
0.0327581244275678	0.44\\
0.032755878063986	0.45\\
0.0327540843112933	0.46\\
0.0327530641777189	0.47\\
0.0327532167818768	0.48\\
0.0327550229431003	0.49\\
0.032759044305181	0.5\\
0.0327659161837905	0.51\\
0.0327763320274978	0.52\\
0.0327910171101857	0.53\\
0.0328106888605566	0.54\\
0.0328360011229833	0.55\\
0.0328674696835803	0.56\\
0.0329053766465169	0.57\\
0.0329496517786566	0.58\\
0.0329997298348606	0.59\\
0.0330543842179096	0.6\\
0.0331115392055537	0.61\\
0.0331680654806253	0.62\\
0.0332195669075594	0.63\\
0.0332601704709183	0.64\\
0.033282336059073	0.65\\
0.0332767083241567	0.66\\
0.0332320390995771	0.67\\
0.0331352156464232	0.68\\
0.0329714370601055	0.69\\
0.0327245880956709	0.7\\
0.032377865902256	0.71\\
0.031914719946902	0.72\\
0.0313201678002506	0.73\\
0.0305825482708195	0.74\\
0.0296957671962528	0.75\\
0.0286620783864268	0.76\\
0.0274954209177622	0.77\\
0.026225302199083	0.78\\
0.0249011718921577	0.79\\
0.0235971728476715	0.8\\
0.0224170798987196	0.81\\
0.0214991442667943	0.82\\
0.0210204498307946	0.83\\
0.0212002580131475	0.84\\
0.022301672467782	0.85\\
0.0246307970116091	0.86\\
0.0285323967492242	0.87\\
0.0343809126402262	0.88\\
0.0425655371154391	0.89\\
0.0534679503461125	0.9\\
0.0674312658366287	0.91\\
0.0847187678107875	0.92\\
0.105461174469834	0.93\\
0.129591468955557	0.94\\
0.156766846760681	0.95\\
0.186278080880678	0.96\\
0.216947652229928	0.97\\
0.247019379466444	0.98\\
0.274044050811342	0.99\\
0.294767741606902	1\\
};

\addplot [color=black, dashdotted, line width=1.0pt]
  table[row sep=crcr]{-0.114531231682908	0\\
-0.114531231798813	0.01\\
-0.114531232118353	0.02\\
-0.114531232556581	0.03\\
-0.114531232970634	0.04\\
-0.114531233158165	0.05\\
-0.114531232855918	0.06\\
-0.114531231739184	0.07\\
-0.114531229423141	0.08\\
-0.114531225467305	0.09\\
-0.11453121938452	0.1\\
-0.114531210656163	0.11\\
-0.114531198755367	0.12\\
-0.114531183180217	0.13\\
-0.114531163498866	0.14\\
-0.114531139408452	0.15\\
-0.114531110809385	0.16\\
-0.114531077896109	0.17\\
-0.114531041264653	0.18\\
-0.114531002036181	0.19\\
-0.114530961994197	0.2\\
-0.114530923731127	0.21\\
-0.114530890797443	0.22\\
-0.114530867843502	0.23\\
-0.114530860740615	0.24\\
-0.114530876663788	0.25\\
-0.114530924113961	0.26\\
-0.114531012852725	0.27\\
-0.114531153717544	0.28\\
-0.114531358280859	0.29\\
-0.114531638312473	0.3\\
-0.114532005001989	0.31\\
-0.114532467897478	0.32\\
-0.114533033518908	0.33\\
-0.114533703611266	0.34\\
-0.114534473013911	0.35\\
-0.114535327140932	0.36\\
-0.114536239093517	0.37\\
-0.114537166461069	0.38\\
-0.114538047914333	0.39\\
-0.114538799752371	0.4\\
-0.114539312636429	0.41\\
-0.114539448827898	0.42\\
-0.114539040343746	0.43\\
-0.114537888549168	0.44\\
-0.114535765820165	0.45\\
-0.114532420022845	0.46\\
-0.114527582663665	0.47\\
-0.114520981654748	0.48\\
-0.11451235969715	0.49\\
-0.11450149929495	0.5\\
-0.114488255353178	0.51\\
-0.114472596158006	0.52\\
-0.114454653259757	0.53\\
-0.114434780346836	0.54\\
-0.114413620578355	0.55\\
-0.114392181001578	0.56\\
-0.114371911586008	0.57\\
-0.114354785033012	0.58\\
-0.114343371851811	0.59\\
-0.114340903227552	0.6\\
-0.114351311964109	0.61\\
-0.114379239310596	0.62\\
-0.114429992860384	0.63\\
-0.114509438074879	0.64\\
-0.114623803517151	0.65\\
-0.114779377834317	0.66\\
-0.114982075228445	0.67\\
-0.115236846012033	0.68\\
-0.115546910351115	0.69\\
-0.115912797039008	0.7\\
-0.116331175780018	0.71\\
-0.116793481724954	0.72\\
-0.117284345659993	0.73\\
-0.117779863080417	0.74\\
-0.118245761100041	0.75\\
-0.118635554346456	0.76\\
-0.118888820035805	0.77\\
-0.118929768325718	0.78\\
-0.11866633633379	0.79\\
-0.117990091736932	0.8\\
-0.116777292626556	0.81\\
-0.11489151119951	0.82\\
-0.112188285523626	0.83\\
-0.108522310106187	0.84\\
-0.103757704648877	0.85\\
-0.0977819016171812	0.86\\
-0.0905236554958538	0.87\\
-0.0819755862437446	0.88\\
-0.0722215110397509	0.89\\
-0.0614685749551381	0.9\\
-0.0500838448209336	0.91\\
-0.0386345634320569	0.92\\
-0.0279306568315947	0.93\\
-0.0190673323473444	0.94\\
-0.0134646912994521	0.95\\
-0.012900208094331	0.96\\
-0.01952870867858	0.97\\
-0.0358831436699326	0.98\\
-0.064848042814328	0.99\\
-0.109596130930499	1\\
};

\addplot [color=black!40, dashdotted, line width=1.0pt]
  table[row sep=crcr]{-0.160035240893983	0\\
-0.160035240577062	0.01\\
-0.160035239604852	0.02\\
-0.160035237914218	0.03\\
-0.160035235404038	0.04\\
-0.160035231941631	0.05\\
-0.16003522737229	0.06\\
-0.160035221532418	0.07\\
-0.160035214266831	0.08\\
-0.160035205450843	0.09\\
-0.1600351950177	0.1\\
-0.160035182991814	0.11\\
-0.160035169528026	0.12\\
-0.160035154956789	0.13\\
-0.160035139834649	0.14\\
-0.160035124998759	0.15\\
-0.160035111623283	0.16\\
-0.160035101274523	0.17\\
-0.160035095960288	0.18\\
-0.160035098167601	0.19\\
-0.160035110881144	0.2\\
-0.160035137573078	0.21\\
-0.160035182152993	0.22\\
-0.160035248864965	0.23\\
-0.160035342117048	0.24\\
-0.160035466227341	0.25\\
-0.160035625070183	0.26\\
-0.160035821606455	0.27\\
-0.160036057283678	0.28\\
-0.160036331295154	0.29\\
-0.16003663969315	0.3\\
-0.160036974359733	0.31\\
-0.16003732185075	0.32\\
-0.160037662144253	0.33\\
-0.160037967344768	0.34\\
-0.160038200419605	0.35\\
-0.160038314073018	0.36\\
-0.160038249898369	0.37\\
-0.160037937986868	0.38\\
-0.160037297213061	0.39\\
-0.160036236460074	0.4\\
-0.160034657089281	0.41\\
-0.160032456995916	0.42\\
-0.160029536619423	0.43\\
-0.160025807289237	0.44\\
-0.160021202275561	0.45\\
-0.160015690872075	0.46\\
-0.160009295752968	0.47\\
-0.160002113709001	0.48\\
-0.159994339664241	0.49\\
-0.159986293593959	0.5\\
-0.159978449592936	0.51\\
-0.159971465871222	0.52\\
-0.159966213873588	0.53\\
-0.159963804026234	0.54\\
-0.159965604813649	0.55\\
-0.159973250992915	0.56\\
-0.159988635787989	0.57\\
-0.160013880914051	0.58\\
-0.160051277322776	0.59\\
-0.160103188717515	0.6\\
-0.160171909273549	0.61\\
-0.16025946675341	0.62\\
-0.160367362503309	0.63\\
-0.160496240859327	0.64\\
-0.160645482518307	0.65\\
-0.160812719702669	0.66\\
-0.160993275755066	0.67\\
-0.161179538429122	0.68\\
-0.161360284877034	0.69\\
-0.161519987417502	0.7\\
-0.161638142773354	0.71\\
-0.161688683661109	0.72\\
-0.161639550295998	0.73\\
-0.161452520224027	0.74\\
-0.161083417292112	0.75\\
-0.160482843531768	0.76\\
-0.159597599819198	0.77\\
-0.158372980402456	0.78\\
-0.15675614014925	0.79\\
-0.154700738368255	0.8\\
-0.152173055250588	0.81\\
-0.149159751564967	0.82\\
-0.145677393691656	0.83\\
-0.141783788240188	0.84\\
-0.137591056766494	0.85\\
-0.133280224744723	0.86\\
-0.12911689350446	0.87\\
-0.125467303742938	0.88\\
-0.122813780550203	0.89\\
-0.12176817136853	0.9\\
-0.123081452511996	0.91\\
-0.127647194627904	0.92\\
-0.136496057518473	0.93\\
-0.150777953506109	0.94\\
-0.171728010126228	0.95\\
-0.200612024114615	0.96\\
-0.238646790692189	0.97\\
-0.286890592446805	0.98\\
-0.346099335325785	0.99\\
-0.41654443760511	1\\
};

\end{axis}
\end{tikzpicture}        \end{minipage}
        \begin{minipage}{0.485\textwidth}
            \centering\hfill
            \setlength{\figurewidth}{0.8\textwidth}
            \setlength{\figureheight}{0.35\textwidth}
                        \begin{tikzpicture}

\begin{axis}[width=\figurewidth,
height=0.667\figureheight,
at={(0\figurewidth,0\figureheight)},
hide axis,
unbounded coords=jump,
xmin=-0.5,
xmax=0.5,
ymin=0,
ymax=1,
axis background/.style={fill=white},
title style={font=\bfseries},
legend style={at={(0.5,-0.03)}, anchor=north, legend columns=3, legend cell align=left, align=left, fill=none, draw=none}
]

\addlegendimage{color=black, line width=1.0pt}
\addlegendentry{~$0.000~s$~~}

\addlegendimage{color=black!40, line width=1.0pt}
\addlegendentry{~$0.156~s$~~}

\addlegendimage{color=black, dashed, line width=1.0pt}
\addlegendentry{~$0.312~s$~~}

\addlegendimage{color=black!40, dashed, line width=1.0pt}
\addlegendentry{~$0.469~s$~~}

\addlegendimage{color=black, dashdotted, line width=1.0pt}
\addlegendentry{~$0.625~s$~~}

\addlegendimage{color=black!40, dashdotted, line width=1.0pt}
\addlegendentry{~$0.781~s$~~}

\end{axis}
\end{tikzpicture}            ~\\[2ex]
            \centering
            \setlength{\figurewidth}{0.8\textwidth}
            \setlength{\figureheight}{0.35\textwidth}
                        \vspace{-0.2cm}
            \begin{tikzpicture}

\begin{axis}[width=\figurewidth,
height=0.667\figureheight,
at={(0\figurewidth,0\figureheight)},
scale only axis,
xmin=-0.5,
xmax=0.5,
xtick={-0.4, -0.2,    0,  0.2,  0.4},
xlabel style={font=\color{white!15!black}},
xlabel={$v_s$},
ymin=1,
ymax=1.2,
ytick={  1, 1.2},
ylabel style={font=\color{white!15!black}},
ylabel={$y$},
axis background/.style={fill=white},
legend style={at={(0.5,-0.03)}, anchor=north, legend columns=7, legend cell align=left, align=left, fill=none, draw=none}
]
\addplot [color=black, line width=1.0pt]
  table[row sep=crcr]{-0.224964474615423	1\\
-0.173667632239446	1.01\\
-0.115890301093788	1.02\\
-0.0538009555309868	1.03\\
0.0102700933953159	1.04\\
0.0739181581791023	1.05\\
0.134754426576272	1.06\\
0.190495617495305	1.07\\
0.239049676137355	1.08\\
0.278594292112013	1.09\\
0.307645293620034	1.1\\
0.325112350761498	1.11\\
0.330339897336401	1.12\\
0.323131735277965	1.13\\
0.303758398275489	1.14\\
0.272946998218356	1.15\\
0.231853935540177	1.16\\
0.182021497686875	1.17\\
0.125319974636656	1.18\\
0.0638774639676896	1.19\\
1.07092407185212e-17	1.2\\
};

\addplot [color=black!40, line width=1.0pt]
  table[row sep=crcr]{0.19093897538784	1\\
0.216672086647148	1.01\\
0.239338052673492	1.02\\
0.258086184017756	1.03\\
0.272212833744857	1.04\\
0.28118780641286	1.05\\
0.284674257144036	1.06\\
0.282541333944324	1.07\\
0.274869088785022	1.08\\
0.26194547312606	1.09\\
0.244255530643614	1.1\\
0.222463192776027	1.11\\
0.197386360329903	1.12\\
0.169966206373023	1.13\\
0.141231852524868	1.14\\
0.112261744399333	1.15\\
0.0841431758401234	1.16\\
0.0579314810681466	1.17\\
0.0346104263239421	1.18\\
0.0150552875692779	1.19\\
-2.84606438789094e-18	1.2\\
};

\addplot [color=black, dashed, line width=1.0pt]
  table[row sep=crcr]{0.437124496712916	1\\
0.414420755574245	1.01\\
0.381828496482328	1.02\\
0.340570958318757	1.03\\
0.292196601553797	1.04\\
0.238520992083024	1.05\\
0.181558660176117	1.06\\
0.12344749197893	1.07\\
0.066368491275005	1.08\\
0.0124639229741369	1.09\\
-0.0362430894680351	1.1\\
-0.0779244950637744	1.11\\
-0.11101592492965	1.12\\
-0.134275405517729	1.13\\
-0.146829971841426	1.14\\
-0.148208431228107	1.15\\
-0.138359047923164	1.16\\
-0.117651484814473	1.17\\
-0.0868629294012554	1.18\\
-0.0471489247216079	1.19\\
-1.38716180288599e-17	1.2\\
};

\addplot [color=black!40, dashed, line width=1.0pt]
  table[row sep=crcr]{0.294767741606902	1\\
0.243807584837938	1.01\\
0.184927040854931	1.02\\
0.120335989327965	1.03\\
0.0524586342807008	1.04\\
-0.0161574801095947	1.05\\
-0.0829370828624922	1.06\\
-0.145373830175485	1.07\\
-0.201124372459394	1.08\\
-0.248096303943901	1.09\\
-0.284526693965827	1.1\\
-0.309048252537059	1.11\\
-0.320740646894008	1.12\\
-0.319165043037596	1.13\\
-0.304380575865274	1.14\\
-0.276942129745071	1.15\\
-0.237879512830208	1.16\\
-0.188658806735104	1.17\\
-0.131127342180383	1.18\\
-0.0674443657576927	1.19\\
-1.25672517334143e-17	1.2\\
};

\addplot [color=black, dashdotted, line width=1.0pt]
  table[row sep=crcr]{-0.109596130930499	1\\
-0.143516282133526	1.01\\
-0.176348578123529	1.02\\
-0.206860771055592	1.03\\
-0.233907690211359	1.04\\
-0.256474222062019	1.05\\
-0.273713409079878	1.06\\
-0.284978237390022	1.07\\
-0.289845920221378	1.08\\
-0.288133765760967	1.09\\
-0.279906033865749	1.1\\
-0.265471524288855	1.11\\
-0.245371986937874	1.12\\
-0.220361789146487	1.13\\
-0.191379603078796	1.14\\
-0.159513175882721	1.15\\
-0.125958504824173	1.16\\
-0.0919749496235712	1.17\\
-0.0588379666995382	1.18\\
-0.0277912392781133	1.19\\
-9.23639190380561e-20	1.2\\
};

\addplot [color=black!40, dashdotted, line width=1.0pt]
  table[row sep=crcr]{-0.41654443760511	1\\
-0.403274333451998	1.01\\
-0.38087508213646	1.02\\
-0.350187362883906	1.03\\
-0.312362934192304	1.04\\
-0.268821406519439	1.05\\
-0.221196962063703	1.06\\
-0.171277021329096	1.07\\
-0.120935158414951	1.08\\
-0.0720607828253711	1.09\\
-0.0264882269509458	1.1\\
0.0140720993186025	1.11\\
0.048097902974893	1.12\\
0.0743121419481346	1.13\\
0.0917309545099028	1.14\\
0.0997005851553504	1.15\\
0.0979219210782754	1.16\\
0.0864617183464371	1.17\\
0.0657500964410594	1.18\\
0.0365643951944028	1.19\\
1.24646224453692e-17	1.2\\
};

\end{axis}
\end{tikzpicture}        \end{minipage}
        \vfill
    \end{minipage}
    \caption{The analytic solution for the transient linear FSI case in two dimensions
        with density $\rho_f = \rho_s = 1$, fluid viscosity $\mu_f = 0.01$ and solid stiffness $\mu_s = 0.1$:
        Fluid and solid velocity, $v_f$ (left) and $v_s$ (right)
        along the $y$-axis over one cycle $T = 1.024$.}
    \label{2D-linear-tf-ts-solution-along-y-fig}
\end{figure}
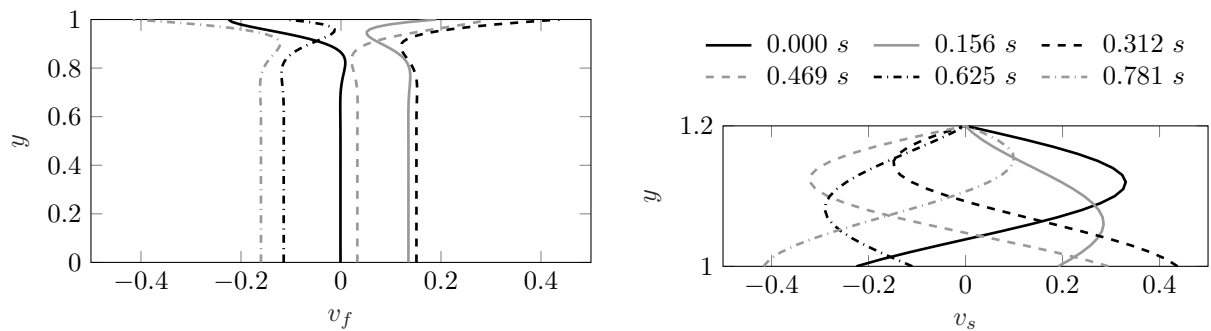

The coupled linear FSI problem is driven by time-periodic boundary conditions
with a time-periodic steady-state solution.
For the given parameter setting, the flow is pulsatile
with strong gradients near the fluid-solid coupling surface,
see Figure~\ref{2D-linear-tf-ts-solution-along-y-fig},
whereas the deformation rate is nonlinear along the $y$-axis.
The flow and deformation rate are further equal at the interface
(as enforced by the coupling conditions)
and constant along the main flow direction (i.e.\ the $x$-axis),
see \cite{HessenthalerBalmusRoehrleNordsletten2020}.
\paragraph{Time-periodic steady-state}
In this section, we compare the error in the fluid velocity variable\footnote{Here,
we choose to track the error in the fluid velocity variable
since the fluid flow field exhibits larger spatial gradients than the solid deformation rate,
see Figure~\ref{2D-linear-tf-ts-solution-along-y-fig}.}
at the end of each cycle using sequential time-stepping
and the end of a single time-periodic cycle for each iteration using time-periodic MGRIT.
Using sequential time-stepping, a time-periodic steady-state can be achieved
by running $10$ cycles sequentially,
see Figure~\ref{2D-linear-tf-ts-mgrit-fsi-error-per-cycle-per-iteration-fig}.
A similar space-time error can be achieved by running $10$ time-periodic MGRIT iterations.

\setlength{\figurewidth}{0.5\textwidth}
\setlength{\figureheight}{0.25\textwidth}
\begin{figure}[ht!]
    \centering
        \begin{tikzpicture}

\begin{axis}[width=0.951\figurewidth,
height=\figureheight,
at={(0\figurewidth,0\figureheight)},
scale only axis,
clip=false,
xmin=1,
xmax=10,
xtick={1, 2, 3, 4, 5, 6, 7, 8, 9, 10},
xminorticks=true,
xlabel style={font=\color{white!15!black}},
xlabel={Cycle / Iteration},
ymin=1e-3,
ymode=log,
yminorticks=true,
ylabel style={font=\color{white!15!black}},
ylabel={Space-time error},
axis background/.style={fill=white},
legend pos={outer north east},
legend style={draw=none, at={(0.3,0.7)}, anchor=west},
legend cell align=left,
]

\addplot [color=black, line width=1.0pt, mark=*]
table[row sep=crcr]{1    0.0316125443929\\ 2   0.0161414913606\\
3   0.00845317609743\\
4   0.00494474159608\\
5    0.00393364726628\\
6   0.00279629808121\\
7   0.00232282134331\\
8   0.00223857970899\\
9   0.00216952366096\\
10   0.00210322831617\\ };
\addlegendentry{Sequential time-stepping}

\addplot [color=black!40, line width=1.0pt, mark=square]
table[row sep=crcr]{1   0.0208302556013\\ 2   0.0123784368321\\
3   0.00776303433437\\
4   0.00558965481546\\
5   0.00402431227496\\
6   0.0030489082723\\
7   0.00260065935554\\
8   0.00237386248605\\ 9   0.00221357151421\\
10   0.00213881198971\\
};
\addlegendentry{MGRIT}

\end{axis}
\end{tikzpicture}    \caption{Space-time error reduction (velocity) for sequential time-stepping over $10$ cycles
        and time-periodic MGRIT over $10$ iterations.}
    \label{2D-linear-tf-ts-mgrit-fsi-error-per-cycle-per-iteration-fig}
\end{figure}
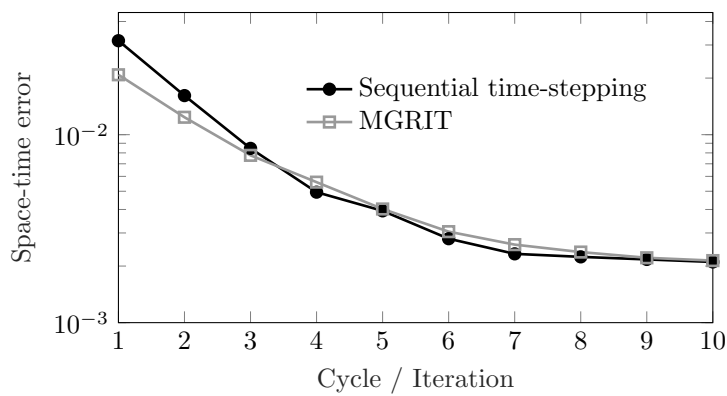

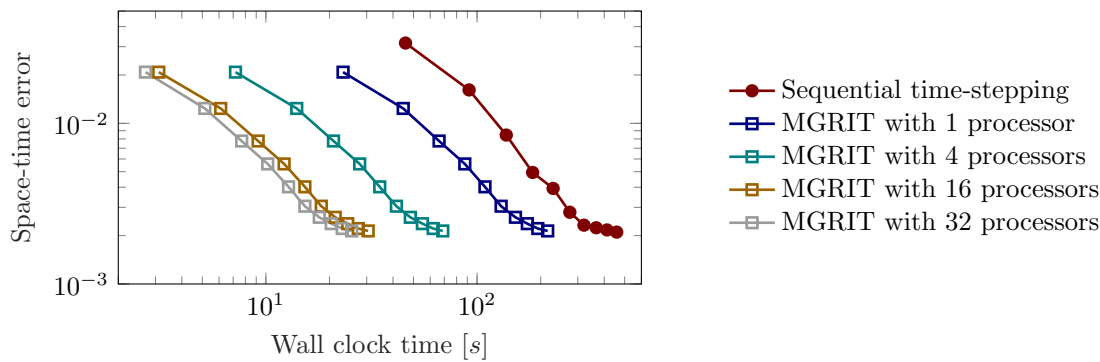
\begin{figure}[ht!]
    \centering
    \begin{minipage}{\textwidth}
        \hfill
        \begin{minipage}{0.55\textwidth}
            \centering
            \setlength{\figurewidth}{0.8\textwidth}
            \setlength{\figureheight}{0.4\textwidth}
                                  \begin{tikzpicture}

\begin{axis}[width=0.951\figurewidth,
height=\figureheight,
at={(0\figurewidth,0\figureheight)},
scale only axis,
clip=false,
xmin=2e0,
xmax=6e2,
xmode=log,
xminorticks=true,
xlabel style={font=\color{white!15!black}},
xlabel={Wall clock time $[s]$},
ymin=1e-3,
ymax=5e-2,
ymode=log,
yminorticks=true,
ylabel style={font=\color{white!15!black}},
ylabel={Space-time error},
axis background/.style={fill=white},
legend pos={outer north east},
legend style={draw=none, at={(1.05,0.5)}, anchor=west},
legend cell align=left,
]

\addplot [color=myred, line width=1.0pt, mark=*]
table[row sep=crcr]{45.838    0.0316125443929\\ 91.676    0.0161414913606\\
137.514   0.00845317609743\\
183.352   0.00494474159608\\
229.19    0.00393364726628\\
275.028   0.00279629808121\\
320.866   0.00232282134331\\
366.704   0.00223857970899\\
412.542   0.00216952366096\\
458.38    0.00210322831617\\ };

\addplot [color=myblue, line width=1.0pt, mark=square]
table[row sep=crcr]{2.32e+01   0.0208302556013\\ 4.47e+01   0.0123784368321\\
6.61e+01   0.00776303433437\\
8.75e+01   0.00558965481546\\
1.09e+02   0.00402431227496\\
1.30e+02   0.0030489082723\\
1.52e+02   0.00260065935554\\
1.73e+02   0.00237386248605\\ 1.94e+02   0.00221357151421\\
2.16e+02   0.00213881198971\\
};

\addplot [color=mygreen, line width=1.0pt, mark=square]
table[row sep=crcr]{7.18e+00   0.0208302556013\\ 1.40e+01   0.0123784368321\\
2.09e+01   0.00776303433437\\
2.78e+01   0.00558965481546\\
3.46e+01   0.00402431227496\\
4.15e+01   0.0030489082723\\
4.83e+01   0.00260065935554\\
5.51e+01   0.00237386248605\\ 6.20e+01   0.00221357151421\\
6.88e+01   0.00213881198971\\
};

\addplot [color=myorange, line width=1.0pt, mark=square]
table[row sep=crcr]{3.11e+00   0.0208302556013\\ 6.11e+00   0.0123784368321\\
9.19e+00   0.00776303433437\\
1.22e+01   0.00558965481546\\
1.53e+01   0.00402431227496\\
1.83e+01   0.0030489082723\\
2.13e+01   0.00260065935554\\
2.44e+01   0.00237386248605\\ 2.74e+01   0.00221357151421\\
3.05e+01   0.00213881198971\\
};

\addplot [color=mygray, line width=1.0pt, mark=square]
table[row sep=crcr]{2.69e+00   0.0208302556013\\ 5.14e+00   0.0123784368321\\
7.68e+00   0.00776303433437\\
1.02e+01   0.00558965481546\\
1.28e+01   0.00402431227496\\
1.53e+01   0.0030489082723\\
1.79e+01   0.00260065935554\\
2.04e+01   0.00237386248605\\ 2.30e+01   0.00221357151421\\
2.55e+01   0.00213881198971\\
};

\end{axis}
\end{tikzpicture}        \end{minipage}
        \begin{minipage}{0.4\textwidth}
            \centering
            \setlength{\figurewidth}{0.9\textwidth}
            \setlength{\figureheight}{0.9\textwidth}
                                  \begin{tikzpicture}

\begin{axis}[width=\figurewidth,
height=\figureheight,
at={(0\figurewidth,0\figureheight)},
hide axis,
unbounded coords=jump,
xmin=0,
xmax=1,
ymin=0,
ymax=1,
axis background/.style={fill=white},
title style={font=\bfseries},
legend style={at={(0.0,0.6)}, anchor=west, legend columns=1, legend cell align=left, align=left, fill=none, draw=none}
]

\addlegendimage{color=myred, line width=1.0pt, mark=*}
\addlegendentry{Sequential time-stepping}
\addlegendimage{color=myblue, line width=1.0pt, mark=square}
\addlegendentry{MGRIT with $1$ processor}
\addlegendimage{color=mygreen, line width=1.0pt, mark=square}
\addlegendentry{MGRIT with $4$ processors}
\addlegendimage{color=myorange, line width=1.0pt, mark=square}
\addlegendentry{MGRIT with $16$ processors~~}
\addlegendimage{color=mygray, line width=1.0pt, mark=square}
\addlegendentry{MGRIT with $32$ processors~~}

\end{axis}
\end{tikzpicture}        \end{minipage}
        \vfill\hfill
    \end{minipage}
    \caption{Space-time error (velocity) compared to wall clock time
        of sequential time stepping and time-periodic MGRIT with zero initial condition.
        Better accuracy can be achieved by simulating additional cycles or performing more MGRIT iterations,
        In general, time-periodic MGRIT can achieve the same order of accuracy at a lower cost.}
    \label{2D-linear-tf-ts-mgrit-fsi-per-cycle-per-iteration-cost-zeroIC-fig}
\end{figure}

Figure~\ref{2D-linear-tf-ts-mgrit-fsi-error-per-cycle-per-iteration-fig}
further illustrates that MGRIT yields a smaller error after the first iteration
compared to sequential time-stepping after the first cycle.
This is due to the fact that the initial setup step for MGRIT is a coarse-grid solve,
and thus, MGRIT starts off from a better starting position for the first iteration.
\paragraph{Parallel performance}
Next, we compare the wall clock times (when using ORCA, see Section~\ref{HPC-machines-sec})
for sequential time-stepping over $10$ cycles
and for the time-periodic MGRIT algorithm over $10$ iterations
(see Table~\ref{2D-linear-tf-ts-mgrit-fsi-nl2-cf08-1FCF-iter8-runtimes-tab})
with no data export (i.e.\ only measuring elapsed times for computation and communication).
The wall clock time of the sequential time-stepping algorithm was $458.4$~seconds for $10$~cycles (approximately
$45.8$ seconds per cycle).
On the other hand, the time-periodic MGRIT algorithm takes $216.0$~seconds when no parallelism is employed
and only $25.5$~seconds when using $32$~processors in the temporal domain (i.e.\ no spatial parallelism),
yielding a maximum speedup of $17.98$x,
see Table~\ref{2D-linear-tf-ts-mgrit-fsi-nl2-cf08-1FCF-iter8-runtimes-tab}.
Figure~\ref{2D-linear-tf-ts-mgrit-fsi-per-cycle-per-iteration-cost-zeroIC-fig} further highlights
the achieved space-time error for each simulated cycle of sequential time-stepping
or each iteration of time-periodic MGRIT.
From Figure~\ref{2D-linear-tf-ts-mgrit-fsi-per-cycle-per-iteration-cost-zeroIC-fig} it is clear,
that time-periodic MGRIT can achieve the same order of accuracy at a lower runtime.
\FloatBarrier
\begin{table}[ht!]
    \centering
    \begin{tabular}{ c | *{4}{ | c }  H  H | c }
        ~
        & Number of
        & Number of
        & Number of
        & Wall clock
        & Speedup
        & Speedup
        & Speedup \\
        Algorithm
        & cycles
        & iterations
        & processors
        & time
        & vs.\ 1 cycle
        & vs.\ 5 cycles
        & vs.\ 10 cycles \\
        \hline\hline
                Time-stepping   & $10$  & -     & $1$   & $458.4~s$ & - & - & -         \\
        MGRIT           & $1$   & $10$  & $1$   & $216.0~s$ & - & - & $2.12$x   \\
        MGRIT           & $1$   & $10$  & $4$   & $68.8~s$  & - & - & $6.66$x   \\
        MGRIT           & $1$   & $10$  & $16$  & $30.5~s$  & - & - & $15.03$x  \\
        MGRIT           & $1$   & $10$  & $32$  & $25.5~s$  & - & - & $17.98$x  \\
    \end{tabular}
    \caption{Wall clock time and speedups for two-level MGRIT with FCF-relaxation
        and temporal coarsening factor $m = 8$.}
    \label{2D-linear-tf-ts-mgrit-fsi-nl2-cf08-1FCF-iter8-runtimes-tab}
\end{table}
\FloatBarrier
\FloatBarrier
\FloatBarrier
\subsubsection{Transient interaction between flow and nonlinear solid}
\label{results-nonlinear-fsi}
\begin{figure}[ht!]
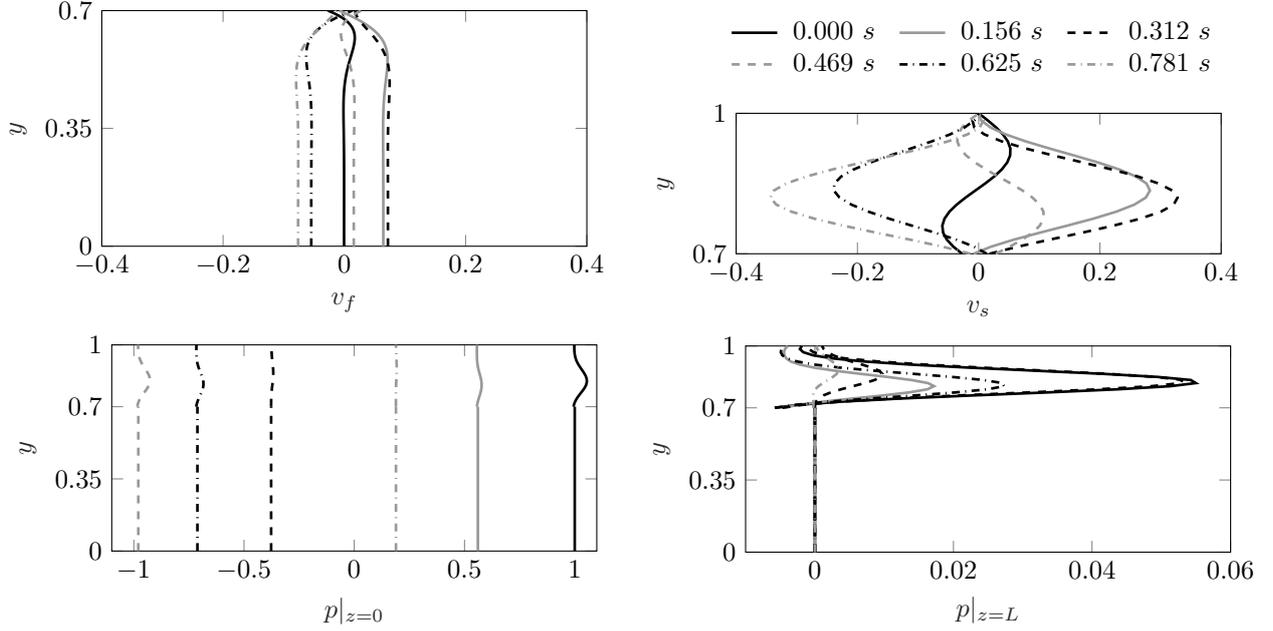

    \centering
    \begin{minipage}{\textwidth}
        \hspace{-10pt}
        \vfill
        \begin{minipage}{0.485\textwidth}
            \centering
            \setlength{\figurewidth}{0.8\textwidth}
            \setlength{\figureheight}{0.4\textwidth}

        \end{minipage}
    \end{minipage}
    \caption{The analytic solution for the transient nonlinear FSI case in three dimensions
        with fluid and solid density $\rho_f = 2.1$ and $\rho_s = 1$, fluid viscosity $\mu_f = 0.03$,
        solid stiffness $\mu_s = 0.1$ and cycle length $T = 1.024$:
        Fluid and solid velocity, $v_f$ (top left) and $v_s$ (top right),
        along the $y$-axis.
        Further, fluid and solid pressure (bottom), $p = p_f$ for $y \in [0, 0.7]$
        and $p = p_s$ for $y \in [0.7, 1]$ over time $t$
        at the inlet ($z = 0$) and the outlet ($z = L$).}
    \label{3D-nonlinear-tf-ts-solution-along-y-fig}
\end{figure}

The coupled nonlinear FSI problem is driven by time-periodic boundary conditions
with a time-periodic steady-state solution.
Similar to the linear case, the flow is pulsatile;
however, with less pronounced gradients near the fluid-solid coupling surface,
see Figure~\ref{3D-nonlinear-tf-ts-solution-along-y-fig},
whereas the deformation rate is nonlinear along the $y$-axis.
The flow and deformation rate are further equal at the interface
(as enforced by the coupling conditions)
and constant along the main flow direction (i.e.\ the $z$-axis),
see \cite{HessenthalerBalmusRoehrleNordsletten2020}.
In contrast to the linear case,
the fluid and solid pressure are discontinuous across the coupling interface
and the solid pressure is varying along the $y$-axis.
\FloatBarrier
\FloatBarrier
\paragraph{Time-periodic steady-state}
First, consider the convergence of the fluid velocity solution
to its time-periodic steady-state for the \emph{coarse} mesh.
Figure~\ref{3D-nonlinear-tf-ts-mgrit-fsi-error-per-cycle-per-iteration-coarse-fig}
illustrates how the fluid velocity error is reduced over $7$ consecutive cycles
using sequential time-stepping.
Starting with a slightly larger error at iteration $1$ (relative to the error
after one sequential time-stepping cycle), time-periodic MGRIT
can reduce the space-time error similarly rapidly
and eventually reduces the error further.
A comparable order of error can be achieved after $6$~time-periodic MGRIT iterations.

Considering a refined space-time discretization,
the same observation holds true although the initial MGRIT error is already smaller
than the error obtained from sequential time-stepping.
\setlength{\figurewidth}{0.5\textwidth}
\setlength{\figureheight}{0.25\textwidth}
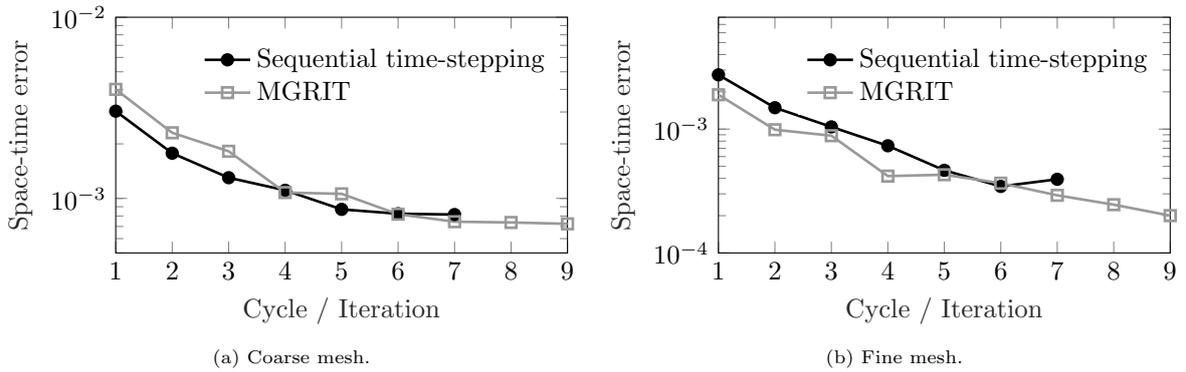
\begin{figure}[ht!]
    \centering
    \begin{subfigure}[b]{0.475\linewidth}
        \centering
        \setlength{\figurewidth}{0.8\textwidth}
        \setlength{\figureheight}{0.4\textwidth}
                        \begin{tikzpicture}

\begin{axis}[width=0.951\figurewidth,
height=\figureheight,
at={(0\figurewidth,0\figureheight)},
scale only axis,
clip=false,
xmin=1,
xmax=9,
xtick={1, 2, 3, 4, 5, 6, 7, 8, 9},
xminorticks=true,
xlabel style={font=\color{white!15!black}},
xlabel={Cycle / Iteration},
ymin=5e-4,
ymax=1e-2,
ymode=log,
yminorticks=true,
ylabel style={font=\color{white!15!black}},
ylabel={Space-time error},
axis background/.style={fill=white},
legend pos={outer north east},
legend style={draw=none, at={(0.175,0.75)}, anchor=west},
legend cell align=left,
]

\addplot [color=black, line width=1.0pt, mark=*]
table[row sep=crcr]{1   0.00303111354044\\ 2   0.00177641201059\\
3   0.00130122009754\\
4   0.00110877673258\\
5   0.000870149945753\\ 6   0.000823389383191\\
7   0.000814674436686\\
};
\addlegendentry{Sequential time-stepping}

\addplot [color=black!40, line width=1.0pt, mark=square]
table[row sep=crcr]{1   0.00400011269713\\ 2   0.00230323229665\\
3   0.00182069900934\\
4   0.00107736033323\\
5   0.00105986871228\\ 6   0.000818227730972\\
7   0.00074385477802\\
8   0.000736856927146\\
9   0.000724005806711\\
};
\addlegendentry{MGRIT}

\end{axis}
\end{tikzpicture}        \caption{Coarse mesh.}
        \label{3D-nonlinear-tf-ts-mgrit-fsi-error-per-cycle-per-iteration-coarse-fig}
    \end{subfigure}
    \begin{subfigure}[b]{0.475\linewidth}
        \centering
        \setlength{\figurewidth}{0.8\textwidth}
        \setlength{\figureheight}{0.4\textwidth}
                        \begin{tikzpicture}

\begin{axis}[width=0.951\figurewidth,
height=\figureheight,
at={(0\figurewidth,0\figureheight)},
scale only axis,
clip=false,
xmin=1,
xmax=9,
xtick={1, 2, 3, 4, 5, 6, 7, 8, 9},
xminorticks=true,
xlabel style={font=\color{white!15!black}},
xlabel={Cycle / Iteration},
ymin=1e-4,
ymax=8e-3,
ymode=log,
yminorticks=true,
ylabel style={font=\color{white!15!black}},
ylabel={Space-time error},
axis background/.style={fill=white},
legend pos={outer north east},
legend style={draw=none, at={(0.175,0.75)}, anchor=west},
legend cell align=left,
]

\addplot [color=black, line width=1.0pt, mark=*]
table[row sep=crcr]{1     0.00274356933347\\ 2      0.00148860984911\\
3      0.00104234502788\\
4      0.000733586324122\\
5      0.00046475500582\\ 6      0.000344648616098\\
7      0.000393508453391\\
};
\addlegendentry{Sequential time-stepping}

\addplot [color=black!40, line width=1.0pt, mark=square]
table[row sep=crcr]{1   0.00189405057001\\ 2   0.000988645537711\\
3   0.000884839320955\\
4   0.000417435299701\\
5   0.000428160705087\\ 6   0.000366062965051\\
7   0.000291975926798\\
8    0.00024538532945\\
9   0.000200222146798\\
};
\addlegendentry{MGRIT}

\end{axis}
\end{tikzpicture}        \caption{Fine mesh.}
        \label{3D-nonlinear-tf-ts-mgrit-fsi-error-per-cycle-per-iteration-medium-fig}
    \end{subfigure}
    \caption{Space-time error reduction (velocity) for sequential time-stepping over $7$ cycles
        and time-periodic MGRIT over $9$ iterations.}
    \label{3D-nonlinear-tf-ts-mgrit-fsi-error-per-cycle-per-iteration-fig}
\end{figure}
\paragraph{Parallel performance}
Comparing wall clock times\footnote{Reported runtimes
for ORCA (\emph{coarse} mesh) and TOM (\emph{fine} mesh).}
of sequential time-stepping and MGRIT for the \emph{coarse} mesh
(see Figure~\ref{3D-nonlinear-tf-ts-mgrit-fsi-per-cycle-per-iteration-cost-vel_zeroIC-coarse-mesh-fig}),
it can be seen that achieved speedups are not as large as for the linear FSI case.
However, significant speedups over sequential time-stepping are achieved
when using $16$ processors to parallelize in the temporal domain.
Depending on the required accuracy
(see Figure~\ref{3D-nonlinear-tf-ts-mgrit-fsi-per-cycle-per-iteration-cost-vel_zeroIC-coarse-mesh-fig}
and Table~\ref{3D-nonlinear-tf-ts-mgrit-fsi-nl2-cf08-1FCF-speedup-coarse-mesh-tab}),
the speedup ranges between $2.96$x and $6.20$x using $16$~processors.
The largest achieved speedup is $6.64$x using $32$~processors.
Thus, the benefit of adding more parallelism beyond $16$~processors is negligible,
while in the linear FSI case, a larger additional speedup can be achieved.
This is likely due to the nonlinear solver taking more iterations toward the end of the temporal domain,
and thus, causing less optimal load balancing.
To achieve an even larger speedup at a higher number of processors, a different decomposition
of the temporal domain could be considered that takes the cost of each time step
(with respect to its position along the temporal domain) into account.
\begin{figure}[ht!]
    \centering
    {
        \begin{subfigure}[b]{0.9\linewidth}
            \centering
            \setlength{\figurewidth}{0.4\textwidth}
            \setlength{\figureheight}{0.2\textwidth}
                                    \begin{tikzpicture}

\begin{axis}[width=\figurewidth,
height=0.667\figureheight,
at={(0\figurewidth,0\figureheight)},
hide axis,
unbounded coords=jump,
xmin=-0.5,
xmax=0.5,
ymin=0,
ymax=1,
axis background/.style={fill=white},
title style={font=\bfseries},
legend style={at={(0.95,-0.03)}, anchor=north, legend columns=3, legend cell align=left, align=left, fill=none, draw=none}
]

\addlegendimage{color=myred, line width=1.0pt, mark=*}
\addlegendentry{Sequential time-stepping}
\addlegendimage{color=myblue, line width=1.0pt, mark=square}
\addlegendentry{MGRIT with $1$ processor}
\addlegendimage{color=mygreen, line width=1.0pt, mark=square}
\addlegendentry{MGRIT with $4$ processors}
\addlegendimage{color=myorange, line width=1.0pt, mark=square}
\addlegendentry{MGRIT with $16$ processors~~}
\addlegendimage{color=mygray, line width=1.0pt, mark=square}
\addlegendentry{MGRIT with $32$ processors~~}
\addlegendimage{color=black!80, line width=1.0pt, mark=square}
\addlegendentry{MGRIT with $64$ processors~~}

\end{axis}
\end{tikzpicture}        \end{subfigure}
    }~\\[3ex]
    \begin{subfigure}[b]{0.475\linewidth}
        \centering
        \setlength{\figurewidth}{0.8\textwidth}
        \setlength{\figureheight}{0.4\textwidth}
                        \begin{tikzpicture}

\begin{axis}[width=0.951\figurewidth,
height=\figureheight,
at={(0\figurewidth,0\figureheight)},
scale only axis,
clip=false,
xmode=log,
xminorticks=true,
xlabel style={font=\color{white!15!black}},
xlabel={Wall clock time $[s]$},
ymin=5e-4,
ymax=6e-3,
ytick={1e-3},
ymode=log,
yminorticks=true,
ylabel style={font=\color{white!15!black}},
ylabel={Space-time error},
axis background/.style={fill=white},
legend pos={outer north east},
legend style={draw=none, at={(1.05,0.5)}, anchor=west},
legend cell align=left,
]

\addplot [color=myred, line width=1.0pt, mark=*]
table[row sep=crcr]{2053.1   0.00303111354044\\ 3939.1   0.00177641201059\\
5999.2   0.00130122009754\\
8032.6   0.00110877673258\\
9744.0   0.000870149945753\\ 11917   0.000823389383191\\
13951   0.000814674436686\\
};

\addplot [color=myblue, line width=1.0pt, mark=square]
table[row sep=crcr]{6.12e+03   0.00400011269713\\ 1.21e+04   0.00230323229665\\
1.80e+04   0.00182069900934\\
2.36e+04   0.00107736033323\\
2.91e+04   0.00105986871228\\ 3.42e+04   0.000818227730972\\
3.92e+04   0.00074385477802\\
4.44e+04   0.000736856927146\\
4.96e+04   0.000724005806711\\
};

\addplot [color=mygreen, line width=1.0pt, mark=square]
table[row sep=crcr]{1.27e+03   0.00400011269713\\ 2.48e+03   0.00230323229665\\
3.58e+03   0.00182069900934\\
4.67e+03   0.00107736033323\\
5.79e+03   0.00105986871228\\ 6.88e+03   0.000818227730972\\
7.94e+03   0.00074385477802\\
8.99e+03   0.000736856927146\\
1.00e+04   0.000724005806711\\
};

\addplot [color=myorange, line width=1.0pt, mark=square]
table[row sep=crcr]{6.40e+02   0.00400011269713\\ 9.78e+02   0.00230323229665\\
1.33e+03   0.00182069900934\\
1.64e+03   0.00107736033323\\
1.95e+03   0.00105986871228\\ 2.25e+03   0.000818227730972\\
2.54e+03   0.00074385477802\\
2.82e+03   0.000736856927146\\
3.11e+03   0.000724005806711\\
};

\addplot [color=mygray, line width=1.0pt, mark=square]
table[row sep=crcr]{6.69e+02   0.00400011269713\\ 9.68e+02   0.00230323229665\\
1.28e+03   0.00182069900934\\
1.58e+03   0.00107736033323\\
1.85e+03   0.00105986871228\\ 2.10e+03   0.000818227730972\\
2.35e+03   0.00074385477802\\
2.60e+03   0.000736856927146\\
2.84e+03   0.000724005806711\\
};

\end{axis}
\end{tikzpicture}        \caption{Coarse mesh:
            The largest achieved speedup is $6.20$x using $16$~processors and $6.64$x using $32$~processors.
            See~Table~\ref{3D-nonlinear-tf-ts-mgrit-fsi-nl2-cf08-1FCF-speedup-coarse-mesh-tab}.}
        \label{3D-nonlinear-tf-ts-mgrit-fsi-per-cycle-per-iteration-cost-vel_zeroIC-coarse-mesh-fig}
    \end{subfigure}\hfill    \begin{subfigure}[b]{0.475\linewidth}
        \centering
        \setlength{\figurewidth}{0.8\textwidth}
        \setlength{\figureheight}{0.4\textwidth}
                        \begin{tikzpicture}

\begin{axis}[width=0.951\figurewidth,
height=\figureheight,
at={(0\figurewidth,0\figureheight)},
scale only axis,
clip=false,
xmin=1e4,
xmax=1e6,
xmode=log,
xminorticks=true,
xlabel style={font=\color{white!15!black}},
xlabel={Wall clock time $[s]$},
ymin=1e-4,
ymax=5e-3,
ymode=log,
yminorticks=true,
ylabel style={font=\color{white!15!black}},
ylabel={Space-time error},
axis background/.style={fill=white},
legend pos={outer north east},
legend style={draw=none, at={(1.05,0.5)}, anchor=west},
legend cell align=left,
]

\addplot [color=myred, line width=1.0pt, mark=*]
table[row sep=crcr]{60170.4     0.00274356933347\\ 130972      0.00148860984911\\
220986      0.00104234502788\\
314441      0.000733586324122\\
391791      0.00046475500582\\ 478072      0.000344648616098\\
563618      0.000393508453391\\
};

\addplot [color=myorange, line width=1.0pt, mark=square]
table[row sep=crcr]{2.29e+04   0.00189405057001\\ 3.35e+04   0.000988645537711\\
4.42e+04   0.000884839320955\\
5.45e+04   0.000417435299701\\
6.48e+04   0.000428160705087\\ 7.47e+04   0.000366062965051\\
8.46e+04   0.000291975926798\\
9.44e+04    0.00024538532945\\
1.04e+05   0.000200222146798\\
};

\addplot [color=mygray, line width=1.0pt, mark=square]
table[row sep=crcr]{2.54e+04   0.00189405057001\\ 3.39e+04   0.000988645537711\\
4.26e+04   0.000884839320955\\
5.07e+04   0.000417435299701\\
5.88e+04   0.000428160705087\\ 6.64e+04   0.000366062965051\\
7.38e+04   0.000291975926798\\
8.11e+04   0.00024538532945\\
8.83e+04   0.000200222146798\\
};

\addplot [color=black!80, line width=1.0pt, mark=square]
table[row sep=crcr]{1.99e+04   0.00189405057001\\ 2.69e+04   0.000988645537711\\
3.44e+04   0.000884839320955\\
4.08e+04   0.000417435299701\\
4.70e+04   0.000428160705087\\ 5.28e+04   0.000366062965051\\
5.84e+04   0.000291975926798\\
6.40e+04   0.00024538532945\\
6.95e+04   0.000200222146798\\
};

\end{axis}
\end{tikzpicture}        \caption{Fine mesh:
            The largest achieved speedup is $7.73$x using $32$~processors and $9.60$x using $64$~processors.
            See~Table~\ref{3D-nonlinear-tf-ts-mgrit-fsi-nl2-cf08-1FCF-speedup-medium-mesh-tab}.}
        \label{3D-nonlinear-tf-ts-mgrit-fsi-per-cycle-per-iteration-cost-vel_zeroIC-medium-mesh-fig}
    \end{subfigure}
    \caption{Space-time error (velocity) compared to wall clock time
        of sequential time stepping and time-periodic MGRIT for two different mesh sizes
        when using a zero initial condition for the fluid velocity
        and the analytic solution for all other variables.
        Note, that the MGRIT run with $1$~processor was omitted for the fine mesh.
        }
    \label{3D-nonlinear-tf-ts-mgrit-fsi-per-cycle-per-iteration-cost-vel_zeroIC-fig}
\end{figure}
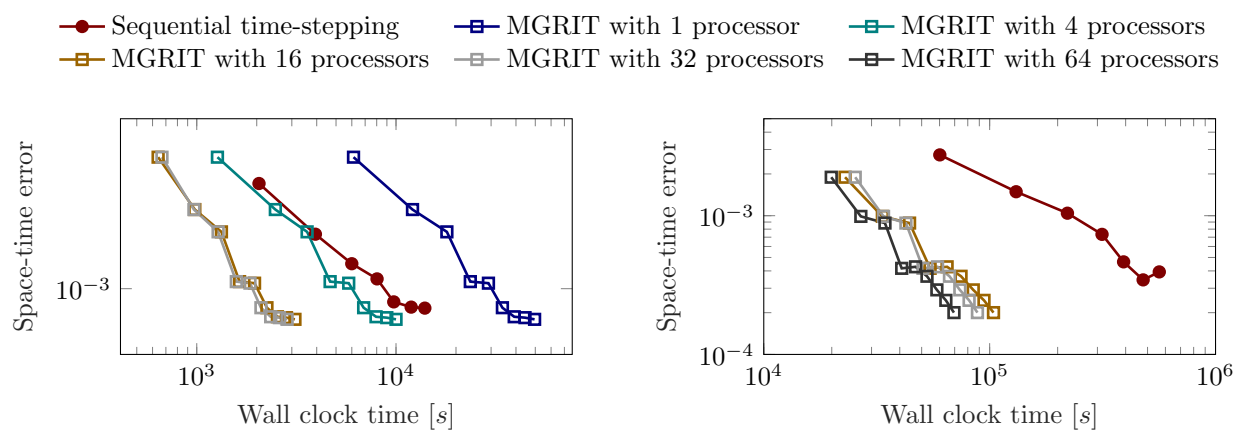

Now considering the \emph{fine} mesh, a larger number of processors can be employed.
Figure~\ref{3D-nonlinear-tf-ts-mgrit-fsi-per-cycle-per-iteration-cost-vel_zeroIC-medium-mesh-fig}
illustrates accuracy (for the velocity approximation) compared to associated cost
for sequential time-stepping, and time-periodic MGRIT with $16$, $32$ and $64$ processors in time.
Again, it is possible to rapidly reduce the space-time error with time-periodic MGRIT,
giving the new algorithm a significant edge over sequential time-stepping in terms of wall clock time.
Considering a comparable space-time error after $6$~cycles (sequential time-stepping)
and $6$~iterations (time-periodic MGRIT),
observed speedups range from $6.40$x using $16$ processors to $9.05$x using $64$ processors,
see Table~\ref{3D-nonlinear-tf-ts-mgrit-fsi-nl2-cf08-1FCF-speedup-medium-mesh-tab}.
It is further noted, that a small increase
of the space-time error is observed between cycle $5$ and $6$
for sequential time-stepping, which similarly occurs for the time-periodic MGRIT algorithm
between iteration $4$ and $5$. After iteration~$5$, however, the time-periodic MGRIT algorithm
can further reduce the space-time error, while the wall clock time is still smaller
than for sequential time-stepping.
\subsection{Flow in a left atrium / left ventricle geometry}
\label{lv-results-sec}
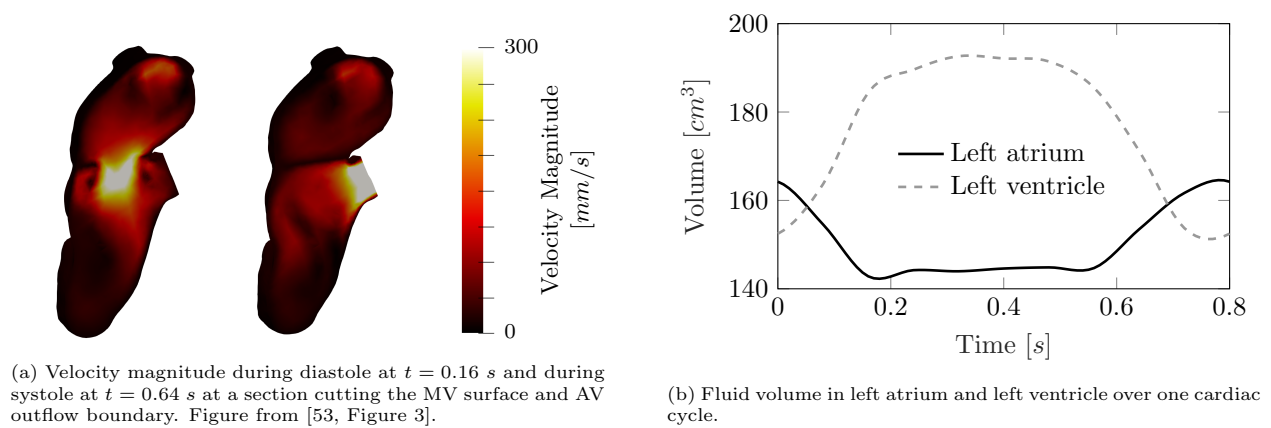
\begin{figure}[ht!]
    \centering
    \begin{subfigure}[b]{0.475\linewidth}
        \includegraphics[height=0.525\linewidth]{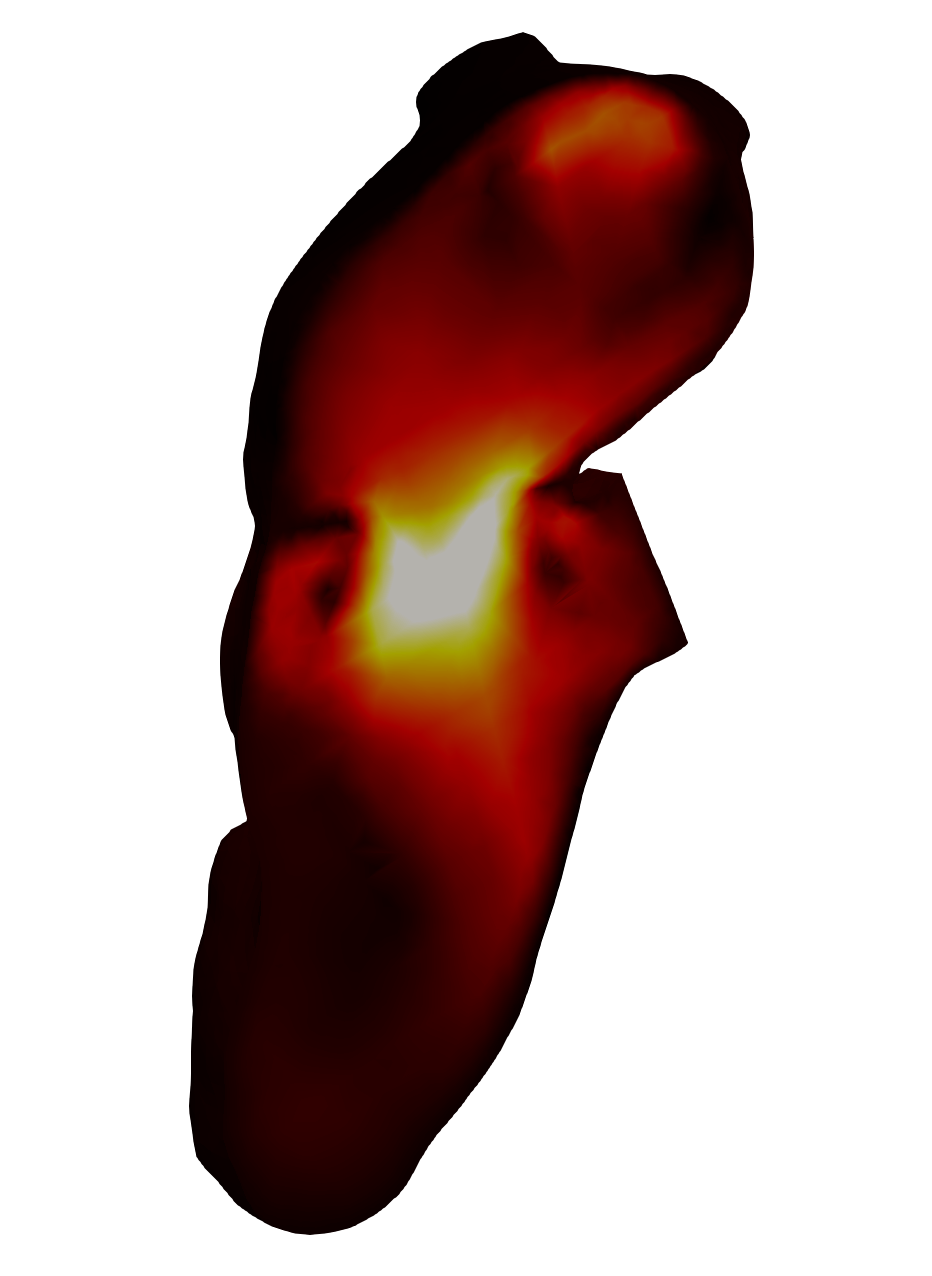}\hspace{-0.5cm}
        \includegraphics[height=0.525\linewidth]{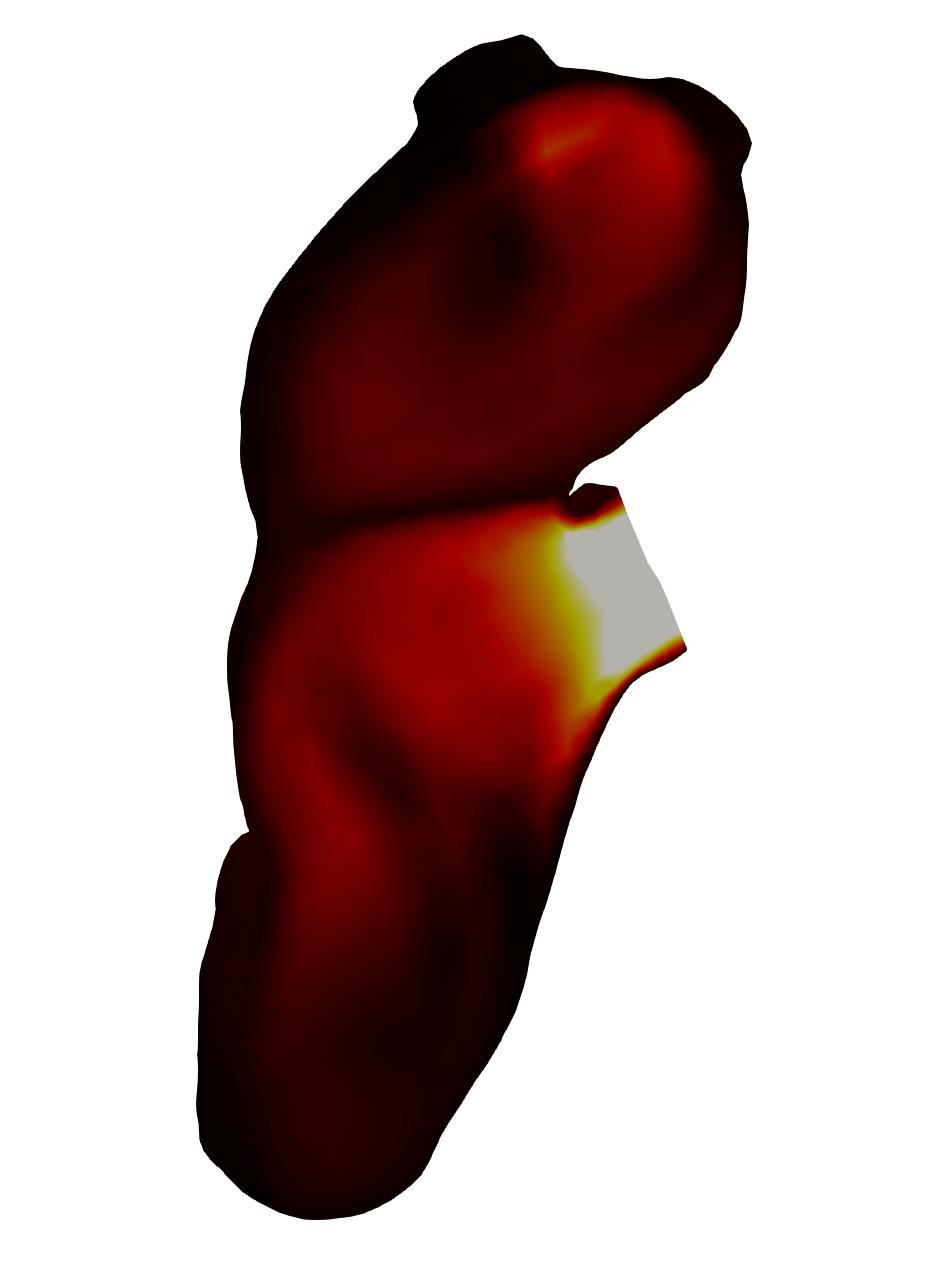}
            \begin{tikzpicture}
                                \node [anchor=south west] at (0,0) { \includegraphics[height=0.5\linewidth]{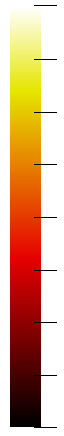} };
                                \draw[black, right] (0.65,4.0) node {\footnotesize{$300$}};
                \draw[black] (1.4,2.1) node[rotate=90] {\small{Velocity Magnitude}};
                \draw[black] (1.8,2.1) node[rotate=90] {\small{[$mm/s$]}};
                \draw[black, right] (0.65,0.225) node {\footnotesize{$0$}};
            \end{tikzpicture}
        \caption{Velocity magnitude during diastole at $t = 0.16~s$
        and during systole at $t = 0.64~s$
        at a section cutting the MV surface
        and AV outflow boundary.
        Figure from~\cite[Figure 3]{MarleviEtAl2021}.}
        \label{lv-flow-application-solution-fig}
    \end{subfigure}\hfill
    \begin{subfigure}[b]{0.475\linewidth}
        \centering
        \setlength{\figurewidth}{0.8\textwidth}
        \setlength{\figureheight}{0.45\textwidth}
                        \begin{tikzpicture}

\begin{axis}[width=0.951\figurewidth,
height=\figureheight,
at={(0\figurewidth,0\figureheight)},
scale only axis,
clip=false,
xmin=0.0,
xmax=0.8,
xlabel style={font=\color{white!15!black}},
xlabel={Time $[s]$},
ymin=1.4e-4,                                                       ymax=2.0e-4,
scaled y ticks=false,
ytick={1.4e-4, 1.6e-4, 1.8e-4, 2.0e-4},
yticklabels={140, 160, 180, 200},
ylabel={Volume $[cm^3]$},                                          ylabel style={font=\color{white!15!black}},
axis background/.style={fill=white},
legend style={draw=none, at={(0.5,0.3)}, anchor=south},
legend cell align=left,
]

\addplot [color=black, line width=1.0pt]
table[]{0.0      0.000164232868736331
0.01       0.00016362403817857
0.02      0.000162727712303482
0.03       0.00016160094506436
0.04      0.000160302722802355
0.05      0.000158892094778506
0.06      0.000157427928781289
0.07      0.000155968560530745
0.08      0.000154571331820822
0.09      0.000153118336109829
0.1      0.000151479992960902
0.11      0.000149747901828039
0.12      0.000148012941791585
0.13      0.000146364652293729
0.14      0.000144891631293565
0.15      0.000143681946922404
0.16      0.000142823553667367
0.17      0.000142373864029779
0.18      0.000142273418759842
0.19       0.00014243634862399
0.2       0.00014277740365558
0.21      0.000143211342909928
0.22      0.000143652572854231
0.23      0.000144015019677071
0.24      0.000144212236812228
0.25      0.000144258238116379
0.26      0.000144251868160426
0.27      0.000144208678847864
0.28      0.000144143717065949
0.29      0.000144071959746487
0.3      0.000144008304851622
0.31      0.000143967584190291
0.32      0.000143964598045941
0.33      0.000144001782439786
0.34      0.000144065062555019
0.35      0.000144147136374617
0.36      0.000144240785427741
0.37      0.000144338798087046
0.38      0.000144433948204478
0.39       0.00014451898026998
0.4      0.000144586601294734
0.41      0.000144639058920042
0.42      0.000144685286791306
0.43      0.000144725482286987
0.44      0.000144759803907052
0.45      0.000144788426529964
0.46      0.000144811541842531
0.47      0.000144829350159068
0.48      0.000144842043487096
0.49       0.00014479306615796
0.5      0.000144647218920423
0.51      0.000144456635071019
0.52      0.000144274137100129
0.53      0.000144152930630349
0.54      0.000144146681409367
0.55      0.000144309619285351
0.56      0.000144696671382259
0.57      0.000145351283162205
0.58      0.000146244694847266
0.59      0.000147321869005737
0.6      0.000148527210164271
0.61      0.000149804534552245
0.62      0.000151096937757021
0.63      0.000152346584979051
0.64      0.000153494418219769
0.65      0.000154588909241837
0.66      0.000155720766593123
0.67      0.000156861595291587
0.68       0.00015798226330967
0.69      0.000159053825511907
0.7      0.000160047736858346
0.71      0.000160935926585642
0.72      0.000161690732060801
0.73      0.000162362292670277
0.74      0.000163009689482125
0.75      0.000163594739265363
0.76      0.000164078670841042
0.77      0.000164422729548107
0.78       0.00016458832291623
0.79      0.000164537144709529
0.8      0.000164232868736331
};
\addlegendentry{Left atrium}

\addplot [color=black!40, dashed, line width=1.0pt]
table[]{0.0      0.000152459934334174
0.01      0.000153207046864446
0.02       0.00015417623170631
0.03      0.000155350967680835
0.04       0.00015671499355073
0.05      0.000158251130815637
0.06      0.000159939673695711
0.07      0.000161757546293106
0.08       0.00016367821571688
0.09      0.000165857133483444
0.1      0.000168425980479143
0.11      0.000171256602743182
0.12      0.000174214149683906
0.13      0.000177157289931342
0.14      0.000179938077164964
0.15       0.00018240260773657
0.16      0.000184392504038253
0.17      0.000185858497692981
0.18       0.00018692478896103
0.19      0.000187672144417955
0.2      0.000188183454821566
0.21      0.000188543219389102
0.22      0.000188836884342073
0.23      0.000189150442883863
0.24      0.000189570293796265
0.25      0.000190075671863497
0.26      0.000190563532158437
0.27      0.000191024729741486
0.28      0.000191450631871712
0.29        0.0001918326743391
0.3      0.000192162388496204
0.31      0.000192431393801451
0.32      0.000192631362015483
0.33      0.000192744423342526
0.34      0.000192768177634282
0.35       0.00019272184579852
0.36      0.000192624871833926
0.37      0.000192496876561559
0.38       0.00019235766127856
0.39       0.00019222718391003
0.4      0.000192125506767008
0.41       0.00019207823650762
0.42       0.00019208049675893
0.43      0.000192102329942701
0.44       0.00019211344727089
0.45      0.000192083294759655
0.46      0.000191981118928434
0.47      0.000191776105357655
0.48      0.000191437588445879
0.49      0.000190998814336925
0.5      0.000190512371108839
0.51      0.000189958358487132
0.52      0.000189316628552903
0.53      0.000188567082825252
0.54       0.00018768968864869
0.55      0.000186664558962109
0.56      0.000185472096642822
0.57      0.000184094283276349
0.58      0.000182543832144701
0.59      0.000180849162603478
0.6      0.000179037953104186
0.61      0.000177137063368078
0.62      0.000175172607718818
0.63      0.000173170142550503
0.64      0.000171154966158628
0.65      0.000168996241761188
0.66      0.000166595116387202
0.67      0.000164074514831248
0.68      0.000161554124498691
0.69      0.000159148914482772
0.7      0.000156969111772736
0.71      0.000155120659964412
0.72      0.000153706133589915
0.73      0.000152692997649133
0.74      0.000151953766049112
0.75      0.000151479250992537
0.76      0.000151256594615545
0.77      0.000151267854017373
0.78      0.000151490202174005
0.79      0.000151897152834357
0.8      0.000152459934334174
};
\addlegendentry{Left ventricle}

\end{axis}
\end{tikzpicture}        \caption{Fluid volume in left atrium and left ventricle over one cardiac cycle.}
        \label{lv-flow-application-la-lv-volume-vel_zeroIC-coarse-mesh-fig}
    \end{subfigure}
    \caption{Flow magnitude during diastole and systole (left)
        and fluid volume over time (right) in the left atrium and left ventricle, respectively.}
\end{figure}

Figure~\ref{lv-flow-application-solution-fig}
illustrates the flow in the left atrium and ventricle during diastole and systole
for cycle~$10$, as obtained from sequential time-stepping.
During diastole, flow through the pulmonary veins causes a slight spiraling of flow
in the left atrium and a filling of the left ventricle through the mitral valve.
Peak velocities occur at the pulmonary vein boundaries and the mitral valve section.
The peak flow through the MV is approximately~$46.18~cm / s$
with respective Reynolds number $Re \approx 3147.74$,
which underpins the choice of a stabilized Galerkin scheme.

Further, the strong jet-like flow through the mitral valve opening causes a circular flow structure
to develop around the jet with reflow regions between the jet and the endocardial wall,
which moves towards the apex of the ventricle after the MV closes.
During systole, the vortices near the MV decelerate and blood is ejected through the aortic valve
boundary.

Considering the volume of the LV over time,
see Figure~\ref{lv-flow-application-la-lv-volume-vel_zeroIC-coarse-mesh-fig},
the end-diastolic volume~\cite{Holt1956} at~$t = 0.37~s$ is approximately,
$EDV \approx 192.50~cm^3$,
and the end-systolic volume at~$t = 0.8~s$ is approximately,
$ESV \approx 152.46~cm^3$.
Thus, the stroke volume is,
$SV = EDV - ESV \approx 40.04~cm^3$,
and the left ventricular ejection fraction is,
$LVEF = 100~SV / EDV \approx 20.80~\%$.

Therefore, according to the LVEF categories defined in the
\emph{European Society of Cardiology Guidelines 2016}~\cite[Section~3.2.1]{ESC2016},
the LVEF is significantly reduced.
Further, the SV is reduced and the EDV and ESV are high
compared to the normal ranges~\cite[Table~1]{LomskyJohanssonGjertssonBjoerkEdenbrandt2008}
in healthy individuals.

\subsubsection{Time-periodic steady-state}
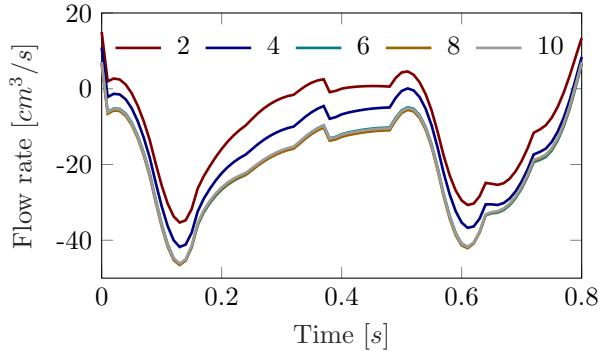
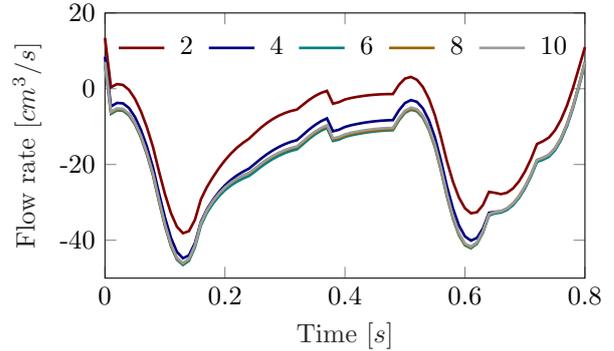
\begin{figure}[ht!]
    \centering
    \begin{subfigure}[b]{0.475\linewidth}
                \centering
        \setlength{\figurewidth}{0.85\textwidth}
        \setlength{\figureheight}{0.45\textwidth}
                        \begin{tikzpicture}

\begin{axis}[width=0.951\figurewidth,
height=\figureheight,
at={(0\figurewidth,0\figureheight)},
scale only axis,
clip=false,
xmin=0.0,
xmax=0.8,
xlabel style={font=\color{white!15!black}},
xlabel={Time $[s]$},
ymin=-5.0e-5,                                                       ymax=2.0e-5,
scaled y ticks=false,
ytick={-4e-5, -2e-5, 0, 2e-5},
yticklabels={-40, -20, 0, 20},
ylabel={Flow rate $[cm^3 / s]$},                                    ylabel style={font=\color{white!15!black}},
axis background/.style={fill=white},
legend style={at={(0.5,0.95)}, anchor=north, legend columns=5, legend cell align=left, align=left, fill=none, draw=none}
]

\addplot [color=myred, line width=1.0pt]
table[]{0      1.49862390310128e-05
0.01       1.9510055116754e-06
0.02      2.70897873673517e-06
0.03      2.44841909545458e-06
0.04      1.12907463243576e-06
0.05     -1.18120653038236e-06
0.06     -4.42837900920683e-06
0.07     -8.54903518424346e-06
0.08     -1.34892320244408e-05
0.09     -1.97221703737694e-05
0.1     -2.55184257406127e-05
0.11     -3.03193181720937e-05
0.12     -3.37277455162292e-05
0.13     -3.53186853015114e-05
0.14     -3.47278738457533e-05
0.15     -3.17449988162515e-05
0.16     -2.63578621725643e-05
0.17     -2.31245823737987e-05
0.18     -2.04935131737312e-05
0.19     -1.80584823993454e-05
0.2     -1.58725901568005e-05
0.21     -1.39551292513393e-05
0.22     -1.22974060010457e-05
0.23     -1.08792483184391e-05
0.24     -9.68614896171653e-06
0.25     -8.07995927639603e-06
0.26     -6.54038285768926e-06
0.27     -5.21601395179385e-06
0.28     -4.10441245031072e-06
0.29     -3.19506340468179e-06
0.3     -2.47798862558237e-06
0.31     -1.94515958999297e-06
0.32     -1.59030137262411e-06
0.33     -3.75641765788337e-07
0.34       7.4946525697181e-07
0.35      1.60776043205067e-06
0.36      2.19415299846518e-06
0.37      2.51112464797337e-06
0.38     -8.99079304040688e-07
0.39     -5.72080449779411e-07
0.4       2.2209953063477e-08
0.41      3.05577748614845e-07
0.42      4.81516975498525e-07
0.43      6.06916570675422e-07
0.44      6.84375916065644e-07
0.45      7.16370906375093e-07
0.46      7.05276255936907e-07
0.47      6.53712268026707e-07
0.48      5.65959415792448e-07
0.49      2.85254422227252e-06
0.5      4.30912187483823e-06
0.51      4.56949751095501e-06
0.52      3.68403427566357e-06
0.53      1.68247308372675e-06
0.54     -1.44874034029326e-06
0.55     -5.78290851920922e-06
0.56     -1.14214363213129e-05
0.57     -1.76628047174853e-05
0.58     -2.28605708973288e-05
0.59     -2.68551890391372e-05
0.6     -2.95088946515797e-05
0.61     -3.06894761814519e-05
0.62     -3.03081945809755e-05
0.63     -2.83466236934899e-05
0.64     -2.48557958476414e-05
0.65     -2.50870773784997e-05
0.66     -2.53781787004911e-05
0.67     -2.49419231568813e-05
0.68     -2.37369947970704e-05
0.69     -2.17696494169696e-05
0.7     -1.90700240802587e-05
0.71     -1.56755657610857e-05
0.72     -1.16255758078035e-05
0.73     -1.07834257039448e-05
0.74     -9.67542421281612e-06
0.75      -7.6795324490055e-06
0.76     -4.78950741300716e-06
0.77     -1.04548213327124e-06
0.78       3.4343916574561e-06
0.79      8.43397083751684e-06
0.8      1.33828680412884e-05
};
\addlegendentry{~$2$~}

\addplot [color=myblue, line width=1.0pt]
table[]{0      1.08906376256969e-05
0.01     -2.19448594048525e-06
0.02     -1.33920794265611e-06
0.03     -1.50187743841093e-06
0.04     -2.77532212573376e-06
0.05     -5.12951277641262e-06
0.06     -8.51519271740664e-06
0.07     -1.28637619886315e-05
0.08     -1.81192264915347e-05
0.09     -2.47545281164067e-05
0.1     -3.10158051846018e-05
0.11     -3.62682325691569e-05
0.12     -4.00173983289387e-05
0.13     -4.17665110414931e-05
0.14     -4.11404095412362e-05
0.15     -3.79909027637559e-05
0.16     -3.24085283900403e-05
0.17      -2.9067381713029e-05
0.18     -2.64855252596977e-05
0.19     -2.42242367505715e-05
0.2     -2.22994409934912e-05
0.21     -2.06943955569154e-05
0.22     -1.93687951180744e-05
0.23     -1.82763605519865e-05
0.24     -1.73836252407671e-05
0.25     -1.60326535699714e-05
0.26     -1.46913237975273e-05
0.27     -1.35120202894887e-05
0.28     -1.24969584161671e-05
0.29       -1.164027570768e-05
0.3     -1.09370577642367e-05
0.31      -1.0385077792495e-05
0.32     -9.98489816653576e-06
0.33     -8.66107009539843e-06
0.34     -7.33133078072564e-06
0.35      -6.1889406558415e-06
0.36     -5.25982917678198e-06
0.37     -4.56554022402726e-06
0.38      -7.9260829634969e-06
0.39     -7.49371263515391e-06
0.4     -6.76422212230751e-06
0.41     -6.32442189492279e-06
0.42     -5.98661925328938e-06
0.43     -5.69544127957054e-06
0.44     -5.45093770289684e-06
0.45     -5.25330662715551e-06
0.46     -5.10568898187356e-06
0.47     -5.00435423962907e-06
0.48     -4.94783134315565e-06
0.49     -2.41453760531634e-06
0.5     -5.77274380528569e-07
0.51       8.9099857216651e-08
0.52      -4.4928022658333e-07
0.53     -2.22291557812259e-06
0.54     -5.29063922738525e-06
0.55     -9.73053471026815e-06
0.56     -1.56200289885281e-05
0.57     -2.22364820329429e-05
0.58     -2.78802264061606e-05
0.59     -3.23204509333915e-05
0.6     -3.53369076613796e-05
0.61      -3.6723921396201e-05
0.62       -3.636090509376e-05
0.63     -3.42557579502699e-05
0.64     -3.05191223194311e-05
0.65     -3.05236612787773e-05
0.66     -3.06790505017427e-05
0.67     -3.01788671904878e-05
0.68     -2.89745823134098e-05
0.69     -2.70681051426927e-05
0.7     -2.44805121158148e-05
0.71     -2.12327159058074e-05
0.72     -1.73401627549083e-05
0.73     -1.66925060015383e-05
0.74     -1.58267725979046e-05
0.75     -1.40643924535413e-05
0.76     -1.13521010721292e-05
0.77     -7.67130941815022e-06
0.78     -3.02612066897731e-06
0.79      2.52450043550025e-06
0.8      8.39380723216244e-06
};
\addlegendentry{~$4$~}

\addplot [color=mygreen, line width=1.0pt]
table[]{0      7.03820440136669e-06
0.01      -6.1193657466829e-06
0.02     -5.20431966914532e-06
0.03      -5.3000714640222e-06
0.04     -6.53245717413954e-06
0.05     -8.89200522015877e-06
0.06     -1.23412335199817e-05
0.07     -1.68209385763418e-05
0.08     -2.22765496554017e-05
0.09     -2.91631364840746e-05
0.1     -3.56666106523965e-05
0.11      -4.1060506931653e-05
0.12     -4.47707281754135e-05
0.13     -4.62884822469922e-05
0.14     -4.53151928875909e-05
0.15     -4.18346732866599e-05
0.16     -3.60460081282505e-05
0.17     -3.26830069173762e-05
0.18     -3.02560734271724e-05
0.19     -2.82550514761681e-05
0.2     -2.66367286297303e-05
0.21     -2.53404001263898e-05
0.22     -2.42969998607221e-05
0.23      -2.3443506825619e-05
0.24     -2.27371551911556e-05
0.25     -2.15151767223601e-05
0.26     -2.02463655412535e-05
0.27     -1.90956869476351e-05
0.28     -1.80787710097543e-05
0.29     -1.71996079180805e-05
0.3     -1.64618689177545e-05
0.31     -1.58707418887495e-05
0.32     -1.54323725509337e-05
0.33     -1.40553510327011e-05
0.34     -1.26497673170707e-05
0.35     -1.14267763354618e-05
0.36     -1.04231808074369e-05
0.37     -9.66807790378974e-06
0.38     -1.30869018818626e-05
0.39     -1.27006644339132e-05
0.4     -1.19964958183511e-05
0.41     -1.15687865807559e-05
0.42      -1.1239752416702e-05
0.43     -1.09552782464104e-05
0.44      -1.0716483410642e-05
0.45     -1.05260761162773e-05
0.46     -1.03828140773654e-05
0.47     -1.02851934030461e-05
0.48     -1.02312540053959e-05
0.49     -7.64990070331771e-06
0.5     -5.69156567423648e-06
0.51     -4.87764888717193e-06
0.52     -5.29569540880016e-06
0.53     -7.02328043675829e-06
0.54      -1.0136416540131e-05
0.55     -1.47101411219689e-05
0.56     -2.08211754212059e-05
0.57     -2.77145972592254e-05
0.58     -3.35833000949068e-05
0.59     -3.80741624070523e-05
0.6     -4.08760582368551e-05
0.61      -4.1767051226664e-05
0.62     -4.07066639865492e-05
0.63     -3.78375851515773e-05
0.64     -3.34080343428269e-05
0.65     -3.28905717106968e-05
0.66     -3.27121867697021e-05
0.67     -3.20257415041802e-05
0.68     -3.07497016136496e-05
0.69     -2.88553330476612e-05
0.7     -2.63318040070886e-05
0.71     -2.31720387269871e-05
0.72     -1.93661750941715e-05
0.73     -1.88074568481301e-05
0.74     -1.80343911399023e-05
0.75     -1.63455863809971e-05
0.76     -1.36728955470817e-05
0.77     -9.98777253528187e-06
0.78     -5.28518343743805e-06
0.79      4.01574431651439e-07
0.8      6.54043101399988e-06
};
\addlegendentry{~$6$~}

\addplot [color=myorange, line width=1.0pt]
table[]{0      6.58539934847304e-06
0.01     -6.62172511094136e-06
0.02     -5.74548091618273e-06
0.03      -5.8786982986446e-06
0.04     -7.14862678360342e-06
0.05     -9.54642966127173e-06
0.06     -1.30354034368226e-05
0.07      -1.7553367930942e-05
0.08      -2.3041216360861e-05
0.09     -2.99394063063011e-05
0.1     -3.64093598080018e-05
0.11     -4.17036325122867e-05
0.12     -4.52443039747692e-05
0.13     -4.65456487383773e-05
0.14     -4.53541067267115e-05
0.15     -4.16977478042522e-05
0.16     -3.57996881275799e-05
0.17     -3.23891024790484e-05
0.18      -2.9956547590762e-05
0.19     -2.79759412164931e-05
0.2     -2.63928348200131e-05
0.21     -2.51399332093333e-05
0.22     -2.41448897164268e-05
0.23     -2.33433957599475e-05
0.24     -2.26922529357328e-05
0.25     -2.15268094329226e-05
0.26     -2.03144962029239e-05
0.27     -1.92204106819279e-05
0.28     -1.82593244659056e-05
0.29       -1.743527205514e-05
0.3     -1.67515751715173e-05
0.31     -1.62124613170762e-05
0.32     -1.58242558812719e-05
0.33     -1.44947257952903e-05
0.34     -1.31327408952513e-05
0.35     -1.19502528699976e-05
0.36     -1.09854899247013e-05
0.37     -1.02680939435497e-05
0.38     -1.37253236214208e-05
0.39     -1.33735169244535e-05
0.4     -1.26945855797527e-05
0.41     -1.22846116643449e-05
0.42     -1.19683574315289e-05
0.43      -1.1692580072381e-05
0.44     -1.14590709991409e-05
0.45     -1.12713480775624e-05
0.46     -1.11280437821176e-05
0.47     -1.10283120781622e-05
0.48     -1.09704415797387e-05
0.49     -8.38045999515867e-06
0.5     -6.40877761613511e-06
0.51     -5.58234196701407e-06
0.52     -5.99281581960066e-06
0.53     -7.72011762901255e-06
0.54     -1.08397476237088e-05
0.55     -1.54241337609589e-05
0.56     -2.15450008926783e-05
0.57     -2.84383659137286e-05
0.58     -3.42786193410601e-05
0.59     -3.86950421693886e-05
0.6     -4.13721015646851e-05
0.61     -4.21016948279099e-05
0.62     -4.08706232957914e-05
0.63     -3.78490465740987e-05
0.64     -3.33018231863035e-05
0.65      -3.2702947080376e-05
0.66     -3.24705008273593e-05
0.67     -3.17486130444008e-05
0.68      -3.0450077444862e-05
0.69     -2.85407151964943e-05
0.7     -2.60058990220066e-05
0.71     -2.28360615869335e-05
0.72     -1.90200637641477e-05
0.73     -1.84502290695662e-05
0.74     -1.76649816556791e-05
0.75     -1.59643747045516e-05
0.76     -1.32816184350333e-05
0.77     -9.59086960555077e-06
0.78     -4.89297987376681e-06
0.79      7.69033087880744e-07
0.8      6.84884724007017e-06
};
\addlegendentry{~$8$~}

\addplot [color=mygray, line width=1.0pt]
table[]{0       7.0234164192073e-06
0.01      -6.1759748829475e-06
0.02      -5.3049697709972e-06
0.03     -5.44481803837853e-06
0.04     -6.72003130023993e-06
0.05     -9.11959261625873e-06
0.06     -1.26052207418646e-05
0.07     -1.71136123485067e-05
0.08     -2.25863624332172e-05
0.09     -2.94679544852069e-05
0.1     -3.59285078299028e-05
0.11     -4.12308546731288e-05
0.12     -4.48025598452014e-05
0.13     -4.61539962548941e-05
0.14     -4.50171888307574e-05
0.15     -4.14048019024345e-05
0.16     -3.55266145150181e-05
0.17     -3.21133954420255e-05
0.18     -2.96598345661092e-05
0.19     -2.76487226655193e-05
0.2     -2.60315827204956e-05
0.21     -2.47450403682914e-05
0.22     -2.37189666256163e-05
0.23     -2.28896795294332e-05
0.24     -2.22160708947345e-05
0.25      -2.1032603552822e-05
0.26     -1.98068512658702e-05
0.27     -1.87027033969218e-05
0.28     -1.77339381501426e-05
0.29     -1.69038718863294e-05
0.3     -1.62152925603784e-05
0.31      -1.5672008730894e-05
0.32     -1.52802701971945e-05
0.33     -1.39489658880033e-05
0.34     -1.25879482226308e-05
0.35     -1.14075384787697e-05
0.36     -1.04445432913331e-05
0.37     -9.72780746704296e-06
0.38      -1.3182867072361e-05
0.39     -1.28250439136944e-05
0.4     -1.21435837391144e-05
0.41     -1.17338347955573e-05
0.42     -1.14193696868915e-05
0.43     -1.11466426482923e-05
0.44     -1.09171865077689e-05
0.45     -1.07337715989009e-05
0.46     -1.05959715087632e-05
0.47     -1.05023821568049e-05
0.48     -1.04512828984174e-05
0.49     -7.87226869147897e-06
0.5     -5.91578486947556e-06
0.51     -5.10430415829853e-06
0.52     -5.52643113423002e-06
0.53     -7.26019184386007e-06
0.54     -1.03805246022857e-05
0.55     -1.49600366866309e-05
0.56     -2.10717430789066e-05
0.57     -2.79562732443532e-05
0.58     -3.37991093849227e-05
0.59     -3.82401511947899e-05
0.6     -4.09673528117827e-05
0.61     -4.17657361809695e-05
0.62     -4.06074064346062e-05
0.63     -3.76474386197453e-05
0.64     -3.31427729886872e-05
0.65     -3.25622039370037e-05
0.66      -3.2330934352436e-05
0.67     -3.15986499773706e-05
0.68     -3.02834694042283e-05
0.69     -2.83553888660778e-05
0.7     -2.58028726544406e-05
0.71     -2.26184400258046e-05
0.72     -1.87923434971469e-05
0.73     -1.82145379279251e-05
0.74     -1.74219304386231e-05
0.75     -1.57165292652879e-05
0.76     -1.30326118959278e-05
0.77     -9.34459382640366e-06
0.78      -4.6530488259758e-06
0.79      9.98236045460348e-07
0.8      7.05716462866384e-06
};
\addlegendentry{~$10$}

\end{axis}
\end{tikzpicture}        \caption{A time-periodic steady-state is achieved by running multiple cycles using sequential time stepping.}
        \label{lv-flow-application-la-pv2-flow-rate-sequential-to-cycle-10-vel_zeroIC-coarse-mesh-fig}
    \end{subfigure}\hfill    \begin{subfigure}[b]{0.475\linewidth}
        \centering
        \setlength{\figurewidth}{0.85\textwidth}
        \setlength{\figureheight}{0.45\textwidth}
                        \begin{tikzpicture}

\begin{axis}[width=0.951\figurewidth,
height=\figureheight,
at={(0\figurewidth,0\figureheight)},
scale only axis,
clip=false,
xmin=0.0,
xmax=0.8,
xlabel style={font=\color{white!15!black}},
xlabel={Time $[s]$},
ymin=-5.0e-5,                                                       ymax=2.0e-5,
scaled y ticks=false,
ytick={-4e-5, -2e-5, 0, 2e-5},
yticklabels={-40, -20, 0, 20},
ylabel={Flow rate $[cm^3 / s]$},                                    ylabel style={font=\color{white!15!black}},
axis background/.style={fill=white},
legend style={at={(0.5,0.95)}, anchor=north, legend columns=5, legend cell align=left, align=left, fill=none, draw=none}
]

\addplot [color=myred, line width=1.0pt]
table[]{0      1.34431520663211e-05
0.01      4.07549877261654e-07
0.02      1.22630138963909e-06
0.03      1.01284943316081e-06
0.04     -2.97004956381769e-07
0.05      -2.6483678578279e-06
0.06     -5.98270925764063e-06
0.07     -1.02334992325677e-05
0.08     -1.53543453850971e-05
0.09     -2.17885248744312e-05
0.1     -2.78130748557578e-05
0.11     -3.28437687048422e-05
0.12     -3.64301966204179e-05
0.13      -3.8138676745669e-05
0.14     -3.75895218819656e-05
0.15     -3.45925792807728e-05
0.16     -2.91744014728574e-05
0.17     -2.59338924526141e-05
0.18     -2.33419539255168e-05
0.19     -2.09824866407964e-05
0.2     -1.89038468204761e-05
0.21      -1.7104438101082e-05
0.22     -1.55734789999171e-05
0.23     -1.42803834950091e-05
0.24     -1.32058003562679e-05
0.25     -1.17123365618883e-05
0.26     -1.02684101162803e-05
0.27     -9.02209690534056e-06
0.28     -7.96466251151807e-06
0.29     -7.09556549183321e-06
0.3      -6.3995664545745e-06
0.31     -5.86948091968423e-06
0.32     -5.49495244175694e-06
0.33     -4.21880148230751e-06
0.34     -2.96771996002729e-06
0.35     -1.92687995051731e-06
0.36     -1.11026529791387e-06
0.37     -5.57156058585861e-07
0.38     -3.94746878584111e-06
0.39     -3.53697726131139e-06
0.4     -2.83182938981269e-06
0.41     -2.43912255286738e-06
0.42     -2.14815682672458e-06
0.43      -1.9055346218255e-06
0.44     -1.71074813117372e-06
0.45     -1.56469488539331e-06
0.46     -1.46527895978065e-06
0.47     -1.41150533348585e-06
0.48     -1.40419572701555e-06
0.49      1.03996600473472e-06
0.5      2.68384029175345e-06
0.51      3.11983335927954e-06
0.52      2.36653300789861e-06
0.53      4.57454746611839e-07
0.54     -2.64342135747395e-06
0.55     -7.02323996704631e-06
0.56     -1.27735366657513e-05
0.57     -1.91584059518876e-05
0.58      -2.4535276659559e-05
0.59     -2.87244898788781e-05
0.6     -3.15560169643687e-05
0.61     -3.28823241168758e-05
0.62     -3.26005144483323e-05
0.63     -3.06932645361469e-05
0.64     -2.72164561045782e-05
0.65      -2.7476700839493e-05
0.66     -2.78201457755844e-05
0.67     -2.74512483630843e-05
0.68     -2.63290829825202e-05
0.69     -2.44498626099375e-05
0.7     -2.18439670746187e-05
0.71     -1.85457442641305e-05
0.72      -1.4585689155644e-05
0.73     -1.38514525791756e-05
0.74     -1.28697316802347e-05
0.75     -1.09936566029822e-05
0.76     -8.18360733619653e-06
0.77     -4.45513741316201e-06
0.78      1.65580999582416e-07
0.79      5.52426651154931e-06
0.8      1.09970490557636e-05
};
\addlegendentry{~$2$~}

\addplot [color=myblue, line width=1.0pt]
table[]{0      8.46576672814127e-06
0.01     -4.67390005526252e-06
0.02     -3.74767301989182e-06
0.03     -3.84436273375537e-06
0.04     -5.08695949241636e-06
0.05     -7.44879704796508e-06
0.06      -1.0883550919023e-05
0.07     -1.53344261058013e-05
0.08     -2.07574239726192e-05
0.09     -2.75703573469993e-05
0.1     -3.39997846341723e-05
0.11     -3.93611978281336e-05
0.12     -4.30924913339463e-05
0.13     -4.47534575211635e-05
0.14     -4.39879219097192e-05
0.15     -4.07182726694529e-05
0.16      -3.5107502302548e-05
0.17     -3.18035652714677e-05
0.18     -2.93387951305701e-05
0.19     -2.72383083408086e-05
0.2     -2.54913682173379e-05
0.21     -2.40538661414673e-05
0.22     -2.28773225375667e-05
0.23      -2.1903988558077e-05
0.24     -2.10871961118925e-05
0.25     -1.97906303724517e-05
0.26     -1.84709015966432e-05
0.27     -1.72870658902597e-05
0.28     -1.62438086424544e-05
0.29     -1.53529095104551e-05
0.3     -1.46059816893904e-05
0.31     -1.40087656890736e-05
0.32     -1.35653582852159e-05
0.33     -1.21864794035808e-05
0.34     -1.07892864176565e-05
0.35     -9.57533509638661e-06
0.36     -8.57876509866944e-06
0.37     -7.82330797296642e-06
0.38     -1.12305778748676e-05
0.39     -1.08109321106186e-05
0.4      -1.0079720744573e-05
0.41     -9.63927697788993e-06
0.42     -9.30412750345146e-06
0.43     -9.01516504140022e-06
0.44     -8.77424629804968e-06
0.45     -8.58041185468476e-06
0.46     -8.43416832107263e-06
0.47     -8.33513503353902e-06
0.48     -8.28300261459112e-06
0.49     -5.70381949928986e-06
0.5      -3.7708770574757e-06
0.51     -2.99138116415067e-06
0.52     -3.45207795068992e-06
0.53     -5.20516825803761e-06
0.54      -8.3200480893907e-06
0.55      -1.2869621460842e-05
0.56     -1.89427141530228e-05
0.57     -2.57674735903387e-05
0.58     -3.15855808099979e-05
0.59     -3.60910119829486e-05
0.6     -3.89720379156405e-05
0.61     -4.01030844784357e-05
0.62      -3.9369579446829e-05
0.63     -3.68521359285149e-05
0.64     -3.27583303640501e-05
0.65     -3.24791657801916e-05
0.66     -3.24458288021694e-05
0.67     -3.18303585190275e-05
0.68     -3.05867185615372e-05
0.69     -2.86664039788025e-05
0.7     -2.60950105624155e-05
0.71     -2.28799757021403e-05
0.72     -1.90205012439109e-05
0.73     -1.84154534391134e-05
0.74     -1.75993310086335e-05
0.75     -1.58759817716432e-05
0.76     -1.31730603348013e-05
0.77     -9.47433106300256e-06
0.78     -4.76726341572351e-06
0.79      9.12101975014819e-07
0.8      7.05209414318967e-06
};
\addlegendentry{~$4$~}

\addplot [color=mygreen, line width=1.0pt]
table[]{0      6.54512602346563e-06
0.01     -6.62900283660967e-06
0.02     -5.70709140592319e-06
0.03     -5.80063450457344e-06
0.04     -7.03673887739626e-06
0.05     -9.41154240898525e-06
0.06     -1.28805540526942e-05
0.07      -1.7389072810277e-05
0.08     -2.28740153042217e-05
0.09     -2.97600727362337e-05
0.1     -3.62272112275739e-05
0.11     -4.15411872446287e-05
0.12      -4.5126257064886e-05
0.13     -4.65163093557564e-05
0.14     -4.54304406443431e-05
0.15     -4.18787979906376e-05
0.16     -3.60816557136622e-05
0.17     -3.27277703557796e-05
0.18     -3.03271106160155e-05
0.19      -2.8361820822272e-05
0.2     -2.67838736264154e-05
0.21     -2.55274810651185e-05
0.22     -2.45233119147797e-05
0.23     -2.37062994792307e-05
0.24     -2.30333122295784e-05
0.25     -2.18488511446805e-05
0.26     -2.06171333911632e-05
0.27     -1.95024636578499e-05
0.28     -1.85194123087614e-05
0.29     -1.76734624633143e-05
0.3     -1.69668148096181e-05
0.31     -1.64060270762083e-05
0.32     -1.59962834429738e-05
0.33     -1.46474664434271e-05
0.34     -1.32669672863875e-05
0.35     -1.20650384160997e-05
0.36     -1.10799080323075e-05
0.37     -1.03420856387674e-05
0.38     -1.37809246121584e-05
0.39     -1.34127488738858e-05
0.4     -1.27177982976073e-05
0.41     -1.22973127181757e-05
0.42     -1.19744480350092e-05
0.43     -1.16937140409546e-05
0.44     -1.14551703775738e-05
0.45     -1.12612829148918e-05
0.46     -1.11120144188849e-05
0.47     -1.10071437237359e-05
0.48     -1.09451077729681e-05
0.49      -8.3492802455051e-06
0.5       -6.373496786357e-06
0.51     -5.54458136158436e-06
0.52     -5.95498036719827e-06
0.53     -7.68142683648955e-06
0.54     -1.08002051023689e-05
0.55     -1.53854245878899e-05
0.56     -2.15026677912533e-05
0.57     -2.83950278270147e-05
0.58     -3.42348772351403e-05
0.59     -3.86548722802522e-05
0.6     -4.13408283894021e-05
0.61     -4.21094266174852e-05
0.62      -4.0928404894449e-05
0.63      -3.7960250011512e-05
0.64     -3.34690736020825e-05
0.65     -3.29049523969198e-05
0.66     -3.26981272265224e-05
0.67     -3.19945107425836e-05
0.68     -3.07107501979505e-05
0.69     -2.88086033573514e-05
0.7     -2.62789569067793e-05
0.71     -2.31143325242772e-05
0.72     -1.93018611303949e-05
0.73     -1.87382970089433e-05
0.74     -1.79617864681578e-05
0.75     -1.62703039723565e-05
0.76     -1.35951433057581e-05
0.77     -9.90905975847337e-06
0.78     -5.20901915423996e-06
0.79      4.63163358590871e-07
0.8      6.57798654282628e-06
};
\addlegendentry{~$6$~}

\addplot [color=myorange, line width=1.0pt]
table[]{0      6.84542434508292e-06
0.01     -6.33412269638282e-06
0.02     -5.43926934775794e-06
0.03     -5.56030874094486e-06
0.04     -6.82026393293961e-06
0.05     -9.21551012192097e-06
0.06     -1.26988722487352e-05
0.07      -1.7214365135697e-05
0.08     -2.26999848732724e-05
0.09     -2.95800466744199e-05
0.1       -3.603176562612e-05
0.11     -4.13186394655534e-05
0.12     -4.48712605187003e-05
0.13     -4.61952031248432e-05
0.14     -4.50316060157781e-05
0.15     -4.14025847759613e-05
0.16     -3.55241918583488e-05
0.17     -3.21208725273341e-05
0.18     -2.96807791351417e-05
0.19     -2.76842835322772e-05
0.2     -2.60815137155513e-05
0.21     -2.48074216566872e-05
0.22     -2.37910023879314e-05
0.23     -2.29680321898374e-05
0.24      -2.2298487003714e-05
0.25      -2.1123183261414e-05
0.26     -1.99068793598241e-05
0.27     -1.88123725940669e-05
0.28     -1.78538469622481e-05
0.29     -1.70335172564463e-05
0.3      -1.6353976472471e-05
0.31     -1.58208457997413e-05
0.32     -1.54393124134398e-05
0.33     -1.41197987979402e-05
0.34     -1.27703287449385e-05
0.35     -1.15984343443827e-05
0.36     -1.06416440066212e-05
0.37      -9.9296842418649e-06
0.38     -1.33895627139302e-05
0.39     -1.30395810453882e-05
0.4     -1.23647057927053e-05
0.41     -1.19603265336238e-05
0.42     -1.16520743568687e-05
0.43     -1.13847514997768e-05
0.44      -1.1159301199279e-05
0.45     -1.09767259024858e-05
0.46     -1.08378266782822e-05
0.47     -1.07424716567903e-05
0.48      -1.0689374091845e-05
0.49     -8.10621822717405e-06
0.5     -6.14328335628548e-06
0.51     -5.32587566704725e-06
0.52     -5.74349267822583e-06
0.53     -7.47580582880869e-06
0.54     -1.05975085270784e-05
0.55     -1.51818786188465e-05
0.56     -2.12965807163964e-05
0.57     -2.81852120878745e-05
0.58     -3.40257493807738e-05
0.59     -3.84505856468586e-05
0.6     -4.11498660195757e-05
0.61     -4.19077371105711e-05
0.62     -4.07035507151226e-05
0.63     -3.77029630113778e-05
0.64      -3.3163750205026e-05
0.65     -3.25657463950716e-05
0.66     -3.23259497532452e-05
0.67     -3.15923972336915e-05
0.68     -3.02803730059078e-05
0.69     -2.83567888532093e-05
0.7     -2.58080976580689e-05
0.71      -2.2626908290374e-05
0.72     -1.88027945736307e-05
0.73     -1.82243534504598e-05
0.74     -1.74317851309932e-05
0.75     -1.57269175472975e-05
0.76     -1.30444644747239e-05
0.77     -9.35669828961668e-06
0.78     -4.66447970134921e-06
0.79      9.82463372771389e-07
0.8      7.03865933316673e-06
};
\addlegendentry{~$8$~}

\addplot [color=mygray, line width=1.0pt]
table[]{0      7.08072467378223e-06
0.01     -6.09148854934168e-06
0.02     -5.18910320952462e-06
0.03     -5.30354521761394e-06
0.04     -6.55728218249996e-06
0.05     -8.94516918385641e-06
0.06     -1.24196435898736e-05
0.07     -1.69245795229349e-05
0.08     -2.24010230558277e-05
0.09     -2.92738162133277e-05
0.1     -3.57293196394451e-05
0.11     -4.10387431444035e-05
0.12     -4.46366326459853e-05
0.13      -4.6022527386667e-05
0.14     -4.49249466526404e-05
0.15     -4.13515296285714e-05
0.16      -3.5508890445274e-05
0.17     -3.21229438881291e-05
0.18     -2.96847564022659e-05
0.19     -2.76809758096332e-05
0.2     -2.60653182240262e-05
0.21     -2.47763531025054e-05
0.22     -2.37438942332236e-05
0.23     -2.29040700273014e-05
0.24      -2.2215567330143e-05
0.25     -2.10225661032874e-05
0.26     -1.97889105932571e-05
0.27     -1.86772709389077e-05
0.28     -1.77011284374657e-05
0.29     -1.68639763437671e-05
0.3     -1.61675555671792e-05
0.31      -1.5617690865373e-05
0.32     -1.52196588776486e-05
0.33     -1.38842165015966e-05
0.34     -1.25203082362708e-05
0.35     -1.13347651873195e-05
0.36     -1.03648262570221e-05
0.37     -9.63950400510087e-06
0.38     -1.30873743216274e-05
0.39     -1.27223789330024e-05
0.4     -1.20354566177651e-05
0.41      -1.1621856568046e-05
0.42     -1.13064370222222e-05
0.43     -1.10331476367965e-05
0.44     -1.08031216742039e-05
0.45     -1.06171727920821e-05
0.46     -1.04759736074283e-05
0.47     -1.03793621441498e-05
0.48     -1.03261851863406e-05
0.49     -7.74542189764991e-06
0.5     -5.78837130235192e-06
0.51     -4.97713832470839e-06
0.52     -5.39910281939324e-06
0.53     -7.13187569930865e-06
0.54     -1.02501435759334e-05
0.55       -1.482799547558e-05
0.56     -2.09355045978622e-05
0.57     -2.78183327075856e-05
0.58     -3.36645130120921e-05
0.59     -3.81157103669397e-05
0.6     -4.08655516744775e-05
0.61     -4.16930536113566e-05
0.62     -4.05642645514404e-05
0.63     -3.76309025590927e-05
0.64     -3.31426189236247e-05
0.65     -3.25770673612374e-05
0.66     -3.23546438311714e-05
0.67     -3.16282875201108e-05
0.68     -3.03171280957731e-05
0.69     -2.83914324422609e-05
0.7      -2.5839817499022e-05
0.71     -2.26563044666723e-05
0.72     -1.88306588012561e-05
0.73     -1.82528542928345e-05
0.74     -1.74610090001781e-05
0.75     -1.57571925782345e-05
0.76     -1.30752641845375e-05
0.77     -9.38843890525491e-06
0.78     -4.69515260662841e-06
0.79      9.57973462437592e-07
0.8      7.03201809759012e-06
};
\addlegendentry{~$10$}

\end{axis}
\end{tikzpicture}        \caption{A time-periodic steady-state is achieved by running multiple iterations using time-periodic MGRIT.}
        \label{lv-flow-application-la-pv2-flow-rate-pint-to-iter-10-vel_zeroIC-coarse-mesh-fig}
    \end{subfigure}
    \caption{Flow rate at pulmonary vein $2$ for even-numbered sequential time-stepping cycles
        and time-periodic MGRIT iterations. Note, a negative flow rate corresponds to inflow.}
    \label{todo-2-fig}
\end{figure}

Figure~\ref{lv-flow-application-la-pv2-flow-rate-sequential-to-cycle-10-vel_zeroIC-coarse-mesh-fig}
illustrates the flow rate at pulmonary vein~$2$ over the duration of one cycle
for all even numbered cycles, as obtained with sequential time-stepping.
The graph highlights that it is necessary to run multiple cycles
to minimize cycle-to-cycle variations and obtain a time-periodic steady-state.
Furthermore, the flow rate at pulmonary vein~$1$ converges to a periodic steady-state quicker
than the flow rate at pulmonary vein~$4$; however, a periodic steady-state is obtained at each vein,
see Figure~\ref{lv-flow-application-la-pv1-pv2-pv3-pv4-flow-rate-sequential-to-cycle-10-vel_zeroIC-coarse-mesh-suppfig}.

On the other hand,
Figure~\ref{lv-flow-application-la-pv2-flow-rate-pint-to-iter-10-vel_zeroIC-coarse-mesh-fig}
highlights a similar convergence behavior when the time-periodic MGRIT algorithm is employed
to obtain a periodic steady-state solution.
In contrast to the sequential time-stepping solution,
the initial error in the flow rates can be reduced slightly more quickly,
compare Figure~\ref{lv-flow-application-la-pv1-pv2-pv3-pv4-flow-rate-sequential-to-cycle-10-vel_zeroIC-coarse-mesh-suppfig}
and Figure~\ref{lv-flow-application-la-pv1-pv2-pv3-pv4-flow-rate-pint-to-iter-10-vel_zeroIC-coarse-mesh-suppfig}.
It seems reasonable to assume that this is achieved by providing a better initial guess
of the initial condition by means of XBraid's \emph{skip-first-down} option
and by updating the initial condition multiple times during each time-periodic MGRIT iteration
(see Section~\ref{mgrit-extension-for-time-periodic-problems-sec}).

\begin{figure}[ht!]
    \centering
    \begin{subfigure}[b]{0.475\textwidth}
        \centering
        \setlength{\figurewidth}{0.85\textwidth}
        \setlength{\figureheight}{0.45\textwidth}
                        \begin{tikzpicture}

\begin{axis}[width=0.951\figurewidth,
height=\figureheight,
at={(0\figurewidth,0\figureheight)},
scale only axis,
clip=false,
xmin=1,
xmax=10,
xtick={1, 2, 3, 4, 5, 6, 7, 8, 9, 10},
xminorticks=true,
xlabel style={font=\color{white!15!black}},
xlabel={Cycle},
ymin=0.0,
ymax=1.5e-5,
yminorticks=true,
ylabel style={font=\color{white!15!black}},
ylabel={Space-time error},
axis background/.style={fill=white},
legend pos={outer north east},
legend style={draw=none, at={(0.55,0.65)}, anchor=west},
legend cell align=left,
]

\addplot [color=myred, line width=1.0pt, mark=*]
table[]{1       7.40714559477597e-06
2       6.89029489685533e-06
3       3.26085027931392e-06
4       1.35841407270671e-06
5       2.04139059867151e-06
6       1.821027495572e-06
7       8.4916311766841e-07
8       2.88924135924516e-07
9       1.86837123314581e-07
10      0
};
\addlegendentry{Vein $1$}

\addplot [color=myblue, line width=1.0pt, mark=*]
table[]{1      9.11562163955612e-06
2      8.10575999694193e-06
3      6.04290251503555e-06
4      3.30349616618596e-06
5      1.03717223074678e-06
6       2.4922631109512e-07
7      4.28257485583391e-07
8       3.2442290791766e-07
9       1.1996854450883e-07
10      0
};
\addlegendentry{Vein $2$}

\addplot [color=mygreen, line width=1.0pt, mark=*]
table[]{1       1.2883148243825e-05
2      1.00200550679952e-05
3      5.75317769960711e-06
4      2.10302901165589e-06
5      8.59410331013139e-07
6       1.2512018912054e-06
7      6.82620464235982e-07
8      2.03053664855647e-07
9      9.95899154956463e-08
10      0
};
\addlegendentry{Vein $3$}

\addplot [color=myorange, line width=1.0pt, mark=*]
table[]{1      1.45916130221215e-05
2      1.12354950493881e-05
3      8.63062240091435e-06
4      5.39223857830549e-06
5      2.22645111968501e-06
6      4.60488330058002e-07
7      3.57843430340149e-07
8      4.11617275196954e-07
9       2.1653672528246e-07
10      0
};
\addlegendentry{Vein $4$}

\end{axis}
\end{tikzpicture}        \caption{Per-cycle error compared to cycle~$10$.}
        \label{lv-flow-application-la-pv1-pv2-pv3-pv4-per-cycle-error-vel_zeroIC-coarse-mesh-fig}
    \end{subfigure}\qquad
    \begin{subfigure}[b]{0.475\textwidth}
        \centering
        \setlength{\figurewidth}{0.85\textwidth}
        \setlength{\figureheight}{0.45\textwidth}
                        \begin{tikzpicture}

\begin{axis}[width=0.951\figurewidth,
height=\figureheight,
at={(0\figurewidth,0\figureheight)},
scale only axis,
clip=false,
xmin=1,
xmax=10,
xtick={1, 2, 3, 4, 5, 6, 7, 8, 9, 10},
xminorticks=true,
xlabel style={font=\color{white!15!black}},
xlabel={{\color{white}y}Iteration{\color{white}y}},
ymin=0.0,
ymax=1.5e-5,
yminorticks=true,
ylabel style={font=\color{white!15!black}},
ylabel={Space-time error},
axis background/.style={fill=white},
legend pos={outer north east},
legend style={draw=none, at={(0.55,0.65)}, anchor=west},
legend cell align=left,
]

\addplot [color=myred, line width=1.0pt, mark=*]
table[]{1      6.80058330067324e-06
2      3.29519438496176e-06
3      1.41750615108761e-06
4      2.08082063276423e-06
5      1.81515493986268e-06
6      8.18493223921087e-07
7      2.95849223507047e-07
8      2.29222804333365e-07
9       2.9541463239674e-08
10      2.65065895227757e-07
};
\addlegendentry{Vein $1$}

\addplot [color=myblue, line width=1.0pt, mark=*]
table[]{1      8.09138835837442e-06
2      6.02784205845835e-06
3      3.28074710028657e-06
4      1.06116477587452e-06
5      2.88874099205029e-07
6      4.13782093286418e-07
7      3.05773883933203e-07
8      9.32323648800457e-08
9      3.57817972980104e-08
10      6.59197936263453e-08
};
\addlegendentry{Vein $2$}

\addplot [color=mygreen, line width=1.0pt, mark=*]
table[]{1      9.99946315252152e-06
2      5.76058362590908e-06
3      2.11539614232969e-06
4      8.97048949075693e-07
5      1.31716771309972e-06
6      7.17637707170145e-07
7      1.90571257763217e-07
8      1.31832326559082e-07
9      4.29466362019918e-08
10      1.78100528002417e-07
};
\addlegendentry{Vein $3$}

\addplot [color=myorange, line width=1.0pt, mark=*]
table[]{1       1.1248922803582e-05
2      8.55294447890268e-06
3      5.34303826234753e-06
4      2.18448866893651e-06
5      3.52449301351087e-07
6      4.15842722870563e-07
7      4.59208580294704e-07
8      2.31520297575689e-07
9      3.59934168762298e-08
10      9.69303798049428e-08
};
\addlegendentry{Vein $4$}

\end{axis}
\end{tikzpicture}        \caption{Per-iteration error compared to cycle~$10$.}
        \label{lv-flow-application-la-pv1-pv2-pv3-pv4-per-iter-vs-cycle-10-error-vel_zeroIC-coarse-mesh-fig}
    \end{subfigure}
    \caption{Convergence of flow rate errors at veins~$1 - 4$ for sequential time-stepping (left)
        and time-periodic MGRIT (right).
        Reference data were selected as the sequential time-stepping solution over cycle~$10$.}
    \label{lv-flow-application-la-pv1-pv2-pv3-pv4-per-cycle-per-iteration-error-vel_zeroIC-coarse-mesh-fig}
\end{figure}
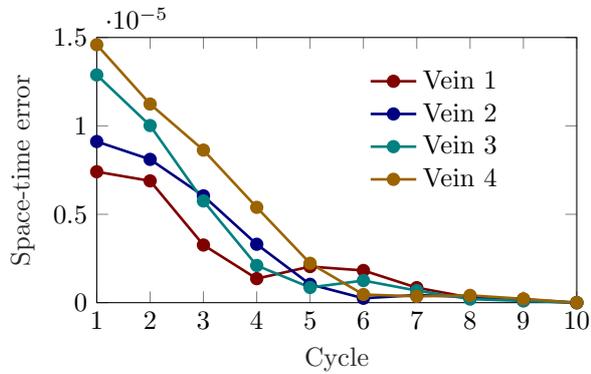
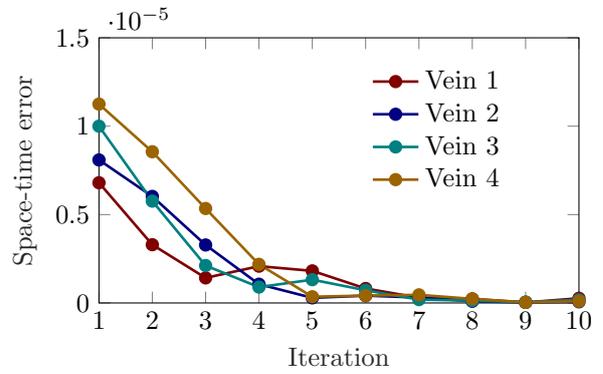

Selecting the space-time solution for cycle~$10$ as the time-periodic steady-state solution,
the error for each cycle and each iteration can be quantified.
Figure~\ref{lv-flow-application-la-pv1-pv2-pv3-pv4-per-cycle-per-iteration-error-vel_zeroIC-coarse-mesh-fig}
highlights that a time-periodic steady-state can be achieved
by running sequential time-stepping for $7$~cycles
or time-periodic MGRIT for $6$~iterations.
Thus, in terms of convergence to a periodic steady-state,
one time-periodic MGRIT iteration roughly corresponds to running one cycle using sequential time-stepping
similar to Section~\ref{stenosed-valve-results} and Section~\ref{results-fsi}.

\subsubsection{Parallel performance}
Here, runtimes and errors for sequential time-stepping and time-periodic MGRIT are compared.
Timing results were obtained on TheoSim (see Appendix~\ref{HPC-machines-sec})
with~$8$~or $16$~allocated processors per node.\footnote{Although each TheoSim node has $20$~processors,
preliminary experiments showed that performance deteriorates beyond $16$~processors per node.}
Space-time errors are those reported
in Figure~\ref{lv-flow-application-la-pv1-pv2-pv3-pv4-per-cycle-per-iteration-error-vel_zeroIC-coarse-mesh-fig}
for vein~$2$, which are representative for all veins, see~\cite{Hessenthaler2020_PhD}.

As Figure~\ref{lv-flow-application-vein-2-per-cycle-per-iteration-cost-vel_zeroIC-coarse-mesh-fig} illustrates,
spatial parallelism can reduce the wall clock time for sequential time-stepping.
For example, the speedup of switching from $8$~processors to $16$~processors in space
is $1.58$x, see Table~\ref{lv-flow-application-runtimes-speedup-coarse-mesh-tab}.
Spatial parallelism, however, saturates with a subsequent increase of the \emph{time-to-solution}
for $32$~processors compared to $16$~processors in space, which is also related
to induced \emph{node-to-node} communication for the employed hardware.

For any required space-time error and the considered distribution of processors for time-
or space-time parallelism (see Table~\ref{lv-flow-application-runtimes-speedup-coarse-mesh-tab}),
time-periodic MGRIT yields a shorter time-to-solution than sequential time-stepping.
The results reported in Figure~\ref{lv-flow-application-vein-2-per-cycle-per-iteration-cost-vel_zeroIC-coarse-mesh-fig}
and Table~\ref{lv-flow-application-runtimes-speedup-coarse-mesh-tab}
further highlight that it is beneficial to assign more processors for parallelization in the temporal domain
instead of the spatial domain.
For example, when using a total number of $80$~processors,
the wall clock time for time-periodic MGRIT can be reduced when more processors are assigned
to the temporal component than the spatial component.
\setlength{\figurewidth}{0.5\textwidth}
\setlength{\figureheight}{0.2\textwidth}
\begin{figure}[ht!]
    \centering
            \begin{tikzpicture}

\begin{axis}[width=0.951\figurewidth,
height=\figureheight,
at={(0\figurewidth,0\figureheight)},
scale only axis,
clip=false,
xmode=log,
xminorticks=true,
xlabel style={font=\color{white!15!black}},
xlabel={Wall clock time $[s]$},
ymode=log,
yminorticks=true,
ylabel style={font=\color{white!15!black}},
ylabel={Space-time error},
axis background/.style={fill=white},
legend pos={outer north east},
legend style={draw=none, at={(1.05,0.5)}, anchor=west},
legend cell align=left,
]

\addplot [color=myred, line width=1.0pt, mark=*]
table[]{2244.8      9.06438575741398e-06
4490.1       8.0545241147998e-06
6738.2      5.99166663289342e-06
8992.3      3.25226028404383e-06
11238.7      9.75444369854959e-07
13517.1      2.07667681389602e-07
15814.2      4.79493367725523e-07
18074.6      3.75658790059791e-07
20352.2      1.73451982258975e-07
22595.9      6.59197936263453e-08
};
\addlegendentry{Time-stepping, $p_x = 8$}

\addplot [color=myblue, line width=1.0pt, mark=*]
table[]{1403.7      9.06438575741398e-06
2812.9       8.0545241147998e-06
4214.2      5.99166663289342e-06
5629.7      3.25226028404383e-06
7057.7      9.75444369854959e-07
8563.8      2.07667681389602e-07
 10017      4.79493367725523e-07
 11439      3.75658790059791e-07
12900.6      1.73451982258975e-07
14365.9      6.59197936263453e-08
};
\addlegendentry{Time-stepping, $p_x = 16$}

\addplot [color=mygreen, line width=1.0pt, mark=*]
table[]{1721      9.06438575741398e-06
3470.8       8.0545241147998e-06
5222.3      5.99166663289342e-06
6934.1      3.25226028404383e-06
8676.8      9.75444369854959e-07
10429.2      2.07667681389602e-07
12198.7      4.79493367725523e-07
13974.2      3.75658790059791e-07
15710.3      1.73451982258975e-07
17467.9      6.59197936263453e-08
};
\addlegendentry{Time-stepping, $p_x = 32$}

\addplot [color=myorange, line width=1.0pt, mark=square]
table[]{961       8.0401524762323e-06
1770      5.97660617631621e-06
2600      3.22951121814444e-06
3390      9.99365567036723e-07
4170      2.41945224411079e-07
4940       4.6501797542855e-07
5730      3.57009766075335e-07
6500      1.53182316562767e-07
7260      5.19772977462874e-08
8030                         0
};
\addlegendentry{MGRIT, $p_x = 16$, $p_t = 5$}

\addplot [color=mygreen, line width=1.0pt, mark=square]
table[]{1170       8.0401524762323e-06
1800      5.97660617631621e-06
2440      3.22951121814444e-06
3070      9.99365567036723e-07
3690      2.41945224411079e-07
4310       4.6501797542855e-07
4940      3.57009766075335e-07
5550      1.53182316562767e-07
6160      5.19772977462874e-08
6780                         0
};
\addlegendentry{MGRIT, $p_x = 8$, $p_t = 10$}

\addplot [color=mygray, line width=1.0pt, mark=square]
table[]{1300       8.0401524762323e-06
1730      5.97660617631621e-06
2160      3.22951121814444e-06
2600      9.99365567036723e-07
3050      2.41945224411079e-07
3470       4.6501797542855e-07
3890      3.57009766075335e-07
4320      1.53182316562767e-07
4760      5.19772977462874e-08
5180                         0
};
\addlegendentry{MGRIT, $p_x = 8$, $p_t = 20$}

\end{axis}
\end{tikzpicture}    \caption{Space-time flow rate error at vein~$2$ compared to wall clock time
    for sequential time stepping and time-periodic MGRIT.
    MGRIT is consistently faster than sequential time-stepping
    and using a larger number of processors in time (denoted as $p_t$)
    instead of space (denoted as $p_x$) reduces the wall clock time of time-periodic MGRIT.}
    \label{lv-flow-application-vein-2-per-cycle-per-iteration-cost-vel_zeroIC-coarse-mesh-fig}
\end{figure}
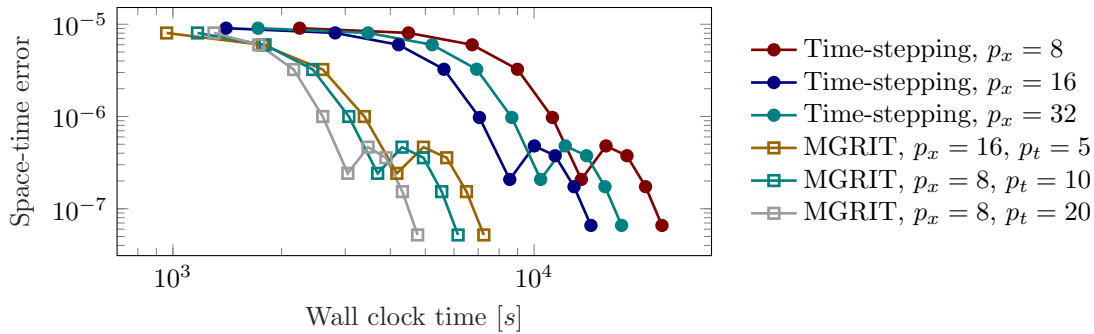
\begin{table}[ht!]
	\centering
    \begin{tabular}{ l || c | c | c | c | c | c | c  H }
                    & Number of & Number of     & \multicolumn{3}{c|}{Number of processors} & Wall clock    & Speedup \\
        Algorithm   & cycles    & iterations    & $p_x$ & $p_t$ & $p_x p_t$                 & time $[s]$    & vs.\ $7$ cycles \\
        \hline\hline
        Time-stepping   & $7$   & -     & $8$   & -     & $8$       & $15814~s$ &         &        \\
                        & $7$   & -     & $16$  & -     & $16$      & $10017~s$ & $1.58$x & $0.79$ \\
                        & $7$   & -     & $32$  & -     & $32$      & $12199~s$ & $1.30$x & $0.33$ \\
        \hline
        MGRIT           & $1$   & $6$   & $16$  & $5$   & $80$      & $4940~s$  & $3.20$x & $0.32$ \\
                        & $1$   & $6$   & $8$   & $10$  & $80$      & $4310~s$  & $3.67$x & $0.37$ \\
                        & $1$   & $6$   & $8$   & $20$  & $160$     & $3470~s$  & $4.56$x & $0.23$ \\
    \end{tabular}
    \caption{Runtimes and respective speedups for sequential time-stepping ($7$~cycles)
        and time-periodic MGRIT ($6$~iterations) with $p_x$~processors in space and $p_t$~processors in time.}
	\label{lv-flow-application-runtimes-speedup-coarse-mesh-tab}
\end{table}
\section{Discussion}
\label{discussion-sec}
In this work, a new time-periodic MGRIT algorithm was proposed
as a modification of the existing non-periodic MGRIT algorithm.
The key ingredient is that only the fine grid is made periodic,
such that the initial condition is continuously updated as an improved
approximation of the time-periodic steady-state solution at the initial / cycle time
becomes available.
This allows the algorithm to naturally solve a time-periodic space-time problem
and accelerates convergence to a time-periodic steady-state solution
across the whole temporal domain,
yielding significant runtime reductions compared to sequential time-stepping.
Existing benefits of the MGRIT algorithm (and its implementations)
remain intact, e.g., its nonintrusiveness,
the ability to solve linear and nonlinear problems,
the use of various cycling strategies and relaxation schemes,
the ability to parallelize in time, in space and in space-time, etc.

The new time-periodic MGRIT algorithm was applied to a variety of application classes
(see Section~\ref{numerical-experiments-sec}):
two-dimensional linear flow in a complex geometry,
linear and nonlinear fluid-structure interaction in two and three dimensions
and coupled nonlinear flow in a three-dimensional left atrium / left ventricle geometry.
For each of the application classes,
it was demonstrated that applying the time-periodic MGRIT algorithm to a single time period
yields the same time-periodic steady-state solution
as sequential time-stepping over multiple time periods, see Section~\ref{results-sec}.
It was repeatedly observed that running one sequential time-stepping cycle
is roughly equivalent to performing one time-periodic MGRIT iteration
with respect to convergence to a time-periodic steady-state,
which yields an approximation
of the required MGRIT iterations for the new algorithm.
That is, many researchers know from experience how many sequential time-stepping cycles
are required to obtain a time-periodic steady-state solution,
which directly translates into the number of MGRIT iterations that need to be performed.

The property that one sequential time-stepping cycle and one time-periodic MGRIT iteration
are equivalent w.r.t.\ steady-state convergence is ideal
in the sense that each MGRIT iteration can update the initial condition
at relatively cheaper cost compared to running one cycle using sequential time-stepping,
see Section~\ref{results-sec},
because the underlying multigrid algorithm achieves convergence in the interior of the temporal domain.
This yielded robust and significant speedups over sequential time-stepping
irrespective of the various employed computing resources,
ultimately enabling more detailed and more complex simulations.

The question why the time-periodic MGRIT algorithm can yield better speedups
compared to those reported in previous works
for similar PDE models~\cite{FalgoutKatzKolevSchroderWissinkYang2015,
HessenthalerNordslettenRoehrleSchroderFalgout2018}
has multiple answers.
For example, exploiting the periodicity reduces the size of the temporal domain
for time-periodic MGRIT, which generally improves the effectiveness of multigrid reduction
within the MGRIT algorithm.
Furthermore, (approximate) updates of the initial condition are communicated as early as possible,
which accelerates the propagation of information across the whole temporal domain.
Additionally, time-periodic MGRIT is jumpstarted by using an initial space-time guess
obtained from a coarse-grid solve during the setup phase,
which means that the first time-periodic MGRIT iteration
starts off from an already improved initial condition at the initial time point.

Another surprising observation is that in some cases,
time-periodic MGRIT without parallelism can outperform sequential time-stepping,
see Section~\ref{results-linear-fsi}.
Accounting for the explanations given in the previous paragraph
and taking into consideration that the time-periodic MGRIT algorithm
is applied to a much smaller temporal domain, one can appreciate
the leeway for MGRIT to outperform sequential time-stepping.
A similar result was reported in~\cite{VandewallePiessens1992},
where a waveform relaxation-based algorithm was able to outperform
the standard algorithm for an initial boundary-value problem
when no parallelism was employed.

For all considered applications, the time-periodicity was boundary driven
and the period length known a priori.
It was, however, not assessed how the time-periodic MGRIT algorithm can be applied
to time-periodic problems where the period length is only known approximately or even unknown.
This aspect is part of future research.

Similarly, the influence of the choice of (e.g., material) parameters on the convergence
and observed speedup of the time-periodic MGRIT algorithm was well beyond the scope of this work,
however, should be investigated in the future.
Related works are, e.g., \cite{SteinerRuprechtSpeckKrause2015,Gander2017,GanderLunet2018}.

All considered problems were relatively small in space,
in part to show speedups without requiring large clusters.
Spatial refinement, and larger spatial problems in general,
will help improve strong scaling
because solving a time step will become
relatively more expensive~\cite{HessenthalerNordslettenRoehrleSchroderFalgout2018}.
Therefore, communication overhead due to time parallelism
will become relatively smaller for larger spatial problems.

While some explorations with respect to analyzing and estimating the convergence rate
of \emph{time-periodic} MGRIT in the nonlinear PDE case were made
as part of developing the proposed method~\cite{Hessenthaler2020_PhD},
no mathematically rigorous a priori convergence estimates are available
and should be part of future research.
It is, however, highlighted that the available convergence bounds
for non-periodic MGRIT~\cite{DobrevKolevPeterssonSchroder2017,Southworth2019,
HessenthalerSouthworthNordslettenRoehrleFalgoutSchroder2020,SouthworthMitchellHessenthalerDanieli}
are a useful tool for improving the algorithm's performance and informing parameter choices a priori.

Traditional (non-periodic) MGRIT is not restricted to the simple two-level case
and neither is the new time-periodic MGRIT variant.
In fact, the modification in the MGRIT algorithm is limited to the fine-grid.
Therefore, the capability of MGRIT to use one or multiple coarse-grids
and thus exploit a greater level of parallelism
to further reduce the time-to-solution is maintained.
Exploring the true multilevel case, however, was beyond the scope of this work
and needs to be investigated in the future.\footnote{Preliminary
results are available in~\cite{Hessenthaler2020_PhD}.}
\section{Conclusion}\label{conclusion-sec}
In this work, we have proposed a time-periodic MGRIT algorithm
as a means to reduce the time-to-solution of numerical algorithms
by exploiting the time periodicity inherent to many applications
in science and engineering.
The time-periodic MGRIT algorithm was proven to be applicable
to a variety of linear and nonlinear single- and multiphysics problems.
It was demonstrated that using the new parallel-in-time algorithm
yields a significant runtime reduction compared to sequential time-stepping.
Robust speedups were observed across the various flow and fluid-structure interaction problems
in two and three dimensions,
for a variety of space-time discretizations and employed computing hardware.
It was further demonstrated that a time-periodic steady-state can be obtained
with respect to a given tolerance
by either running $q$ sequential time-stepping cycles or $q$ time-periodic MGRIT iterations.
Thus,
an intuition about the required number of sequential time-stepping cycles directly
translates into the number of MGRIT iterations that need to be performed to achieve a periodic steady-state.
\section{Acknowledgements}
D.N.\ would like to acknowledge funding from Engineering
and Physical Sciences Research Council (EP/N011554/1 and EP/R003866/1).

O.R.\ and A.H.\ were funded by Deutsche Forschungsgemeinschaft
(DFG, German Research Foundation) under Germany's Excellence Strategy
- EXC 2075 - 390740016.
We acknowledge the support
by the Stuttgart Center for Simulation Science (SimTech).
\section*{Additional Information}
Declarations of interest: none.\\[1ex]

\begin{spacing}{0.5}
    \tiny
    This document was prepared as an account of work sponsored
    by an agency of the United States government.
    Neither the United States government nor Lawrence Livermore National Security, LLC,
    nor any of their employees makes any warranty, expressed or implied,
    or assumes any legal liability or responsibility for the accuracy, completeness,
    or usefulness of any information, apparatus, product, or process disclosed,
    or represents that its use would not infringe privately owned rights.
    Reference herein to any specific commercial product, process,
    or service by trade name, trademark, manufacturer,
    or otherwise does not necessarily constitute or imply its endorsement, recommendation,
    or favoring by the United States government or Lawrence Livermore National Security, LLC.
    The views and opinions of authors expressed herein do not necessarily state
    or reflect those of the United States government
    or Lawrence Livermore National Security, LLC,
    and shall not be used for advertising or product endorsement purposes.
\end{spacing}
\section*{References}
\bibliography{main}
\FloatBarrier
\clearpage

\renewcommand{\thesection}{SM\arabic{section}}
\setcounter{section}{0}\setcounter{subsection}{0}\setcounter{equation}{0}
\setcounter{figure}{0}
\setcounter{table}{0}
\setcounter{page}{1}
\renewcommand{\theequation}{SM\arabic{equation}}
\renewcommand{\thefigure}{SM\arabic{figure}}
\renewcommand{\bibnumfmt}[1]{[SM#1]}
\renewcommand{\citenumfont}[1]{SM#1}

\begin{center}
        \textbf{\large
    Supplementary Material 1:\\[1ex]
    Time-periodic steady-state solution of fluid-structure interaction and cardiac
flow problems through multigrid-reduction-in-time}\\[3ex]
        Andreas Hessenthaler$^{a,b}$,
    Robert D.\ Falgout$^c$,
    Jacob B.\ Schroder$^d$,
    Adelaide de Vecchi$^e$,
    David Nordsletten$^{e,f}$,
    Oliver R\"ohrle$^{a,b}$\\[2ex]
        {\itshape \footnotesize
    ${}^a$Institute for Modelling and Simulation of Biomechanical Systems,
    University of Stuttgart, Pfaffenwaldring 5a,\\
    70569 Stuttgart, Germany\\
    ${}^b$Stuttgart Center for Simulation Technology, University of Stuttgart,
    Pfaffenwaldring 5a, 70569 Stuttgart, Germany\\
    ${}^c$Center for Applied Scientific Computing, Lawrence Livermore National Laboratory, Livermore, CA 94551\\
    ${}^d$Department of Mathematics and Statistics, University of New Mexico, Albuquerque, NM 87131\\
    ${}^e$School of Biomedical Engineering and Imaging Sciences,
    King's College London, 4th FL Rayne Institute,\\
    St Thomas Hospital, London, SE1 7EH\\
    ${}^f$Department of Biomedical Engineering and Cardiac Surgery,
    University of Michigan,
    NCRC B20, 2800 Plymouth Rd,\\
    Ann Arbor, 48109\\
    }
    \end{center}

\label{online-supplement-sec}

\section{Multigrid-reduction-in-time (MGRIT)}
\label{mgrit-time-grid-hierarchy-suppsec}

\begin{align}
    R A_0 P
                            &= R
    \begin{bmatrix}
        I \\
        -\Phi_0 & I \\
              & -\Phi_0 & I \\
              &       & \ddots & \ddots \\
              &         &       & -\Phi_0 & I \\
    \end{bmatrix}
    \begin{bmatrix}
        [I, \Phi_0, \cdots, \Phi_0^{m - 1}]^T \\
        & [I, \Phi_0, \cdots, \Phi_0^{m - 1}]^T \\
        && \ddots \\
        &&& I \\
    \end{bmatrix} \nonumber \\
        &= \begin{bmatrix}
        I \\
        & \Phi_0^{m - 1} & \Phi_0^{m - 2} & \cdots & I \\
        &                        &                                &           &   & \ddots \\
        &                        &                                &           &   &        & \Phi_0^{m - 1} & \Phi_0^{m - 2} & \cdots & I \\
    \end{bmatrix}
    \begin{bmatrix}
        I \\
        0 \\
        \vdots \\
        0 \\
        -\Phi_0^m & I \\
                  & 0 \\
                  & \vdots \\
                  & 0 \\
                  & -\Phi_0^m & I \\
                  && \ddots & \ddots \\
                  &&        & -\Phi_0^m & I \\
    \end{bmatrix} \nonumber \\
            &= \begin{bmatrix}
        I \\
        - \Phi_0^m & I \\
        & - \Phi_0^m & I \\
        && \ddots & \ddots \\
        &&& - \Phi_0^m & I \\
    \end{bmatrix}
    = R_I A_0 P
    = \begin{bmatrix}
        I \\
        & 0 & 0 & \cdots & I \\
        &   &   &        &   & \ddots \\
        &   &   &        &   &        & 0 & 0 & \cdots & 0 & I \\
    \end{bmatrix}
    A_0 P
    \label{RlAlPl-exact-eqn}
\end{align}

\section{Flow in a left atrium / left ventricle geometry: Weak formulation}
\label{lv-flow-application-weak-formulation-suppsec}
Finite element discretizations were constructed using $\mathbb{P}^1 - \mathbb{P}^1$~elements
for fluid velocity and pressure and $\mathbb{P}^1$~elements for the Lagrange multipliers
on the coupling domain,
resulting in $55842$~degrees-of-freedom (DOFs).
The discrete solution at each time step~$n$ can then be written as follows:

Find~$\boldsymbol{s}^n
:= ( \boldsymbol{v}_A^n, \boldsymbol{v}_V^n,
\boldsymbol{\lambda}_A^n, \boldsymbol{\lambda}_V^n,
p_A^n, p_V^n )
\in \boldsymbol{\mathcal{S}}_D^h
:= \boldsymbol{\mathcal{V}}_D^h \times \boldsymbol{\mathcal{U}}_D^h
\times \boldsymbol{\mathcal{M}}_0^h \times \boldsymbol{\mathcal{N}}_0^h
\times \mathcal{W}_A^h \times \mathcal{W}_V^h$,
such that for every~$d
:= ( \boldsymbol{y}_A, \boldsymbol{y}_V, \boldsymbol{\varphi}_A, \boldsymbol{\varphi}_V, q_A, q_V )
\in \boldsymbol{\mathcal{S}}_0^h
:= \boldsymbol{\mathcal{V}}_0^h \times \boldsymbol{\mathcal{U}}_0^h
\times \boldsymbol{\mathcal{M}}_0^h \times \boldsymbol{\mathcal{N}}_0^h
\times \mathcal{W}_A^h \times \mathcal{W}_V^h$:
\begin{alignat}{4}
    &R \left( \boldsymbol{s}^n, \boldsymbol{s}^{n-1},
        \boldsymbol{w}_A^n, \boldsymbol{w}_V^n;
        \boldsymbol{d} \right) \nonumber \\
        &\quad:= \int_{\Omega_A^n} \left[ \frac{\boldsymbol{v}_A^n - \boldsymbol{v}_A^n}{\delta_0}
        + \left( \boldsymbol{v}_A^n - \boldsymbol{w}_A^n \right) \cdot \nabla_{\boldsymbol{x}} \boldsymbol{v}_A^n \right] \cdot \boldsymbol{y}_A~d\boldsymbol{x} \nonumber \\
    &\quad\qquad + \int_{\Omega_A^n} \boldsymbol{\sigma}_A^n : \nabla_{\boldsymbol{x}} \boldsymbol{y}_A + \boldsymbol{\varphi}_A \nabla_{\boldsymbol{x}} \cdot \boldsymbol{v}_A^n~d\boldsymbol{x} \nonumber \\
    &\quad\qquad - \int_{\Gamma_{Vn}^n} \boldsymbol{t}_A^n \cdot \boldsymbol{y}_A~d\boldsymbol{x} \nonumber \\
        &\quad\qquad +\int_{\Omega_V^n} \left[ \frac{\boldsymbol{v}_V^n - \boldsymbol{v}_V^n}{\delta_0}
        + \left( \boldsymbol{v}_V^n - \boldsymbol{w}_V^n \right) \cdot \nabla_{\boldsymbol{x}} \boldsymbol{v}_V^n \right] \cdot \boldsymbol{y}_V~d\boldsymbol{x} \nonumber \\
    &\quad\qquad + \int_{\Omega_V^n} \boldsymbol{\sigma}_V^n : \nabla_{\boldsymbol{x}} \boldsymbol{y}_V + \boldsymbol{\varphi}_V \nabla_{\boldsymbol{x}} \cdot \boldsymbol{v}_A^n~d\boldsymbol{x} \nonumber \\
    &\quad\qquad - \alpha \int_{\Gamma_{AV}^n} \boldsymbol{t}_V^n \cdot \boldsymbol{y}_V~d\boldsymbol{x} \nonumber \\
        &\quad\qquad + \int_{\Omega_{LM}^0} \boldsymbol{\varphi}_A \cdot \left( \boldsymbol{v}_A^n - \alpha \boldsymbol{w}_A^n - (1 - \alpha) \boldsymbol{v}_V^n \right)~d\boldsymbol{x} \nonumber \\
        &\quad\qquad + \int_{\Omega_{LM}^0} \alpha \boldsymbol{\varphi}_V \left( \boldsymbol{v}_A^n - \boldsymbol{v}_V^n \right)~d\boldsymbol{x} \nonumber \\
        &\quad\qquad + \int_{\Omega_{LM}^0} (1 - \alpha) \left( \boldsymbol{\lambda}_A \cdot \boldsymbol{y}_A - \boldsymbol{\lambda}_V \cdot \boldsymbol{y}_V \right)~d\boldsymbol{x} \nonumber \\
        &\quad\qquad + \zeta_1 \int_{\Omega_A^n} \left[
        \left( \boldsymbol{v}_A^n - \boldsymbol{w}_A^n \right) \cdot \nabla_{\boldsymbol{x}} \boldsymbol{v}_A^n
    \right] \cdot \left[
        \boldsymbol{v}_A^n \cdot \nabla_{\boldsymbol{x}} \boldsymbol{y}_A + \nabla_{\boldsymbol{x}} q_A
    \right]~d\boldsymbol{x} \nonumber \\
    &\quad\qquad + \zeta_2 \int_{\Omega_A^n} \left( \nabla_{\boldsymbol{x}} \cdot \boldsymbol{v}_A \right)
    \cdot \left( \nabla_{\boldsymbol{x}} \cdot \boldsymbol{y}_A \right)~d\boldsymbol{x} \nonumber \\
    &\quad\qquad + \zeta_3 \int_{\Omega_A^n} \left( \nabla_{\boldsymbol{x}} p_A^n \right)
    \cdot \left( \boldsymbol{v}_A^n \cdot \nabla_{\boldsymbol{x}} \boldsymbol{y}_A + \nabla_{\boldsymbol{x}} q_A \right)~d\boldsymbol{x} \nonumber \\
        &\quad\qquad + \zeta_1 \int_{\Omega_V^n} \left[
        \left( \boldsymbol{v}_V^n - \boldsymbol{w}_V^n \right) \cdot \nabla_{\boldsymbol{x}} \boldsymbol{v}_V^n
    \right] \cdot \left[
        \boldsymbol{v}_V^n \cdot \nabla_{\boldsymbol{x}} \boldsymbol{y}_V + \nabla_{\boldsymbol{x}} q_V
    \right]~d\boldsymbol{x} \nonumber \\
    &\quad\qquad + \zeta_2 \int_{\Omega_V^n} \left( \nabla_{\boldsymbol{x}} \cdot \boldsymbol{v}_V \right)
    \cdot \left( \nabla_{\boldsymbol{x}} \cdot \boldsymbol{y}_V \right)~d\boldsymbol{x} \nonumber \\
    &\quad\qquad + \zeta_3 \int_{\Omega_V^n} \left( \nabla_{\boldsymbol{x}} p_V^n \right)
    \cdot \left( \boldsymbol{v}_V^n \cdot \nabla_{\boldsymbol{x}} \boldsymbol{y}_V + \nabla_{\boldsymbol{x}} q_V \right)~d\boldsymbol{x}
    \label{lv-weak-form-cG1cG1-eqn}
\end{alignat}
where the stabilization terms with parameters
$\zeta_1 = \rho \delta_x^{max} / v_f^{max}$,
$\zeta_2 = \rho \delta_x^{max} v_f^{max}$
and $\zeta_3 = \delta_x^{max} / v_f^{max}$
with $v_f^{max} = 100~cm / s$
are added according to the cG(1)cG(1) scheme~\cite{HoffmanJanssonDeabreu2011}.

Note, that the domain velocity $\boldsymbol{w}_A^n$ and $\boldsymbol{w}_V^n$
is provided from the CT data, see Section~\ref{lv-flow-application-model-problem-sec}.

The definitions of the function spaces are:
\begin{align}
	S^1 \left( \Omega^0_i \right) &= \{f:\Omega^0_i \rightarrow \mathbb{R}~|~
	f\in\mathcal{C}^0(\bar{\Omega}^0_i),~
	f|_{\tau_e} \in \mathbb{P}^1 (\tau_e),~\forall~\tau_e \subset \mathcal{T}^h_i \},
\end{align}
which represent the first-order piecewise continuous polynomial spaces
defined on $\Omega^0_i$. Consequently, we can define:
\begin{alignat*}{6}
	\boldsymbol{\mathcal{V}}^h &=  \left[S^1 ( \Omega^0_A )\right]^3, &\qquad&
	\boldsymbol{\mathcal{U}}^h &&= \left[S^1 ( \Omega^0_V )\right]^3, \\
	\boldsymbol{\mathcal{M}}^h &=  \left[S^1 (\Gamma_{CA} ) \right]^3, &\qquad&
	\boldsymbol{\mathcal{N}}^h &&= \left[S^1 (\Gamma_{VA} ) \right]^3, \\
	\mathcal{W}_A^h            &=  S^1 ( \Omega^0_A ), &\qquad&
	\mathcal{W}_V^h            &&= S^1 ( \Omega^0_V ).
\end{alignat*}

Further restrictions are applied on the respective spaces for the atrium and ventricle
in order to incorporate the Dirichlet and homogeneous boundary conditions:
\begin{alignat}{4}
	\boldsymbol{\mathcal{V}}^h_D & = \{\boldsymbol{v} \in \boldsymbol{\mathcal{V}}^h
    ~|~&&\boldsymbol{v} = \boldsymbol{w}_A \text{ on } \Gamma_{WA}^h \}, \\
	\boldsymbol{\mathcal{V}}^h_0 & = \{\boldsymbol{v} \in \boldsymbol{\mathcal{V}}^h
    ~|~&&\boldsymbol{v} = \boldsymbol{0}   \text{ on } \Gamma_{WA}^h \}, \\
	\boldsymbol{\mathcal{U}}^h_D & = \{\boldsymbol{v} \in \boldsymbol{\mathcal{U}}^h
    ~|~&&\boldsymbol{v} = \boldsymbol{w}_V \text{ on } \Gamma_{WV}^h, \nonumber \\
    &&&\boldsymbol{v} = \boldsymbol{w}_V \text{ on } \Gamma_{AV}^h \text{ for } t \in (0, 0.37], \nonumber \\
    &&&\boldsymbol{v} = \boldsymbol{w}_V \text{ on } \Gamma_{CV}^h \setminus \Gamma_{CVo}^h \text{ for } t \in (0.37, 0.8] \}, \\
	\boldsymbol{\mathcal{U}}^h_0 & = \{\boldsymbol{v} \in \boldsymbol{\mathcal{U}}^h
    ~|~&&\boldsymbol{v} = \boldsymbol{0} \text{ on } \Gamma_{WV}^h, \nonumber \\
    &&&\boldsymbol{v} = \boldsymbol{0}    \text{ on } \Gamma_{AV}^h \text{ for } t \in (0, 0.37], \nonumber \\
    &&&\boldsymbol{v} = \boldsymbol{0}    \text{ on } \Gamma_{CV}^h \setminus \Gamma_{CVo}^h \text{ for } t \in (0.37, 0.8] \},
\end{alignat}
and similarly for the coupling domains,
\begin{alignat}{4}
	\boldsymbol{\mathcal{M}}^h_0 & = \{\boldsymbol{\lambda} \in \gamma_{\Gamma_{CA}} \boldsymbol{\mathcal{V}}^h_0 \}, \\
	\boldsymbol{\mathcal{N}}^h_0 & = \{\boldsymbol{\lambda} \in \gamma_{\Gamma_{CV}} \boldsymbol{\mathcal{U}}^h_0 \},
\end{alignat}
where $\gamma_{\Gamma_{CA}}$ and $\gamma_{\Gamma_{CV}}$ are the trace operators
on $\Gamma_{CA}$ and $\Gamma_{CV}$, respectively.

\section{Tables}

\FloatBarrier

\subsection{Parallel machines}\label{HPC-machines-sec}
\begin{table}[ht!]
    \centering
\begin{tabular}{ l | l }
        Name                        & LEAD \\
        \hline\hline
        Operating System            & Scientific Linux 6.4 (Carbon) \\
        Compute Nodes               & 1x AMD Opteron 6373 G34 \\
                                    & \qquad 2.3 GHz, 4x 16 cores, Abu Dhabi \\
                                    & 7x AMD Opteron 6328 G34 \\
                                    & \qquad 3.2 GHz, 4x 8 cores, Abu Dhabi \\
        Memory per Node             & 256 GB \\
        Disks per Node              & 4 x 128 GB Samsung SSD \\
        Network                     & 2 x 10 GBit Intel 82599ES Ethernet Dual SFP+ \\
                                    & 1 x Melanox MCX353A-QCBT, \\
                                    & Infiniband single-port QSFP, IB 40 GB/s \\
                                    & 2 x 1 GBit \\
        Total Number of Processors  & 288 \\
    \end{tabular}
    \caption{Hardware and software specifications for LEAD cluster.}
\end{table}
\begin{table}[ht!]
    \centering
    \begin{tabular}{ l | l }
        Name                        & TheoSim \\
        \hline\hline
        Operating System            & Scientific Linux 6.10 (Carbon) \\
        Compute Nodes               & 52x Intel(R) Xeon(R) E5-2650 v2 \\
                                    & \qquad 2.8 GHz, 2x 10 cores, Ivy Bridge \\
                                    & \qquad Available for project: 18 nodes \\
        Memory per Node             & 256 GB \\
        Disks per Node              & 4 x 500 GB Samsung SSD \\
        Network                     & 1 x QDR-Infiniband Interconnect (40 GBit) \\
                                    & 1 x 1 GBit Ethernet \\
        Total Number of Processors  & 1040 \\
                                    & \qquad Available for project: 360 \\
    \end{tabular}
    \caption{Hardware and software specifications for TheoSim cluster.}
\end{table}
\begin{table}[ht!]
    \centering
    \begin{tabular}{ l | l }
        Name                        & ORCA \\
        \hline\hline
        Operating System            & Ubuntu 16.04 \\
        Compute Nodes               & 1x AMD Ryzen Threadripper 2990WX \\
                                    & \qquad 3.0 GHz, 32 cores \\
        Memory per Node             & 128 GB \\
        Disks per Node              & 3 x 1 TB Samsung NVME SSD \\
        Total Number of Processors  & 32 \\
    \end{tabular}
    \caption{Hardware and software specifications for ORCA cluster.}
\end{table}
\begin{table}[ht!]
    \centering
    \begin{tabular}{ l | l }
        Name                        & TOM \\
        \hline\hline
        Operating System            & Suse SLES 11 SP1 \\
        Compute Nodes               & 76x Intel(R) Xeon(R) E7-8837 \\
                                    & \qquad 2.66GHz, 8 cores, Westmere EX \\
        Memory per Node             & 64 GB \\
        Network                     & NUMAlink 5 Interconnect \\
                Total Number of Processors  & 608 \\
    \end{tabular}
    \caption{Hardware and software specifications for TOM cluster.}
\end{table}

\FloatBarrier

\FloatBarrier
\begin{table}
	\centering
    \begin{tabular}{ l || c  c  c | c  c | c  c }
                                            &       &       &           & \multicolumn{2}{c|}{TheoSim}  & \multicolumn{2}{c}{LEAD}  \\
        Algorithm                           & $p_x$ & $p_t$ & $p_x p_t$ & Wall clock        & Speedup   & Wall clock    & Speedup   \\
                                            &       &       &           & time $[s]$        &           & time $[s]$    &           \\
        \hline\hline
        Sequential time-stepping            & $1$   & $1$   & $1$       & $332$             & $1.00$x   & $523$     &           \\
                                            & $8$   & $1$   & $8$       & $64$              & $5.19$x   & $83$      & $6.30$x   \\
                                            & $16$  & $1$   & $16$      & $43$              & $7.72$x   & $56$      & $9.34$x   \\
                                            & $32$  & $1$   & $32$      & $260$             & $1.28$x   &           &           \\
        \hline
        MGRIT, F-relaxation, $m = 2$      & $16$  & $1$   & $16$      & $323$             & $1.03$x   & $446$     & $1.17$x   \\
\end{tabular}
    \caption{Runtimes and respective speedups for sequential time-stepping ($10$~cycles)
        and time-periodic MGRIT ($q$~iterations, such that IC tol $10^{-1}$ is satisfied)
        with $p_x$~processors in space and $p_t$~processors in time.}
	\label{stenosed-valve-runtimes-speedup-LEAD-TheoSim-tab}
\end{table}
\FloatBarrier

\FloatBarrier
\begin{table}[ht!]
    \centering
    \begin{tabular}{ c | c || c | c || c | c | c | c }
        \multicolumn{2}{ c || }{Sequential time-stepping} & \multicolumn{2}{c || }{MGRIT} & \multicolumn{4}{c}{Speedup} \\
                \#cycles & Error & \#iterations
        & Error & $1$ proc & $4$ procs & $16$ procs & $32$ procs \\
                \hline\hline
                $2$ & $1.78 \cdot 10^{-3}$
        & $3$ & $1.82 \cdot 10^{-3}$ & $0.22$x & $1.10$x & $2.96$x& $3.08$x \\
                $4$ & $1.11 \cdot 10^{-3}$
        & $4$ & $1.08 \cdot 10^{-3}$ & $0.34$x & $1.72$x & $4.90$x& $5.08$x \\
                $7$ & $8.15 \cdot 10^{-4}$
        & $6$ & $8.18 \cdot 10^{-4}$ & $0.41$x & $2.03$x & $6.20$x& $6.64$x \\
                && $9$ & $7.24 \cdot 10^{-4}$ & $0.28$x & $1.40$x & $4.49$x& $4.91$x \\
            \end{tabular}
    \caption{Coarse mesh: Speedup depending on required accuracy and number of processors
        for two-level MGRIT with FCF-relaxation and temporal coarsening factor $m = 8$.
        For a graphical presentation,
        see Figure~\ref{3D-nonlinear-tf-ts-mgrit-fsi-per-cycle-per-iteration-cost-vel_zeroIC-coarse-mesh-fig}.}
    \label{3D-nonlinear-tf-ts-mgrit-fsi-nl2-cf08-1FCF-speedup-coarse-mesh-tab}
\end{table}
\FloatBarrier
\FloatBarrier
\begin{table}[ht!]
    \centering
    \begin{tabular}{ c | c || c | c || c | c | c }
        \multicolumn{2}{ c || }{Sequential time-stepping} & \multicolumn{2}{c || }{MGRIT} & \multicolumn{3}{c}{Speedup} \\
                \#cycles & Error & \#iterations
        & Error & $16$ procs & $32$ procs & $64$ procs \\
                \hline\hline
                $3$ & $1.04 \cdot 10^{-3}$
        & $2$ & $9.98 \cdot 10^{-4}$ & $6.60$x & $6.52$x & $8.22$x \\
                $5$ & $4.65 \cdot 10^{-4}$
        & $4$ & $4.17 \cdot 10^{-4}$ & $7.19$x & $7.73$x & $9.60$x \\
                $6$ & $3.93 \cdot 10^{-4}$
        & $6$ & $3.66 \cdot 10^{-4}$ & $6.40$x & $7.20$x & $9.05$x \\
                && $9$ & $2.00 \cdot 10^{-4}$ & $4.60$x & $5.41$x & $6.88$x \\
            \end{tabular}
    \caption{Fine mesh: Speedup depending on required accuracy and number of processors
        for two-level MGRIT with FCF-relaxation and temporal coarsening factor $m = 8$.}
    \label{3D-nonlinear-tf-ts-mgrit-fsi-nl2-cf08-1FCF-speedup-medium-mesh-tab}
\end{table}
\FloatBarrier

\FloatBarrier
\begin{figure}[ht!]
    \centering
    \begin{subfigure}[b]{0.475\linewidth}
        \centering
        \setlength{\figurewidth}{0.85\textwidth}
        \setlength{\figureheight}{0.45\textwidth}
                        %
        \caption{A time-periodic steady-state is achieved by running multiple cycles using sequential time stepping.}
        \label{lv-flow-application-la-pv1-pv2-pv3-pv4-flow-rate-sequential-to-cycle-10-vel_zeroIC-coarse-mesh-suppfig}
    \end{subfigure}\hfill
    \begin{subfigure}[b]{0.475\linewidth}
        \centering
        \setlength{\figurewidth}{0.85\textwidth}
        \setlength{\figureheight}{0.45\textwidth}
                        %
        \caption{A time-periodic steady-state is achieved by running multiple iterations using time-periodic MGRIT.}
        \label{lv-flow-application-la-pv1-pv2-pv3-pv4-flow-rate-pint-to-iter-10-vel_zeroIC-coarse-mesh-suppfig}
    \end{subfigure}
    \caption{Flow rates at pulmonary veins $1 - 4$ (top to bottom) for sequential time-stepping (left)
        and time-periodic MGRIT (right). Note, a negative flow rate corresponds to inflow.}
    \label{lv-flow-application-la-pv1-pv2-pv3-pv4-flow-rate-pint-seq-to-10-vel_zeroIC-coarse-mesh-suppfig}
\end{figure}
\FloatBarrier
\end{document}